\documentclass[twocolumn,twocolappendix]{aastex63}

\usepackage[figuresright]{rotating}
\usepackage{amsmath}
\usepackage{amsfonts}
\usepackage{graphicx}
\usepackage{rotating}
\usepackage{natbib}
\usepackage{txfonts}
\usepackage{color}
\usepackage{wrapfig}
\usepackage{amssymb}
\usepackage{color}
\usepackage{xcolor}
\usepackage{ulem}

\shorttitle{Fate of Phosphorus bearing species in the ISM}
\shortauthors{Sil et al.}

\begin{document}
\title{Chemical complexity of phosphorous bearing species in various regions of the Interstellar medium}
\author[0000-0001-5720-6294]{Milan Sil}
 \email{milansil93@gmail.com}
 \affiliation{Indian Centre for Space Physics, 43 Chalantika, Garia Station Road, Kolkata 700084, India}
  \author{Satyam Srivastav}
 \affiliation{Department of Physics, Institute of Science, Banaras Hindu University, Varanasi, 221005, India}
 \author[0000-0002-5224-3026]{Bratati Bhat}
 \author[0000-0002-7657-1243]{Suman Kumar Mondal}
  \affiliation{Indian Centre for Space Physics, 43 Chalantika, Garia Station Road, Kolkata 700084, India}
 \author[0000-0003-1602-6849]{Prasanta Gorai}
\affiliation{Department of Space, Earth \& Environment, Chalmers University of Technology, SE-412 96 Gothenburg, Sweden}
\affiliation{Indian Centre for Space Physics, 43 Chalantika, Garia Station Road, Kolkata 700084, India}
\author[0000-0003-1745-9718]{Rana Ghosh}
 \affiliation{Indian Centre for Space Physics, 43 Chalantika, Garia Station Road, Kolkata 700084, India}
 \author{Takashi Shimonishi}
\affiliation{Center for Transdisciplinary Research,
Niigata University, Nishi-ku, Niigata 950-2181, Japan}
\affiliation{Environmental Science Program, Department of Science, Faculty
of Science, Niigata University, Nishi-ku, Niigata 950-2181, Japan}
\author[0000-0002-0193-1136]{Sandip K. Chakrabarti}
 \affiliation{Indian Centre for Space Physics, 43 Chalantika, Garia Station Road, Kolkata 700084, India}
\author[0000-0002-2833-0357]{Bhalamurugan Sivaraman}
\affiliation{Physical Research Laboratory, Navrangpura, Ahmedabad 380009, India}
\author{Amit Pathak}
\affiliation{Department of Physics, Institute of Science, Banaras Hindu University, Varanasi, 221005, India}
\author{Naoki Nakatani}
\affiliation{Institute for Catalysis, Hokkaido University, N21W10 Kita-ku, Sapporo, Hokkaido 001-0021, Japan \&
Department of Chemistry, Graduate School of Science and Engineering, Tokyo Metropolitan University, 1-1 Minami-Osawa, Hachioji Tokyo 192-0397, Japan}
\author{Kenji Furuya}
\affiliation{Center for Computational Sciences, University of Tsukuba, Tsukuba, 305-8577, Japan \& National Astronomical Observatory of Japan, Tokyo 181-8588, Japan}
 \author[0000-0003-4615-602X]{Ankan Das}
 \email{ankan.das@gmail.com}
 \affiliation{Indian Centre for Space Physics, 43 Chalantika, Garia Station Road, Kolkata 700084, India}

\begin{abstract}
Phosphorus related species are not known to be as omnipresent in space as hydrogen, carbon, nitrogen, oxygen, and sulfur-bearing species. Astronomers spotted very few P-bearing molecules in the interstellar medium and circumstellar envelopes. Limited discovery of the P-bearing species imposes severe constraints in modeling the P-chemistry. In this paper, we carry out extensive chemical models to follow the fate of P-bearing species in diffuse clouds, photon-dominated or photodissociation regions (PDRs), and hot cores/corinos. We notice a curious correlation between the abundances of PO and PN and atomic nitrogen. Since N atoms are comparatively abundant in diffuse clouds and PDRs than in the hot core/corino region, PO/PN reflects $<1$ in diffuse clouds, $<<1$ in PDRs, and $>1$ in the late warm-up evolutionary phase of the hot core/corino regions. During the end of the post-warm-up phase, we obtain PO/PN $>1$ for hot core and $<1$ for its low mass analog. We employ a radiative transfer model to investigate the transitions of some of the P-bearing species in diffuse cloud and hot core regions and estimate the line profiles. Our study estimates the required integration time to observe these transitions with ground-based and space-based telescopes. We also carry out quantum chemical computation of the infrared features of PH$_3$ along with various impurities. We notice that SO$_2$ overlaps with the PH$_3$ bending-scissoring modes around $\sim (1000-1100)$ cm$^{-1}$. We also find that the presence of CO$_2$ can strongly influence the intensity of the stretching modes around $\sim 2400$ cm$^{-1}$ of PH$_3$.
\end{abstract}


\section{Introduction} \label{sec:intro}
Phosphorus (P) and its compounds play a crucial role in the chemical evolution in galaxies. Phosphorus is an essential element of life and is one of the main biogenic elements. Its origin in the terrestrial system is yet to be fully understood. The Atacama Large Millimeter/submillimeter Array (ALMA) and European Space Agency Probe Rosetta suggest that P-related species might have traveled from star-forming regions to the Earth \citep[e.g.][]{altw16,rivi20} through comets.
The interstellar chemistry of P-bearing molecules has significant astrophysical relevance, and very little has been revealed so far. P-bearing compounds are the major components of any living system, where they carry out numerous biochemical functions. P-bearing molecules play a significant role in forming large bio-molecules or living organisms; they store and transmit the genetic information in nucleic acids and work in nucleotides as precursors in the synthesis of RNA and DNA \citep{maci97}. Moreover, these molecules are the essential components of phospholipids \citep[main characteristic features of cellular membranes,][]{maci05}.

Phosphorus is relatively rare in the interstellar medium (ISM). However, it is ubiquitous in various meteorites \citep{jaro90,pase19,lodd03}. On average, P is the thirteenth most abundant element in a typical meteoritic material and the eleventh most abundant element in the Earth's crust \citep{maci05}.

\cite{jura78} identified P$^+$ with a cosmic abundance of $\sim 2\times10^{-7}$ in hot regions ($\sim 1200$ K). 
P-bearing molecules such as PN, PO, HCP, CP, CCP, PH$_3$ have been observed in circumstellar envelopes around evolved stars \citep{guel90,tene07,agun07,agun08,tene08,half08,mila08,debe13,agun14,ziur18}. PN was detected in several star-forming regions \citep{turn87,ziur87,turn90,caux11,yama11,font16,mini18,font19}, and it remained the only P-containing species identified in dense ISM for many years \citep{ziur87,turn90,font16}. \cite{rivi16} reported the first detection of PO toward two massive star-forming regions W51 1e/2e and W3(OH), along with PN by using the IRAM 30m telescope. Thereafter, both PO and PN were simultaneously observed in several low and high mass star-forming regions and Galactic Centre \citep{lefl16,rivi18,rivi20,berg19,bern21}.

Theoretical investigation on P-chemistry modeling has been extensively reported in several studies \citep{mill91,char94,aota12,lefl16,rivi16,jime18}. However, the P-chemistry of the dense cloud region is not well constrained. The main uncertainty lies in the depletion factor of the initial elemental abundance of P.
The degree of complexity of gas-phase abundance within the gas and the exact depletion of different elements onto the grains are still uncertain \citep{jenk09}. Very recently, \cite{nguy20}
discovered the chemical desorption of phosphine. This kind of study is indeed very crucial in constraining the modeling parameters.

Phosphine (PH$_3$) is the phosphorus cousin of ammonia (NH$_3$) and is a relatively stable molecule that could hold a significant fraction of P in various astronomical environments. Scientists consider PH$_3$ to be a biosignature \citep{sous20}. PH$_3$ was identified in the planets of our solar system containing a reducing atmosphere. The Voyager data confirmed PH$_3$ within Jupiter and Saturn with volume mixing ratios of 0.6 and 2 ppm, respectively \citep{maci05}. It further suggested that the P/H ratio shows harmony with the solar value. \cite{flet09} derived the global distribution of PH$_3$ on Jupiter and Saturn using 2.5 cm$^{-1}$ spectral resolution of Cassini/CIRS observations. PH$_3$ could be produced deep inside the reducing atmospheres of giant planets \citep{breg75,tarr92} at high temperatures and pressures, and dredged upwards by the convection \citep{noll97,viss06}. Very surprisingly, in Venus, there is no such reducing atmosphere, but $\sim 20$ ppb of PH$_3$ was inferred \citep{grea20} in the deck of venus atmosphere. 
They could not account for such a high amount of PH$_3$ with the steady-state chemical models, including the photochemical pathways. They also explored the other abiotic routes for explaining this high abundance. But none of them seems to be suitable.  They speculated some unknown photo or geo-chemical origin of it. More far-reaching, its high abundance by some of the biological means could not be ruled out. \cite{bain20} also could not explain the presence of PH$_3$ in Venus's clouds by any abiotic mechanism based on their state-of-the-art understanding.
This discovery opened up a series of debates. Very recently, \cite{vill20} question about the analysis and interpretation of the spectroscopic data used in \cite{grea20}. These authors claimed that there is no PH$_3$ in Venus's atmosphere. \cite{snell20} also found no statistical evidence for
PH$_3$ in the atmosphere of Venus.

Among the simple P-bearing species, PH$_3$ was tentatively identified (J = 1 $\rightarrow$ 0, $266.9$ GHz) in the envelope of the carbon-rich stars IRC +10216 and CRL 2688 \citep{tene08,agun08}. Later, the presence of PH$_3$ was confirmed by \cite{agun14}. They observed the J = 2 $\rightarrow$ 1 rotational transition of PH$_3$ (at $534$ GHz) in IRC +10216 using the HIFI instrument on board Herschel. They predicted a very high abundance of PH$_3$ in this region. Suppose the gas-phase reactions are responsible for such a high abundance of PH$_3$. In that case, it should occur through a barrierless reaction or with a reaction with a little barrier at low temperatures.
Here, we employ various state-of-the-art chemical models to understand the formation of PH$_3$ under different interstellar conditions like diffuse cloud, interstellar photon dominated regions or photodissociation regions (PDRs), and hot core regions.

Recently, \cite{chan20} attempted to observe HCP (2–1), CP (2–1), PN (2–1), and PO (2–1) with the IRAM 30m telescope along the line of sight to the compact extra-galactic quasar B0355+508. They were unable to detect these transitions along this line of sight. But based on their observations, they predicted 3$\sigma$ upper limits of these transitions. However, they successfully identified HNC (1–0), CN (1–0), and C$^{34}$S (2–1) in absorption and $^{13}$CO (1–0) in emission along the same line of sight.

We organize this paper as follows. In Section \ref{sec:chem_network}, we prepare a currently available chemical network for the P-bearing species, and in Section \ref{sec:bind_energy}, we present their various kinetic information. Section \ref{sec:chem_model} is devoted to the chemical model explaining P-chemistry in different parts of the ISM. In Section \ref{sec:RATRAN}, we describe our results obtained with the radiative transfer model for the diffuse and hot core region. In Section \ref{sec:IR_spectra}, we explain the infrared spectrum of PH$_3$ ice under various circumstances, and finally, in Section \ref{sec:Conclusion}, we draw our conclusion.

\section{The chemical network of phosphorus} \label{sec:chem_network}
We prepare an extensive network to study the chemistry of the P-bearing species. 
Since nitrogen (N) and P have five electrons in the valance shell, P-chemistry is sometimes considered analogous to N-chemistry. For example, successive hydrogenation of N yields NH$_3$, whereas, for P, it yields PH$_3$. However, the presence of NH$_3$ is ubiquitous in the ISM \citep{cheu68,wils79,keen83,maue88}, the presence of PH$_3$ is not so universal \citep{thor83}. On the other hand, P's chemistry can notably differ from the N-related chemistry under the physical conditions prevailing in the different star-forming regions.
In this paper, we prepare an extensive network of P by following the chemical pathways explained in \cite{thor84,adam90,mill91,font16,jime18,sous20,rivi16,chan20}.
\cite{char94} and \cite{anic93} differentiated the formation/destruction mechanism of PH$_n$ (n=1, 2, 3) and their cationic species. \cite{thor84} presented the reaction pathways for P, PO, P$^+$, PO$^+$, PH$^+$, HPO$^+$ and H$_2$PO$^+$. Furthermore, in a recent study, \cite{chan20} extends the gas-phase chemical network of PN and PO in accordance with \cite{mill87} \& \cite{jime18}.  
We use the kinetic data of chemical reactions from the KInetic Database for Astrochemistry \citep{wake15} (hereafter, KIDA) and the UMIST Database for Astrochemistry \citep[UDfA,][]{mcel13}.
In the Appendix Table \ref{tab:reaction_path}, we show all the reactions considered in our P-network. 

\begin{deluxetable*}{clcccccccccc}
\tablecaption{Calculated binding energy (with MP2/aug-cc-pVDZ) and enthalpy of formation (with DFT-B3LYP/6-31G(d,p)) of P-bearing species. \label{tab:binding}}
\tablewidth{0pt}
\tabletypesize{\scriptsize} 
\tablehead{
\colhead{Serial} & \colhead{Species} & \colhead{Astronomical}                               & \colhead{Ground} & \multicolumn{6}{c}{Binding Energy [Kelvin]} & \multicolumn{2}{c}{Enthalpy of formation [kJ/mol]} \\
\colhead{No.} & \colhead{ } & \colhead{status} & \colhead{state} & \colhead{CO monomer} & \colhead{CH$_3$OH monomer} & \colhead{H$_2$O monomer} & \colhead{H$_2$O tetramer} & \colhead{H$_2$ monomer} & \colhead{Available$^d$} & \colhead{Our calculated} & \colhead{Available$^d$}
}
\startdata
1. & P & \nodata & quartet & 170 & 442 & 270 & 616$^c$ & 107 & 1100 & 315.557$^a$, 310.202$^b$ & 315.663$^a$, 316.5$^b$ \\
2. & P$_2$ & not observed & singlet & 378 & 904 & 494 & 1671 & 223 & \nodata &182.260$^a$, 180.434$^b$ & 145.8$^a$ \\
3. & PN & observed & singlet & 324 &2560  & 2326 & 2838 & 399 & 1900 & 218.985$^a$, 217.974$^b$ & 172.48$^a$, 171.487$^b$ \\
4. & PO & observed & doublet & 703 & 4334 & 2818 & 4600 & 508 & 1900 & 13.483$^a$, 12.490$^b$ & -27.548$^a$, -27.344$^b$ \\
5. & HCP & observed & singlet & 572 & 2122 & 1723 & 2468 & 132 & 2350 & 251.466$^a$, 250.054$^b$ & 217.79$^a$, 216.363$^b$ \\
6. & CP & observed & doublet & 300 & 1335 & 1126 & 1699$^c$ & 165 & 1900 & 538.388$^a$, 540.696$^b$ & 517.86$^a$, 520.162$^b$ \\
7. & CCP & observed & doublet & 2181 & 3900 & 2701 & 2868 & 396& 4300 & 660.424$^a$, 664.413$^b$ & \nodata \\
8. & HPO & not observed & singlet & 602 & 4097 & 2838 & 5434 & 521 & 2350 & -38.035$^a$, -41.915$^b$ & -89.9$^a$, -93.7$^b$ \\
9. & PH & not observed & triplet & 270  & 780 & 491 & 944$^c$ & 134& 1550 & 228.785$^a$, 227.878$^b$ & 231.698$^a$, 230.752$^b$ \\
10. & PH$_2$ & not observed & doublet & 285 & 851 & 965 & 1226$^c$ &  138 & 2000 & 128.872$^a$, 125.036$^b$ & 139.333$^a$, 135.474$^b$ \\
11. & PH$_3$ & observed & singlet & 716 & 952 & 606 & 1672 & 545 & \nodata & 12.934$^a$, 5.006$^b$ & 13.4$^a$ \\
\enddata
\tablecomments{
$^a$Enthalpy of formation at T = 0 K and 1 atmosphere pressure, $^b$Enthalpy of formation at T = 298 K and 1 atmosphere pressure, $^c$\cite{das18}, $^d$KIDA (\url{http://kida.astrophy.u-bordeaux.fr})}
\end{deluxetable*}

\section{The binding energy of P-bearing species} \label{sec:bind_energy}
A continuous exchange of chemical ingredients within the gas and grain can determine the chemical complexity of the ISM. The major drawback in constraining this chemical process by astrochemical modeling is the lack of information about the interaction energy of the chemical species with the grain surface. The fate of the chemical complexity on the grain surface depends on the residence time of the incoming species on the grain. Thus, the binding energy ({\it BE}) of interstellar species plays a crucial role in the chemical complexity of the ISM. 

In a periodic treatment of surface adsorption phenomena, {\it BE} is related to the interaction energy ($\Delta$E), as:
\begin{equation}
\label{eqn:1}
    BE = -\Delta E
\end{equation}
For a bounded adsorbate {\it BE} is a positive quantity and is defined as:
\begin{equation}
\label{eqn:2}
    BE = (E_{surface}+E_{species})-E_{ss}
\end{equation}
where $\rm{E_{ss}}$ is the optimized energy
for the complex system where a species is placed at a distance from the grain surface through a weak Van der Waals interaction. $\rm{E_{surface}}$ and $\rm{E_{species}}$ are the optimized energies of
the grain surface and species, respectively.

In dense molecular clouds, a sizeable portion of the grain mantle would cover water, methanol, and CO molecules. 
In more dense regions, gas-phase H$_2$ could easily accrete on the grain and mostly gets back to the gas phase due to their low sticking probability and lower {\it BEs}. But some H$_2$ could be trapped under some accreting species. For example, one H$_2$ may accrete on another H$_2$ before it is desorbed. In this situation, the ``encounter desorption" mechanism is supposed to be an essential means of transportation of H$_2$ to the gas phase \citep{hinc15}. Usually, in the literature, only the encounter desorption of H$_2$ is considered due to its wide presence in the dense interstellar medium. Recently, \cite{chang20} considered the encounter desorption of H atom as well. 
But the {\it BEs} of the interstellar species with the different substrates are unknown.
Looking at these aspects, in Table \ref{tab:binding}, we report the {\it BEs} of some relevant P-bearing species by considering H$_2$O, CO, CH$_3$OH, and H$_2$ as a substrate. The computed {\it BEs} should have a range of values that depend on the molecule's position on the ice \citep{das18,ferr20}. Whenever we have obtained different BE values at other binding sites, we exert the average of some of our calculations and note in Table \ref{tab:binding}. In this paper, we only consider the {\it BE} with water substrate for our simulation. {\it BEs} with the other substrates are provided for future usage.
We perform all the {\it BE} calculations by \textsc{Gaussian} 09 suite of programs \citep{fris13} using Equations \ref{eqn:1} and \ref{eqn:2}. To find the optimized energy of all structures, we use a Second-order M\o ller-Plesset (MP2) method with an aug-cc-pVDZ basis set \citep{dunn89}. We did not include the zero-point vibrational energy (ZPVE) and basis set superposition error (BSSE) correction \citep{das18}. A fully optimized ground-state structure is verified to be a stationary point (having non-imaginary frequency) by harmonic
vibrational frequency analysis. Ground state spin multiplicity of the species is also noted in Table \ref{tab:binding}. To evaluate it, we take the help of the \textsc{Gaussian} 09 suite of programs and run separate calculations (job type ``opt+freq"), each with different spin multiplicities, and then compare the results between them. The lowest energy electronic state solution of the chosen spin multiplicity is the ground state noted.

Due to the similarities between the structure of NH$_3$ and PH$_3$, \cite{chan20} considered the {\it BE} of PH$_3$ same as NH$_3$. The {\it BE} of NH$_3$ is $5800$ K according to the KIDA database. Since they also used this {\it BE} of PH$_3$, they did not obtain much PH$_3$ by thermal desorption process in the cold regions. They pointed out that most of the PH$_3$ came to the gas phase by photo-desorption rather than thermal desorption from the colder region. In our calculation, we found a noticeable decrease in the {\it BE} of PH$_3$, which could enable PH$_3$ to populate the gas phase by thermal desorption even at low temperatures. 
\cite{das18} noted {\it BE} of NH$_3$ with a c-hexamer configuration of water $\sim 5163$ K, 
whereas we have found it $\sim 2395$ K for PH$_3$.
With significant ice constituents (monomer of CO, CH$_3$OH, H$_2$O, and H$_2$), 
our computed {\it BE} of PH$_3$ seems to be $<1000$ K. \cite{sous20} noted that PH$_3$ has very low water solubility (at 17 $^\circ$C only 22.8 ml of gaseous PH$_3$ dissolve in 100 ml of water), and it does not easily stick to aerosols. 
The recent claim of PH$_3$ in the Venusian atmosphere \citep{grea20} prompted us to determine the {\it BE} of PH$_3$ with some other species like sulfuric acid (H$_2$SO$_4$) and Benzene ($\rm{C_6H_6}$). $\rm{H_2SO_4}$ which are the principal constituents of the Venus atmosphere. Our computed {\it BE} with $\rm{H_2SO_4}$ and Benzene is $\sim 2271$ K and $\sim 3094$ K, respectively. 
However, \cite{snell20} and \cite{vill20} questioned about the identification of \cite{grea20}.
Table \ref{tab:binding} also provides our calculated enthalpy of formation for several P-bearing species and the BE values noted in various places.

\begin{figure*}
\centering
\includegraphics[width=\textwidth]{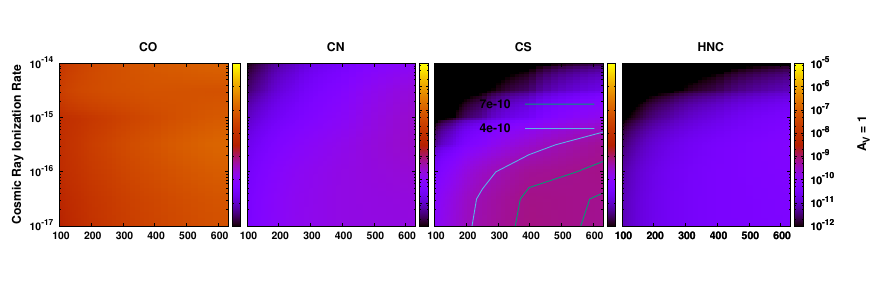}
\vskip -2.6cm
\includegraphics[width=\textwidth]{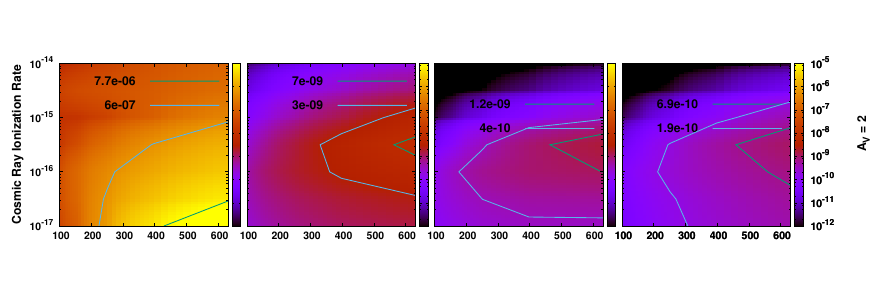}
\vskip -2.6cm
\includegraphics[width=\textwidth]{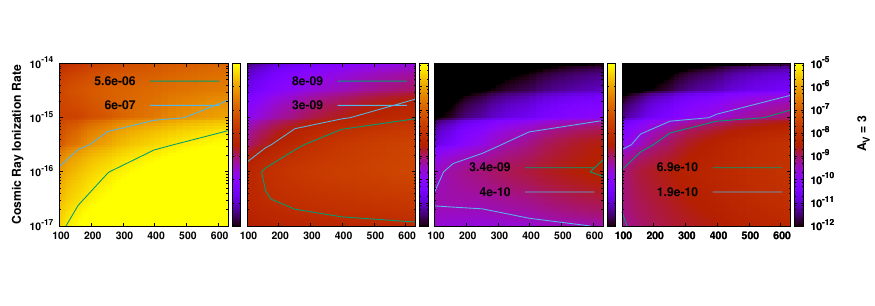}
\vskip -2.6cm
\includegraphics[width=\textwidth]{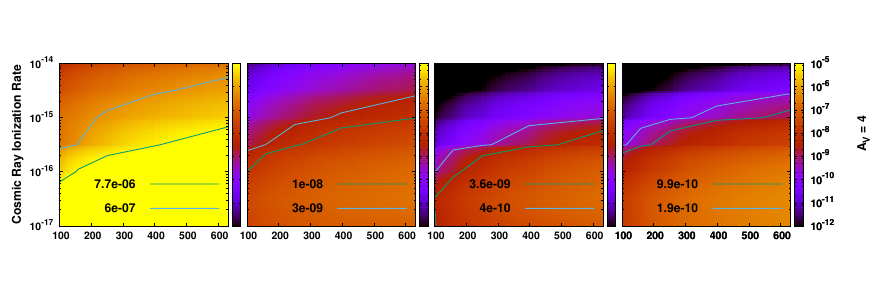}
\vskip -2.6cm
\includegraphics[width=\textwidth]{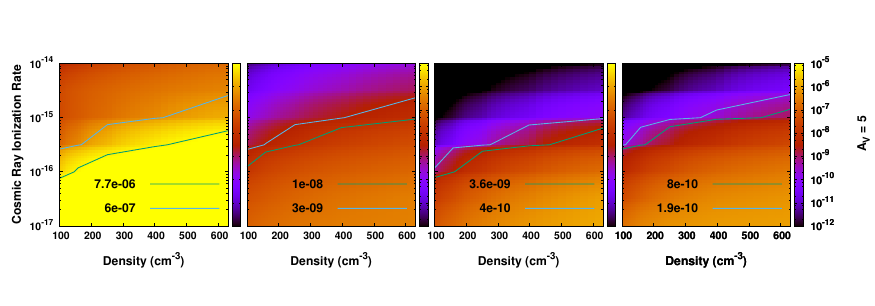}
\caption{Parameter space of the abundances of CO, CN, CS, and HNC for $\rm{A_V}=1,\ 2,\ 3,\ 4,\ 5$ mag for the diffuse cloud model with the \textsc{Cloudy} code. The right side of each panel is marked with color coded values of abundance with respect to total hydrogen nuclei. The contours are highlighted around the previously observed abundance limit \citep{chan20} toward the cloud with $v_{LSR}=-17$ km s$^{-1}$, including the inferred uncertainties.}
    \label{fig:param_space}
\end{figure*}

\begin{figure*}
\centering
\includegraphics[width=8.5cm, height=6cm]{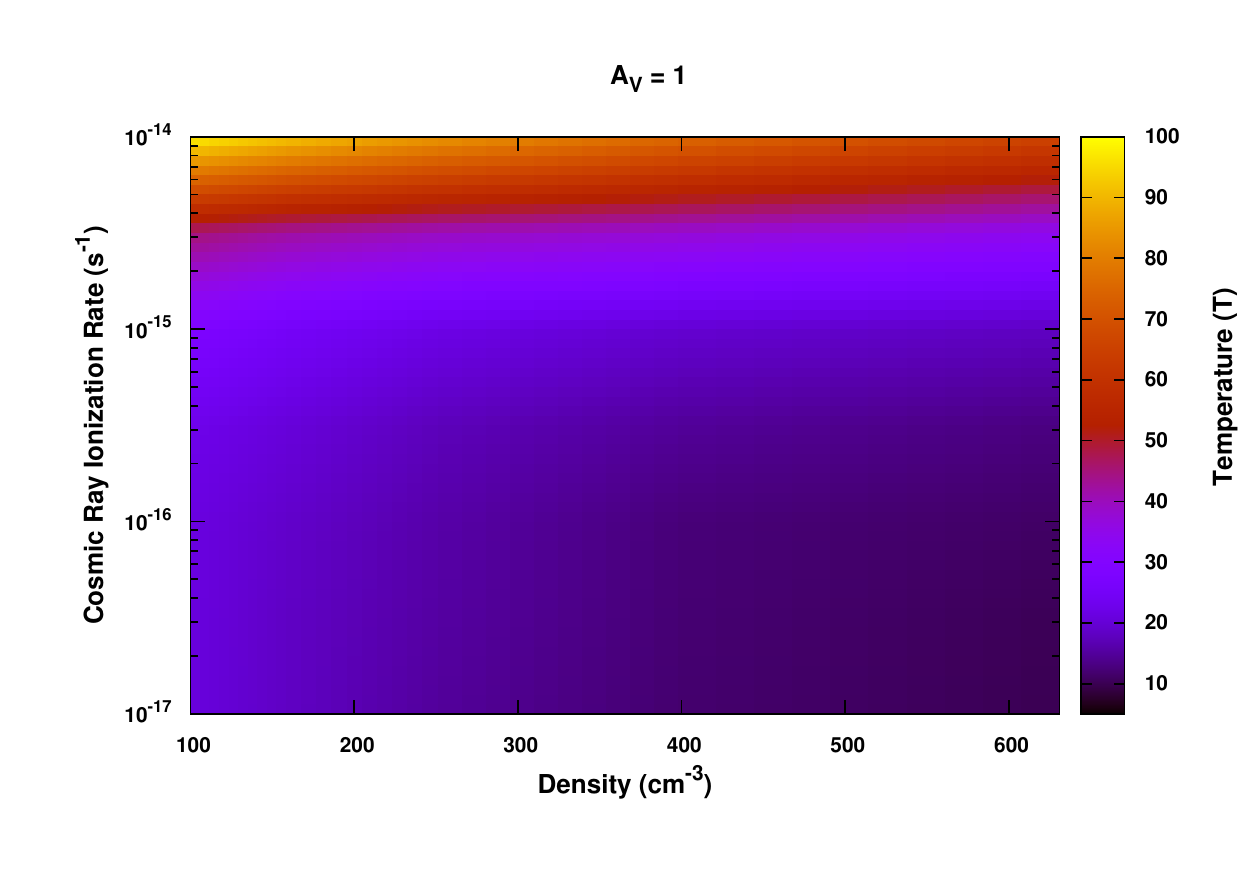}
\includegraphics[width=8.5cm, height=6cm]{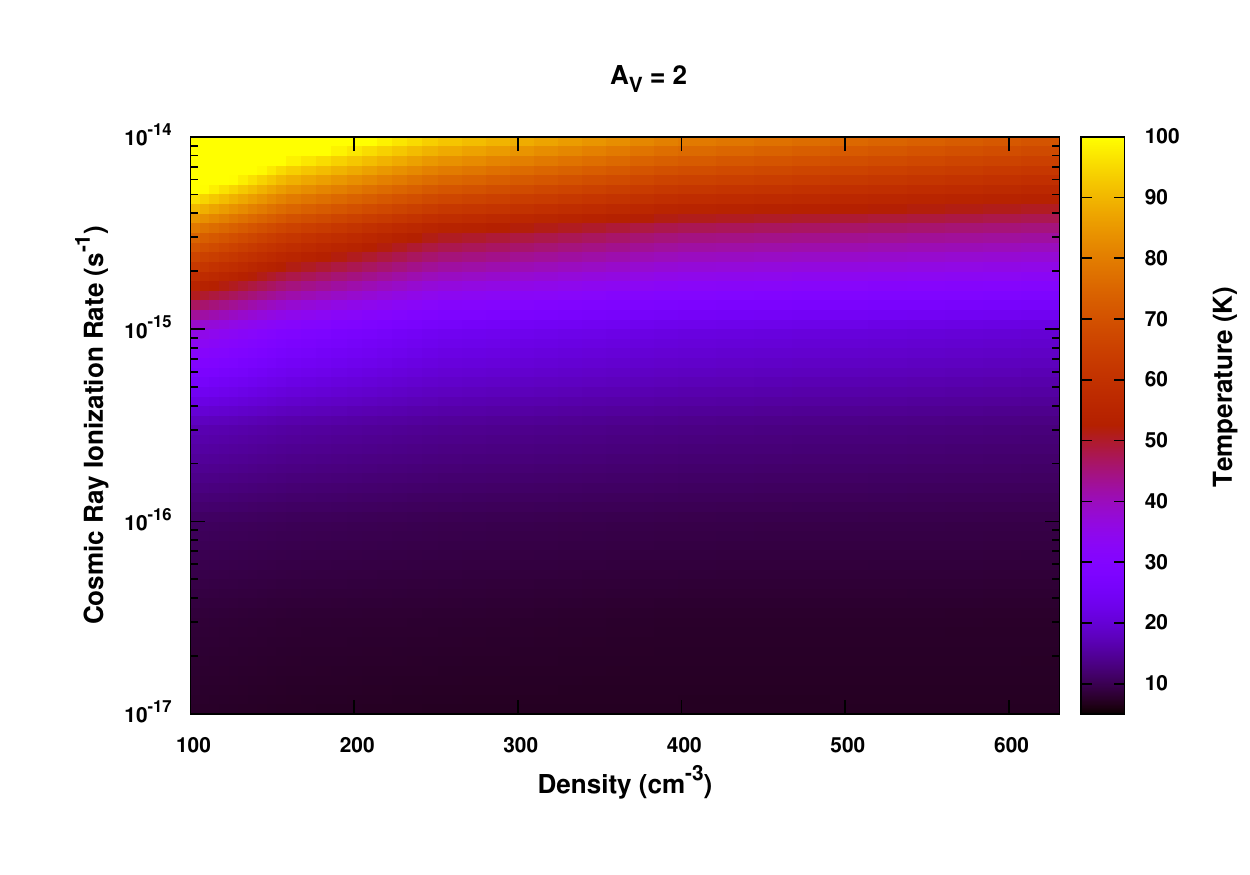}
\includegraphics[width=8.5cm, height=6cm]{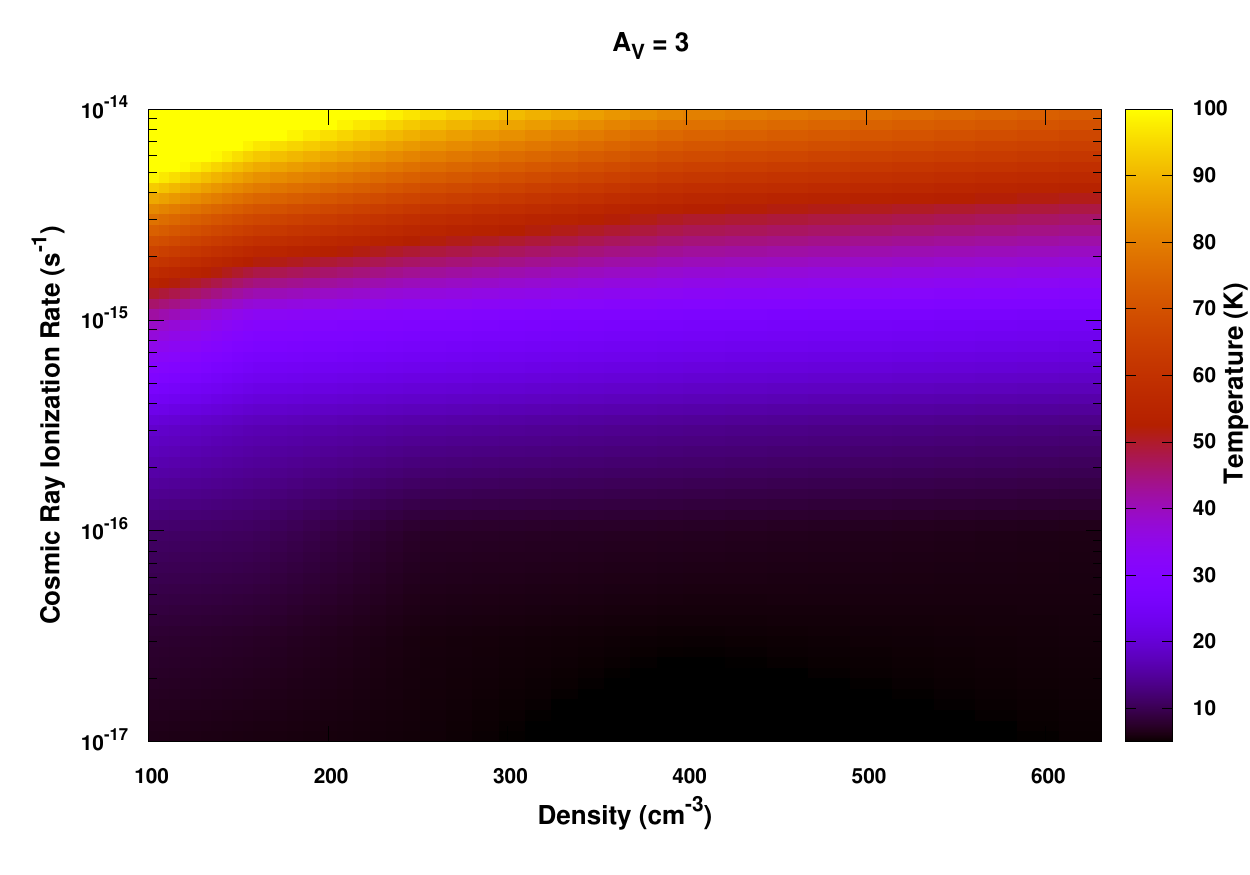}
\caption{Parameter space of the temperature for $\rm{A_V}=1,\ 2,\ 3$ mag for the diffuse cloud model with the \textsc{Cloudy} code. The right side of each panel is marked with color coded values of temperature.}
\label{fig:param_temp}
\end{figure*}

\section{chemical model} \label{sec:chem_model}
We use the reaction pathways shown in the Appendix (see Table \ref{tab:reaction_path}) to check the fate of the P-bearing species in various parts of the ISM (diffuse clouds, PDRs, hot corino, and hot cores). Here, we employ mainly two models to study the chemical evolution of these species; 
a) spectral synthesis code, \textsc{Cloudy} \citep[version 17.02, last described by][]{ferl17} and 
b) Chemical Model for Molecular Cloud (hereafter CMMC) code \citep{das15a,gora17a,gora17b,sil18,gora20a,shim20}.

\begin{figure}
\centering
\includegraphics[width=9cm, height=6cm]{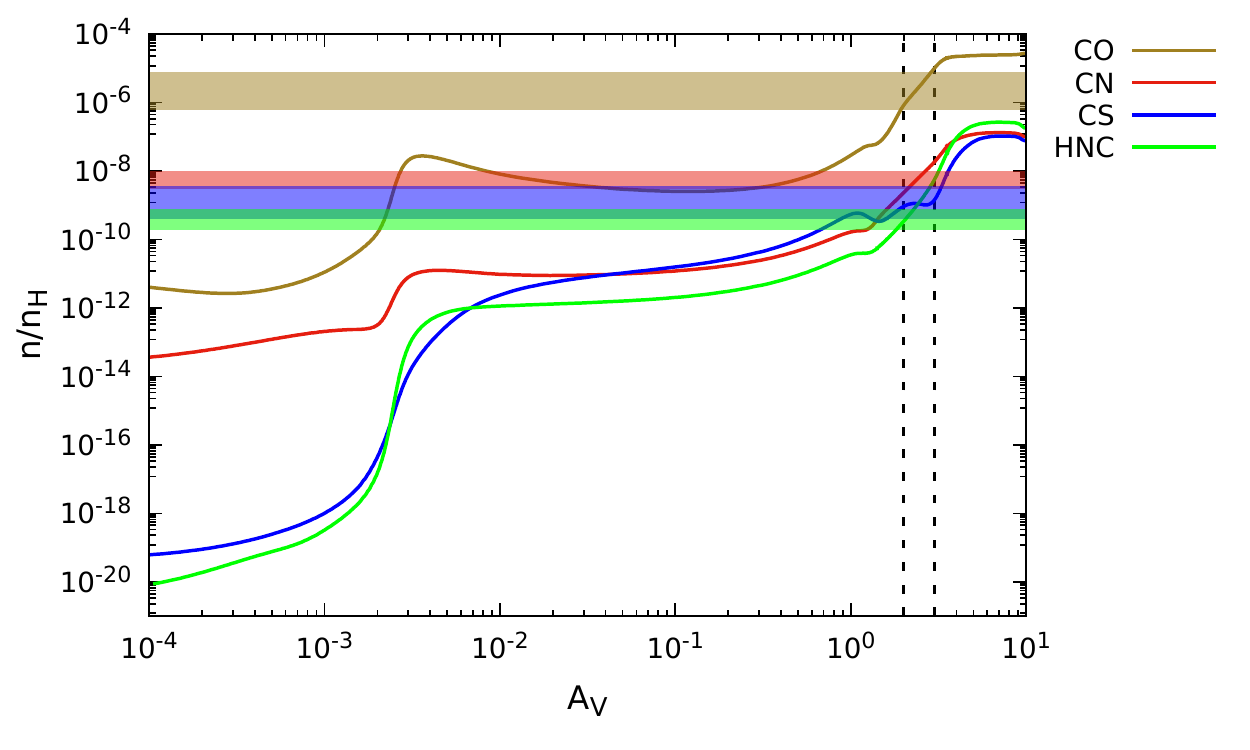}
\caption{Chemical evolution of the abundances of CO, CN, CS, and HNC for the diffuse cloud model ($\rm{n_H}=300$ cm$^{-3}$ and $\rm{\zeta=1.7\times10^{-16}\ s^{-1}}$) with the \textsc{Cloudy} code. The colored horizontal
bands correspond to the observed abundances \citep{chan20} toward the cloud with $v_{LSR}=-17$ km s$^{-1}$, including the inferred uncertainties. The vertical dashed line indicates the visual extinction parameter of best agreement between observation and model results.}
\label{fig:diffuse_obs_CLOUDY}
\end{figure}

\begin{figure*}
\centering
\includegraphics[width=8.5cm, height=6cm]{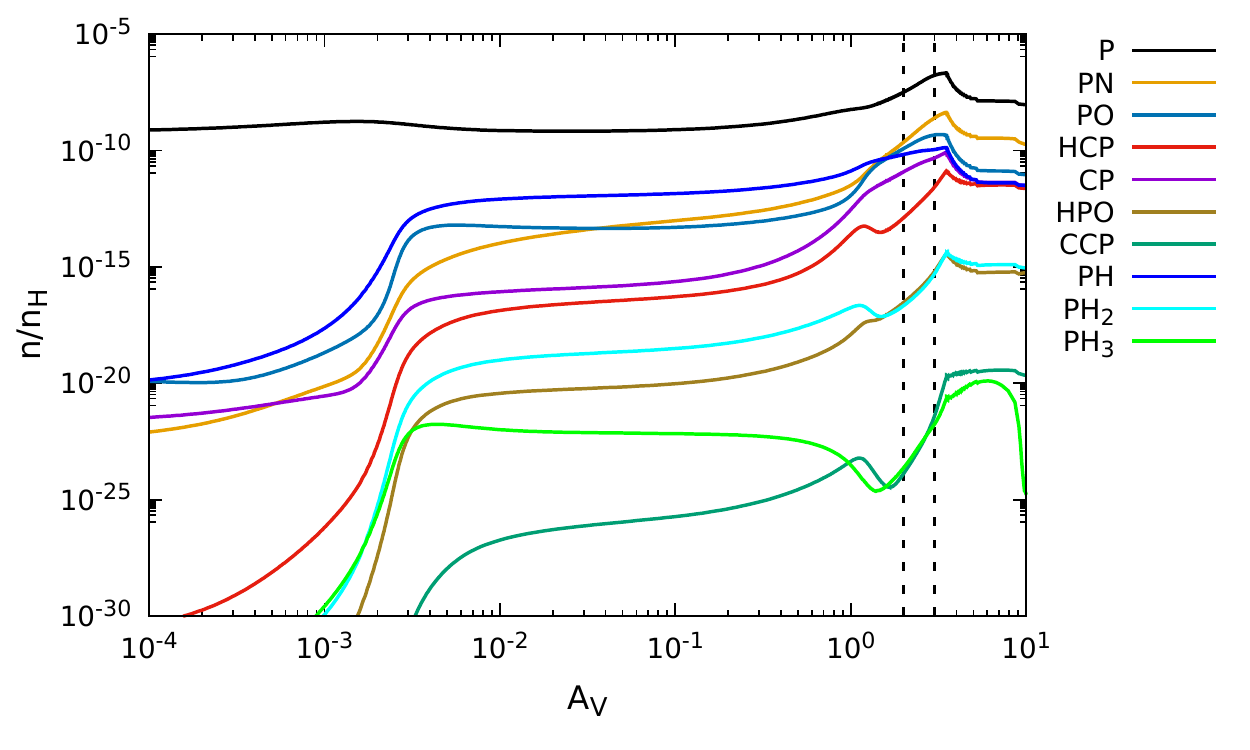}
\includegraphics[width=8.5cm, height=6cm]{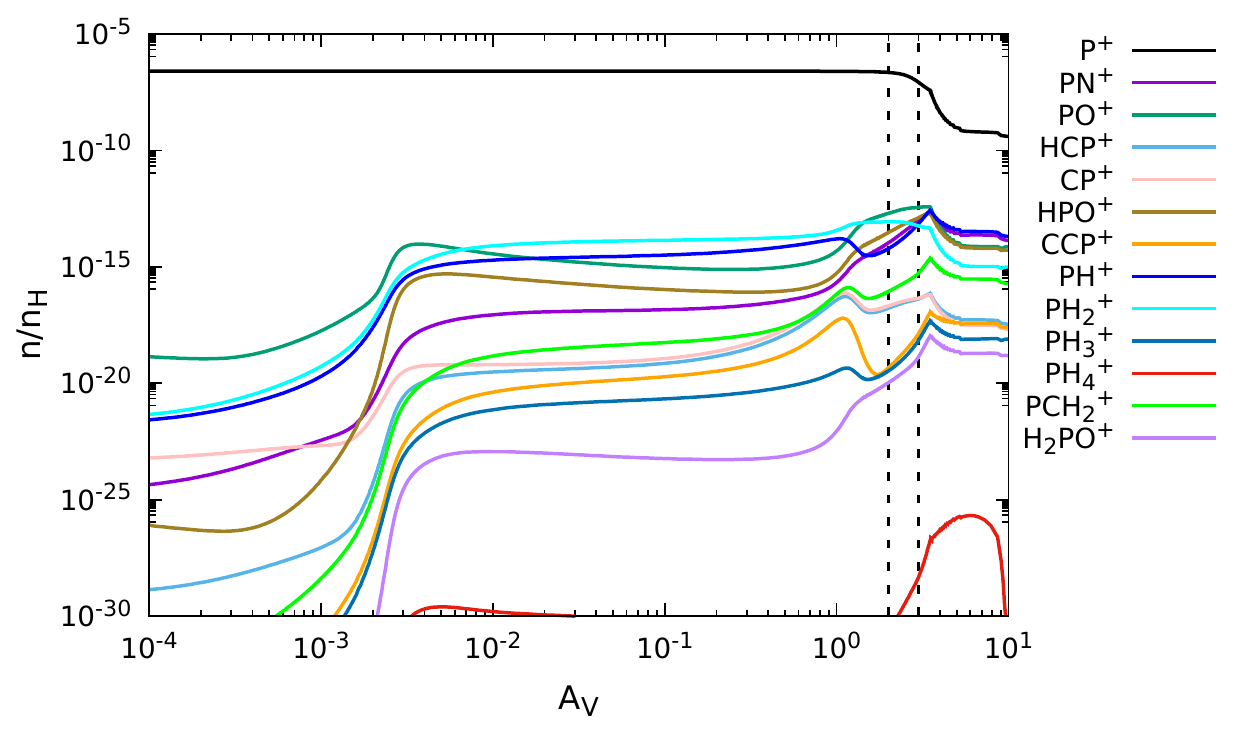}
\caption{Chemical evolution of the abundances of neutral P-bearing species (left panel), and their corresponding cations (right panel) for the diffuse cloud model ($\rm{n_H}=300$ cm$^{-3}$ and $\rm{\zeta=1.7\times10^{-16}\ s^{-1}}$) with the \textsc{Cloudy} code. The vertical dashed line indicates the visual extinction parameter of the best agreement between observation and model results.}
\label{fig:diffuse_phosphorus_CLOUDY}
\end{figure*}

\begin{figure}
\centering
\includegraphics[width=9cm, height=6cm]{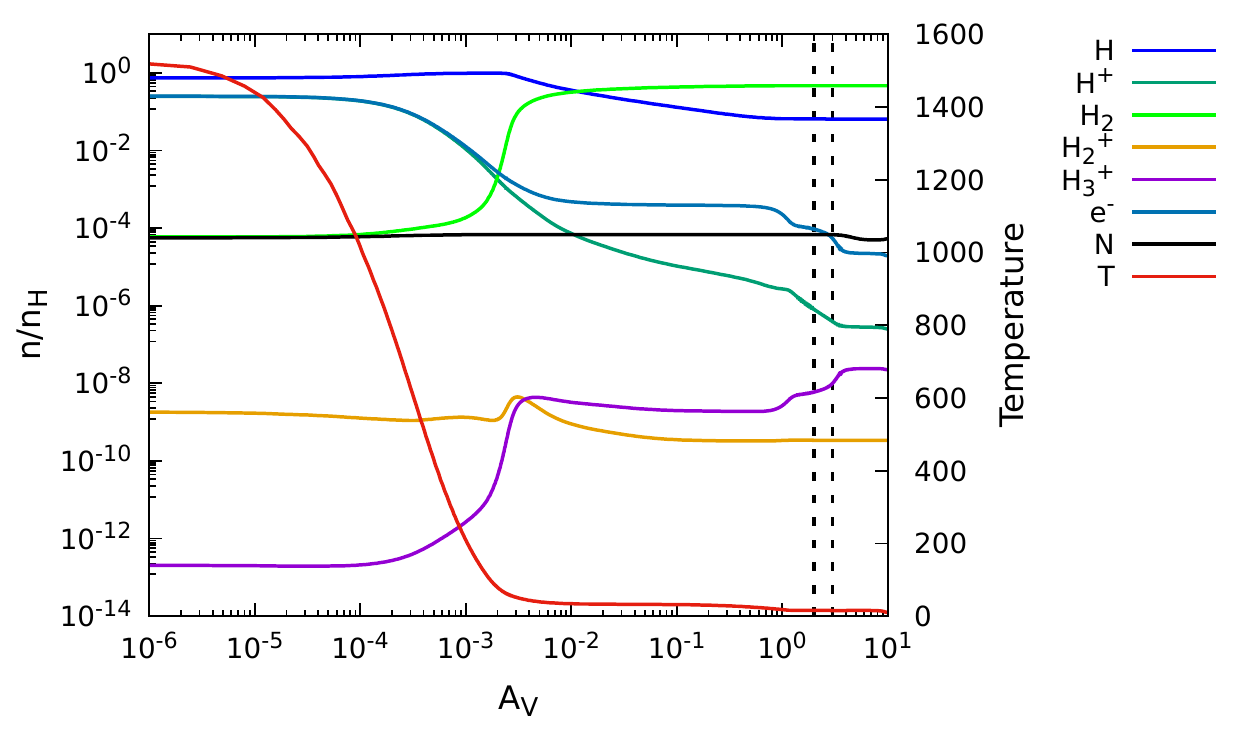}
\caption{Abundance profiles of H, H$^+$, H$_2$, H$_2^+$, H$_3^+$, e$^-$, N, and temperature profile for the diffuse cloud model ($\rm{n_H}=300$ cm$^{-3}$ and $\rm{\zeta=1.7\times10^{-16}\ s^{-1}}$) with the \textsc{Cloudy} code. The vertical dashed line indicates the visual extinction parameter of best agreement between observation and model results.}
\label{fig:DIFF_HH2}
\end{figure}

\subsection{Spectral synthesis code}
We use a photo-ionization code, \textsc{Cloudy}, which simulates matter under a broad range of interstellar conditions. It is for general use under an open-source
license\footnote{\url{https://gitlab.nublado.org/cloudy/cloudy/-/wikis/home}}. Using \textsc{Cloudy} code, we check the fate of P-bearing species in diffuse cloud and PDR environment. 

\subsubsection{Diffuse cloud model}
 Based on the observations carried out by \cite{chan20}, they prepared a diffuse cloud model to explain the observed abundance of HNC, CN, CS, and CO. 
We also employ a similar process to explain the observed abundances of the four molecules and then look at the fate of the P-bearing molecules under these circumstances. 
For this modeling, we consider the initial elemental abundance as given in \cite{chan20}. But in \textsc{Cloudy}, only the atomic elemental abundance is allowed (no ionic or molecular form), so we use these abundances as the initial elemental abundance (see Table \ref{tab:diff_cloud}). For each calculation, we check whether the micro-physics considered is time steady or not. We notice that the largest reaction time scale is much shorter than the diffuse cloud's typical lifetime, $\sim 10^7$ years. We use the grain size distribution, which is appropriate for the $R_V=3.1$ extinction curve of \cite{math77}. This grain size distribution is called ``ISM" in the \textsc{Cloudy}. We use the anisotropic radiation field, which is appropriate for the ISM to specify the intensity of the incident local interstellar radiation field. Additionally, we include the cosmic ray microwave background with a redshift value of $1.52$ \citep{weng00}.

\begin{deluxetable}{cccc}
\tablecaption{Initial elemental abundance for the diffuse cloud and PDR model considered in the \textsc{Cloudy} code. \label{tab:diff_cloud}}
\tablewidth{0pt}
\tabletypesize{\scriptsize} 
\tablehead{
\colhead{\bf Element} &\colhead{\bf Abundance} & \colhead{\bf Element} &\colhead{\bf Abundance}
}
\startdata
H         & 1.0 & Si         & $3.2 \times 10^{-5}$\\
He         & $8.5 \times 10^{-2}$ & Fe         & $3.2 \times 10^{-5}$\\
N         & $6.8 \times 10^{-5}$ & Na         & $1.7 \times 10^{-6}$\\
O         & $4.9 \times 10^{-4}$ & Mg        & $3.9 \times 10^{-5}$\\
C         & $2.7 \times 10^{-4}$ & Cl        & $3.2 \times 10^{-7}$\\ 
S         & $1.3 \times 10^{-5}$ & P         & $2.6 \times 10^{-7}$ \\
& & F         & $3.6 \times 10^{-8}$\\
\enddata
\end{deluxetable}

\begin{deluxetable*}{cccccccc}
\tablecaption{Estimated column density and optical depth of the observed molecules for the diffuse cloud model with the \textsc{Cloudy} code. \label{tab:comp_obs}}
\tablewidth{0pt}
\tabletypesize{\scriptsize} 
\tablehead{
\colhead{Species} & \colhead{Transitions} & \colhead{E$_{up}$ [K]} & \colhead{Frequency [GHz]} & \multicolumn{2}{c}{Optical Depth [$\tau$]} & \multicolumn{2}{c}{Total Column Density [cm$^{-2}$]} \\
\colhead{ } &\colhead{ } &\colhead{ } & \colhead{ } & \colhead{Model} & \colhead{Observation$^a$} & \colhead{Model} & \colhead{Observation$^a$}
}
\startdata
CN & $N=1-0,\ J=1/2-1/2,\ F=3/2-1/2$ & 5.5 & 113.16867 & 0.256522 & $0.23\pm0.07$ & $1.47\times10^{12}$ & $(0.87\pm0.28)\times10^{13}$ \\
HNC & $J=1-0$ & 4.4 & 90.66357 & 0.33962 & $0.50\pm0.10$ & $2.11\times10^{11}$ & $(0.69\pm0.16)\times10^{12}$ \\
C$^{34}$S & $J=2-1$ & 6.9 & 96.41295 & \nodata & $0.04\pm0.02$ & \nodata & $(1.64\pm0.82)\times10^{11}$ \\
$^{13}$CO & $J=1-0$ & 5.3 & 110.20135 & \nodata & $0.154\pm0.004$ & \nodata & $(3.98\pm0.16)\times10^{14}$ \\
\hline
HCP & $J=2-1$ & 5.8 & 79.90329 & $2.56\times10^{-6}$ & $<0.02$ & $1.10\times10^{8}$ & $<2.27\times10^{12}$ \\
PN & $J=2-1$ & 6.8 & 93.97977 & 0.238015 & $<0.02$ & $1.62\times10^{11}$ & $<4.20\times10^{10}$ \\
CP & $N=2-1,\ J=3/2-1/2,\ F=2-1$ & 6.8 & 95.16416 & $4.92\times10^{-4}$ & $<0.02$ & $1.16\times10^{10}$ & $<1.26\times10^{12}$ \\
PO & $J=5/2-3/2,\ \Omega=1/2,\ F=3-2,\ e$ & 8.4 & 108.99845 & 0.0138432 & $<0.02$ & $9.08\times10^{10}$ & $<4.29\times10^{11}$\\
PO & $J=5/2-3/2,\ \Omega=1/2,\ F=2-1,\ e$ & 8.4 & 109.04540 & 0.0125113 & $<0.02$ & $9.08\times10^{10}$ & $<6.70\times10^{11}$ \\
PO & $J=5/2-3/2,\ \Omega=1/2,\ F=3-2,\ f$ & 8.4 & 109.20620 & 0.0055508 & $<0.02$ & $9.08\times10^{10}$ & $<4.34\times10^{11}$ \\
PO & $J=5/2-3/2,\ \Omega=1/2,\ F=2-1,\ f$ & 8.4 & 109.28119 & 0.0182699 & $<0.02$ & $9.08\times10^{10}$ & $<6.69\times10^{11}$ \\
\enddata
\tablecomments{
$^a$\cite{chan20}}
\label{tab:comparison}
\end{deluxetable*}

To check the validity of our model against the observed abundances, we first explain the observed abundance of CO, CN, CS, and HNC as obtained in \cite{chan20}. For this, we use a parameter space for the formation of these species. Our
parameter space consists of a density variation of $100-600$ cm$^{-3}$ and a cosmic ray ionization rate of H$_2$ ($\zeta_{H_2}$) in between $10^{-17}-10^{-14}$ s$^{-1}$. We use a stopping criterion at different $A_V$ values. 
The cases obtained by using stopping criterion $A_V= 1-5$ mag are given in Figure \ref{fig:param_space}. 
The observed abundances of CO, CN, CS, and HNC of \cite{chan20} toward the cloud having a $V_{LSR}=-17$ km s$^{-1}$ are highlighted by the contours. 
Figure \ref{fig:param_temp} shows the variation of temperature at the end of the calculation for $A_V=1-3$ mag. The parameter space also consists of the variation of n$_H$ and $\zeta_{H_2}$. Temperature variation is shown with the color bar.
 
From Figure \ref{fig:param_space} and \ref{fig:param_temp}, we see that the parameters which were adopted by \cite{chan20} for the diffuse cloud model (i.e., $\zeta_{H_2}=1.7 \times 10^{-16} \ s^{-1}$, $n_H=300$ cm$^{-3}$ and $T_{gas}=40$ K) can also reproduce the observed abundances of CO, CN, CS, and
HNC with our model. Figure \ref{fig:diffuse_obs_CLOUDY} shows the obtained abundances of the four molecules when we consider $n_H=300$ cm$^{-3}$ and $\zeta_{H_2}$ = $1.7 \times 10^{-16}\ s^{-1}$. The X-axis shows the visual extinction of the cloud, and the Y-axis shows the abundance of n$_H$. Our results are in agreement with the observation between $A_V=2-3$ mag. We highlight the best-suited zone by the black dashed curve.
Thus, based on Figures \ref{fig:param_space}, \ref{fig:param_temp}, and \ref{fig:diffuse_obs_CLOUDY}, we use, $A_V=2$ mag, $\zeta_{H_2}= 1.7\times 10^{-16}$, and $n_H=300$ cm$^{-3}$ as the best fitted parameters to explain the observed abundances of these species. This yields a $A_V/N(H) = 5.398 \times 10^{-22}$ mag cm$^2$ and dust to gas ratio of $6.594 \times 10^{-3}$.
Table \ref{tab:comparison} compares our obtained optical depth and column densities with the observations. Obtained optical depths of CN and HNC are close to the observed values, but the column densities diverge by a few factors. This slight mismatch is because the \textsc{Cloudy} model deals with steady-state values, but the column densities are time-dependent in reality.

Figure \ref{fig:diffuse_phosphorus_CLOUDY} depicts the abundances of most of the P-bearing species considered in this study. The left panel shows the neutral P-bearing species, whereas the right panel shows the P-bearing ions. It is interesting to note that though we consider a neutral atomic abundance in our model, the abundance of P$^+$ is high due to the strong cosmic ray ionization and presence of a non-extinguished interstellar radiation field. The abundances of the comparatively larger P-bearing species are very low in the diffuse environment, and it would be pretty challenging to observe them. We notice that some simple neutral P-bearing species like PN, PO, HCP, CP, and PH are significantly more abundant than PH$_3$. The abundances of PO and PN appear to be comparable to each other (with PO/PN $<1$ when A$_V \geq $ 2).
Table \ref{tab:comparison} shows that the obtained column densities for this case are listed. There is an excellent agreement between the observed and the modeled optical depths of CN and HNC with the \textsc{Cloudy} code.

In Figure \ref{fig:DIFF_HH2}, the abundances of H, H$_2$, and N are shown along with the major hydrogen-related ions, H$^+$, ${\rm H_2}^+$, ${\rm H_3}^+$. In \textsc{Cloudy}, the cloud temperature is calculated self-consistently, depending on the known excitations. The primary source of such excitation is cosmic ray ionization. Figure \ref{fig:DIFF_HH2} depicts the gas temperature variation with $A_V$ with the solid red curve. Deep inside the cloud, the temperature drops sufficiently and enhances the formation of molecular hydrogen.  For $A_V=2$ mag, we obtained a temperature $\sim 10$ K when we use a cosmic ray ionization rate of $1.7 \times 10^{-16}$ s$^{-1}$. It is a little lower than that used in \cite{chan20}.

\begin{figure*}
\centering
\includegraphics[width=8.5cm, height=6cm]{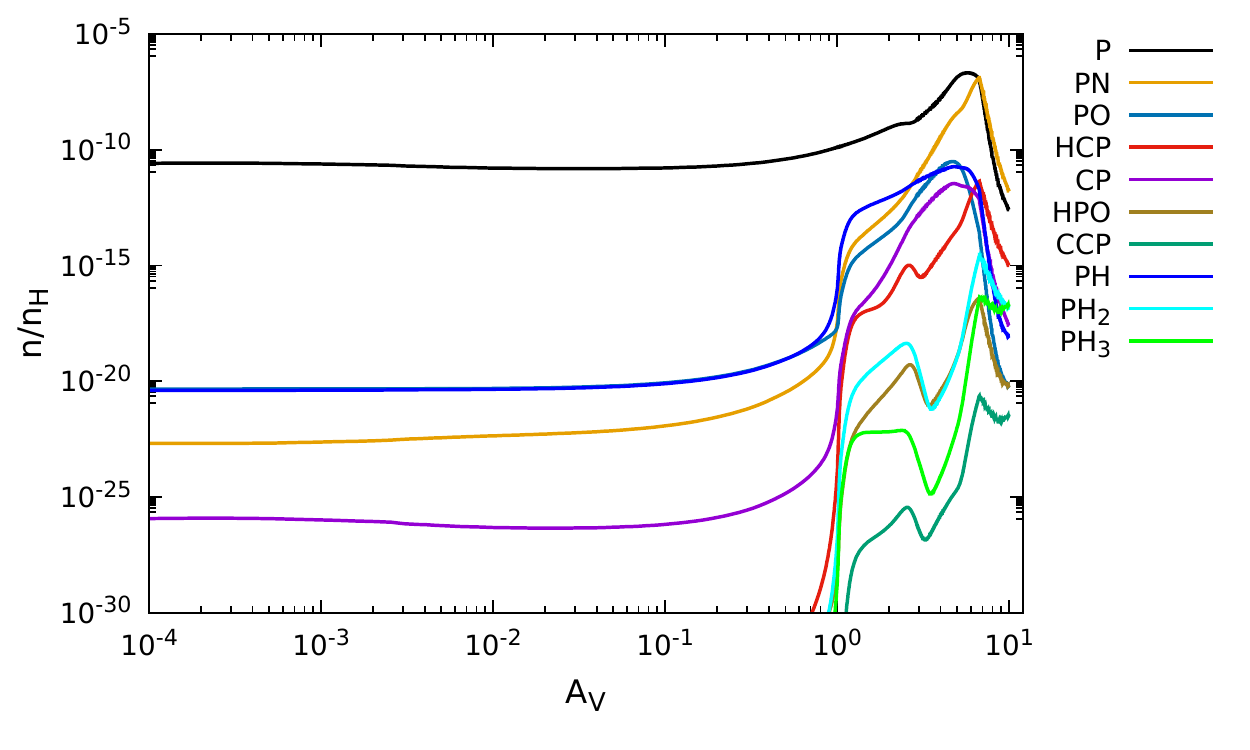}
\includegraphics[width=8.5cm, height=6cm]{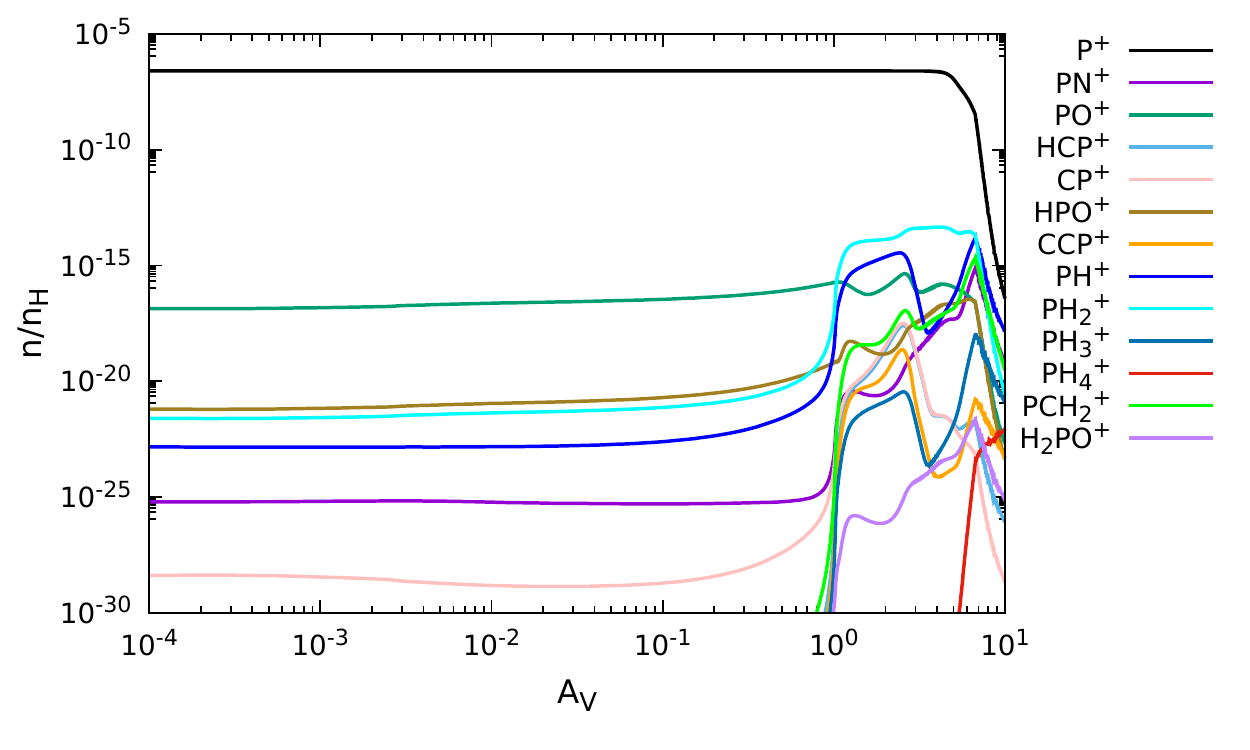}
\caption{Chemical evolution of the abundances of important P-bearing species with the \textsc{Cloudy} code considering \cite{roll07} F4 PDR model.}
\label{fig:PDR_phosphorus_CLOUDY}
\end{figure*}

\begin{figure*}
\centering
\includegraphics[width=9cm, height=6cm]{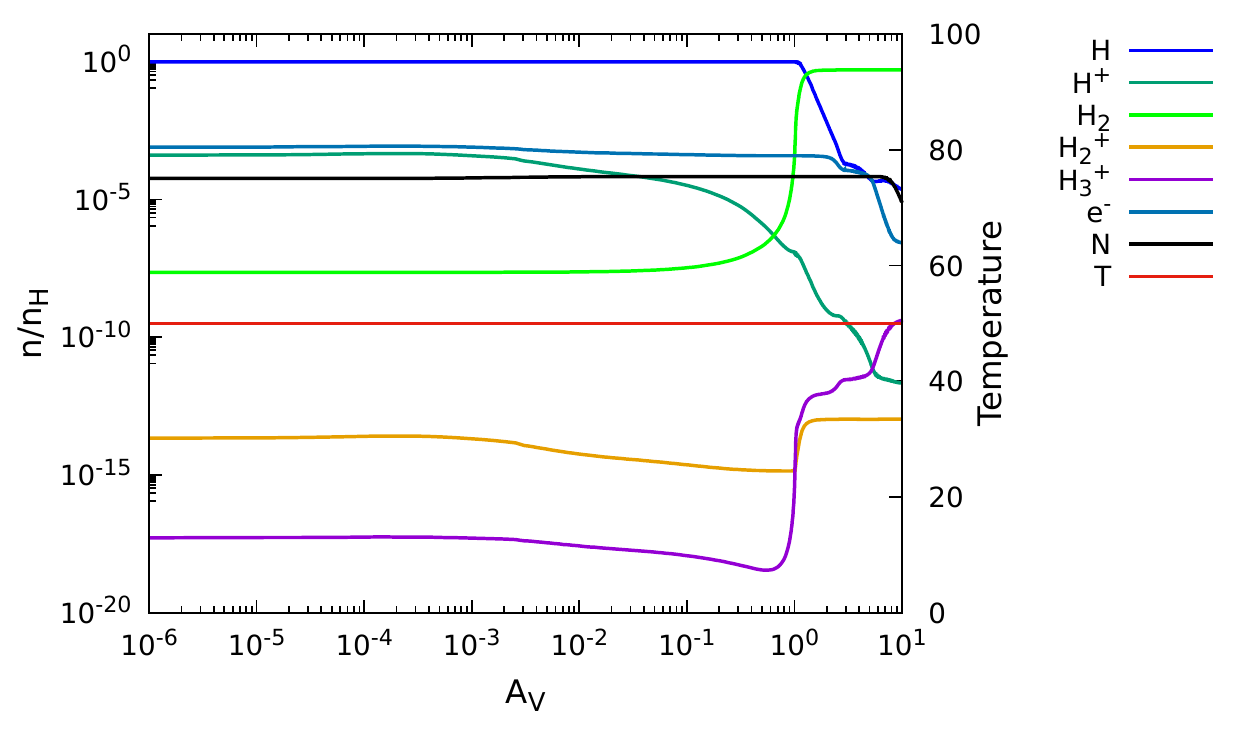}
\caption{Abundance profiles of H, H$^+$, H$_2$, H$_2^+$, H$_3^+$, e$^-$, N, and temperature profile with the \textsc{Cloudy} code considering \cite{roll07} F4 PDR model.}
\label{fig:PDR_HH2}
\end{figure*}

\subsubsection{PDR model}
Here, we use \textsc{Cloudy} code to explain the abundances of the P-bearing species in the PDRs. The \textsc{Cloudy} code is capable of considering realistic physical conditions around this region.  PDRs play an essential role in interstellar chemistry. They are responsible for the emission characteristics of ISM and dominate the infrared and sub-millimeter spectra of star-forming regions and galaxies. In the 
PDRs, H$_2$-non-ionizing
far-ultraviolet photons from stellar sources control gas heating and chemistry. Here, the gas is heated by the
far-ultraviolet radiation (FUV, $6 < h\nu < 13.6$ eV, from the ambient UV field and hot stars) and cooled via the emission of
spectral line radiation of atomic and molecular species and continuum emission by dust. \textsc{Cloudy} includes various PDR benchmark models \citep{roll07}, giving a range of physical conditions associated with PDR environments. We chose the F4 model of \cite{roll07}. This model considered a constant gas  temperature of $50$ K and dust temperature of $20$ K. A  plane-parallel, semi-infinite cloud of the total constant density of $10^{5.5}$ cm$^{-3}$, and standard
UV field as $\chi = 10^5$ times the \cite{drai78} field ($G_0=1.71\chi$) is used. The
cosmic-ray H ionization rate $\zeta = 5\times10^{-17}$ s$^{-1}$, the visual extinction $\rm{A_V = 6.289 \times 10^{-22}N_{H,tot}}$ and a dust to gas ratio $7.646 \times 10^{-3}$ are assumed. We also consider these physical parameters along with the initial elemental abundances as considered in our diffuse cloud model (Table \ref{tab:diff_cloud}). Figure \ref{fig:PDR_phosphorus_CLOUDY} shows the abundances of important P-bearing species. In the diffuse cloud region (see Figure \ref{fig:diffuse_phosphorus_CLOUDY}), we obtained PO and PN comparable abundances. However, for the PDR model, we found a very high abundance of PN compared to PO. Its peak abundance appears to be higher by several orders of magnitude than PO (i.e., PO/PN $<<1$). 
We use the photo-destruction rate of PO and PN from \cite{jime18}. The photodissociation rate of PO is estimated from the photodissociation rate of NO. Similarly, the photodissociation rate of PN is estimated from the photodissociation rate of N$_2$. Using these values reflect a higher photo-dissociation rate of PO (with $\alpha$ $= 3 \times 10^{-10}$ s$^{-1}$ and $\gamma = 2.0$) than PN (with $\alpha$ $= 5 \times 10^{-12}$ s$^{-1}$ and $\gamma = 3.0$). Alike \cite{jime18}, here also, we find that even if PN is photo dissociated, it may be able to revert again via $\rm{P+CN \rightarrow PN+C}$ due to the presence of a large amount of atomic P in the gas phase.     
The rapid conversion of PO to PN by the reaction $\rm{PO + N \rightarrow PN + O}$ is also crucial for maintaining a high abundance of PN than PO.

It is not straightforward to relate the diffuse cloud region cases with the PDR because of different physical circumstances. In the diffuse cloud model, we have obtained the peak abundances of PN and PO at $\rm{A_V=2-3}$ mag, whereas, in the PDR model, we have received these peaks around $\rm{A_V=6-7}$ mag. In both cases, the PO/PN ratio is $<1$, but this ratio seems to be $<<1$ for the PDR. \cite{jime18} also modeled the effect of intense UV-photon-illumination by varying the interstellar radiation field within ranges typical of PDRs. They showed for higher extinctions ($\rm{A_V=7.5}$ mag), and under high-UV radiation fields ($\chi=10^4$ Habing), the abundance of PN always remains above that of PO both for long-lived and short-lived collapse phase. Figure \ref{fig:PDR_HH2} shows the abundances of H, H$^+$, H$_2$, ${\rm H_2}^+$, ${\rm H_3}^+$, N, and electron.

\subsection{CMMC code}
In the earlier section, we have implemented a spectral synthesis code, \textsc{Cloudy}, to study the chemical evolution of the P-bearing species in the diffuse cloud region. Comparatively larger P-bearing species are not very profuse in space, which creates a burden constraining the understanding of the P-bearing species of the ISM. The major drawback in the P-chemistry modeling is the uncertainty of the P depletion factor. The grain surface chemistry plays a significant role in shaping the chemical complexity in these regions.
Since \textsc{Cloudy} only considers the surface reactions of some key species; it would be impractical to apply the \textsc{Cloudy} code in the dense cloud region. Thus, we use our gas-grain CMMC code \citep{das15a,gora17a,gora17b,sil18,gora20a} to explore the fate of P-bearing species in the denser region. In the next section, we first test our model for the diffuse region to validate our results and further extend it for the more evolved phase.

\subsubsection{Diffuse cloud model}

\begin{deluxetable}{cccc}
\tablecaption{Initial elemental abundance for the diffuse cloud model considered in the CMMC code \citep{chan20}. \label{tab:diff_cloud_CMMC}}
\tablewidth{0pt}
\tabletypesize{\scriptsize} 
\tablehead{
\colhead{\bf Element} &\colhead{\bf Abundance} & \colhead{\bf Element} &\colhead{\bf Abundance}
}
\startdata
H         & 1.0 & Si$^+$         & $3.2 \times 10^{-5}$\\
He         & $8.5 \times 10^{-2}$ & Fe$^+$         & $3.2 \times 10^{-5}$\\
N         & $6.8 \times 10^{-5}$ & Na$^+$         & $1.7 \times 10^{-6}$\\
O         & $4.9 \times 10^{-4}$ & Mg$^+$        & $3.9 \times 10^{-5}$\\
C$^+$         & $2.7 \times 10^{-4}$ & Cl$^+$        & $3.2 \times 10^{-7}$\\ 
S$^+$         & $1.3 \times 10^{-5}$ & P$^+$         & $2.6 \times 10^{-7}$ \\
& & F$^+$         & $3.6 \times 10^{-8}$\\
\enddata
\end{deluxetable}

\begin{figure*}
\centering
\includegraphics[width=8cm, height=12cm, angle=270]{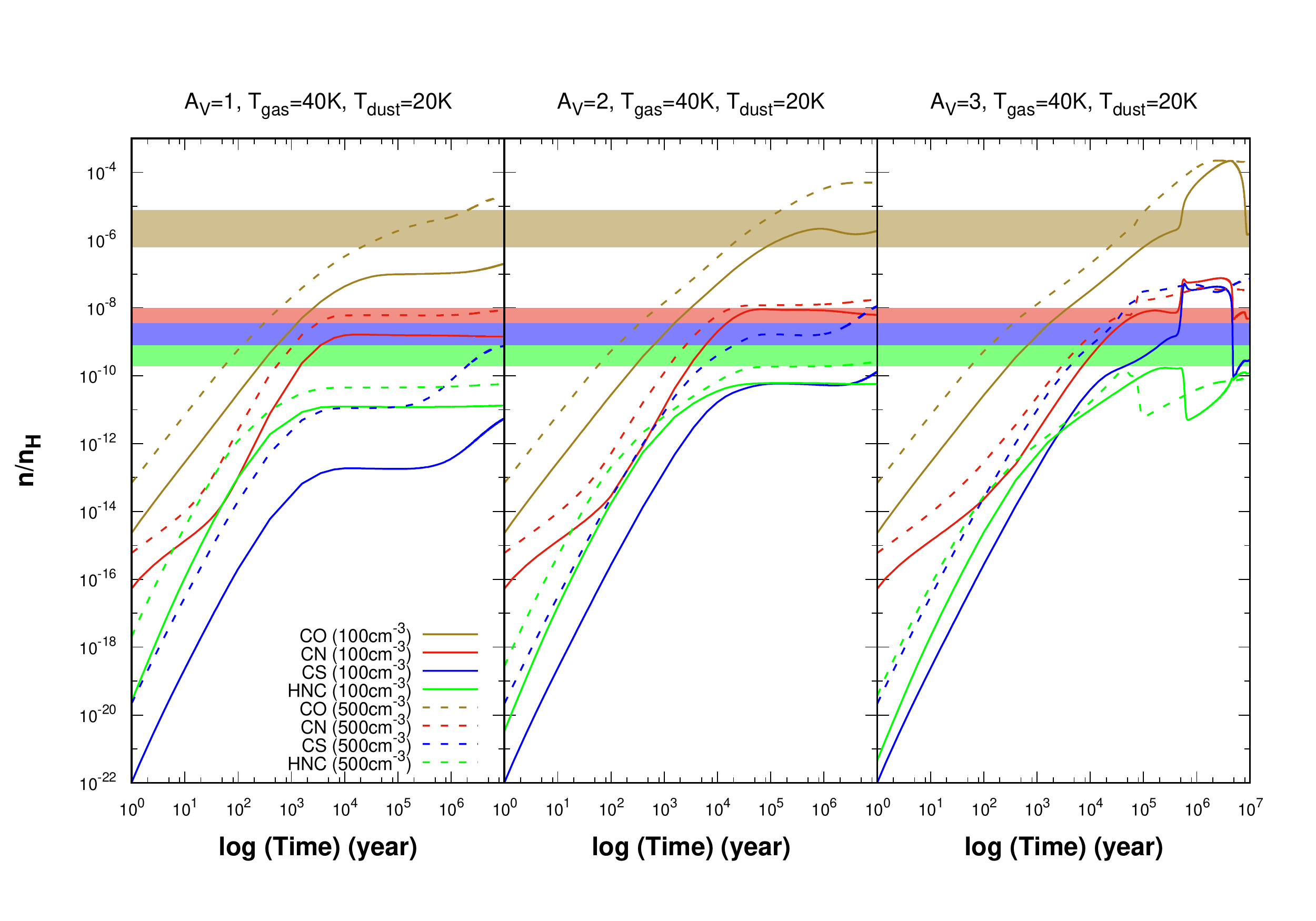}
\caption{Time evolution of the abundances of CO, CN, CS, and HNC with the CMMC code for diffuse cloud. Observed abundances are also highlighted for the better understanding. The solid curves represent the case with $n_H=100$ cm$^{-3}$ and dashed curves represent the case with $n_H=500$ cm$^{-3}$.}
\label{fig:diff_CMMC}
\end{figure*}

\begin{figure*}
\centering
\includegraphics[width=8cm, height=6cm]{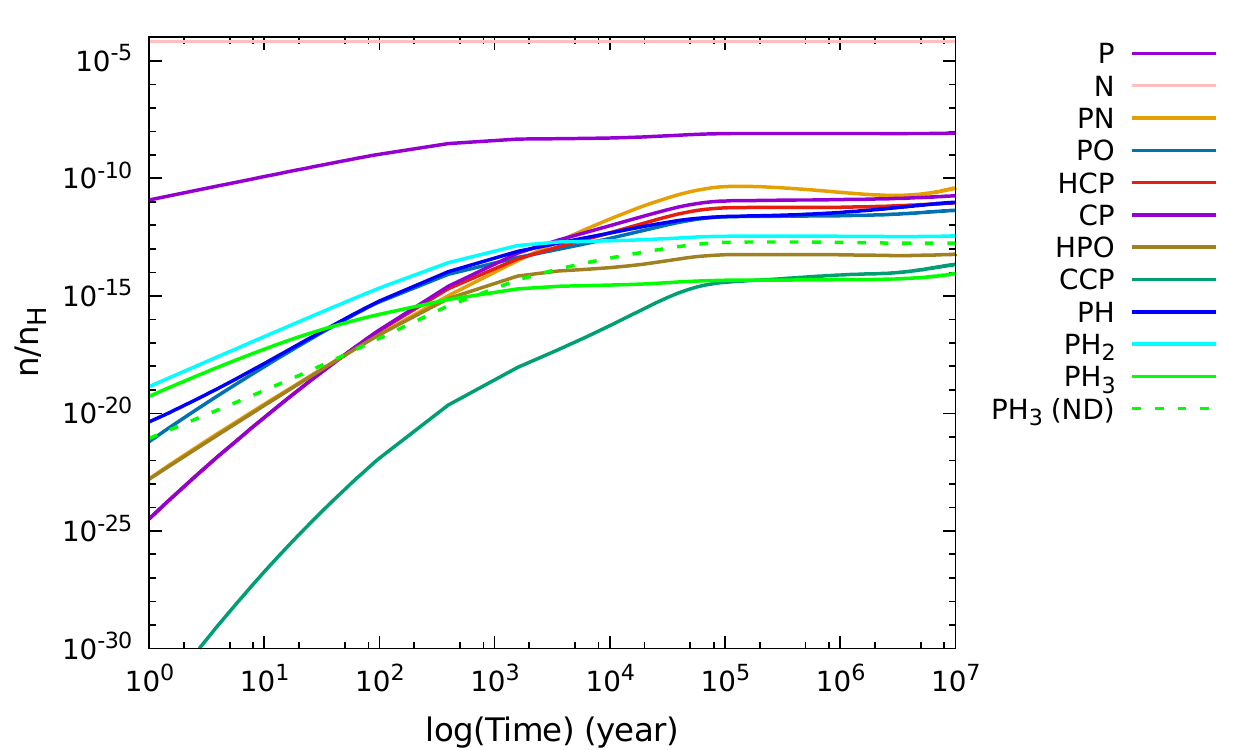}
\includegraphics[width=8cm, height=6cm]{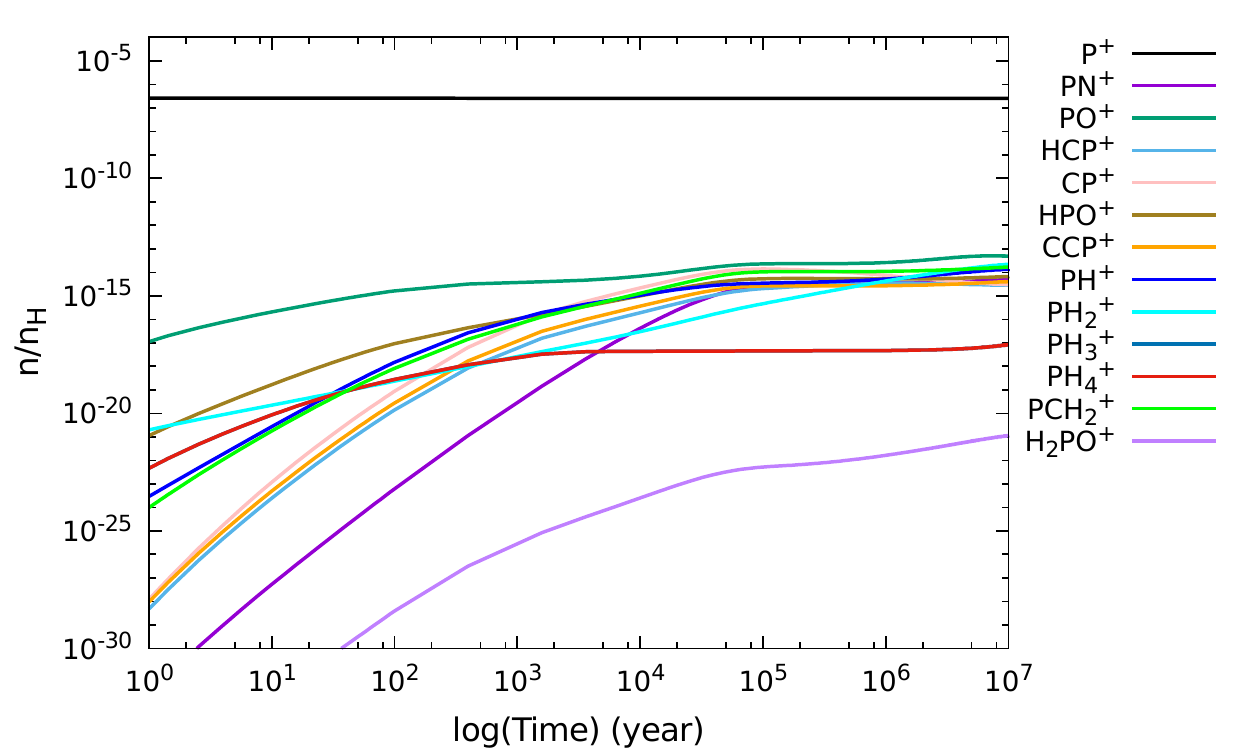}
\caption {Chemical evolution of the P-bearing molecules for $A_V=2$ mag, $n_H=300$ cm$^{-3}$, $T_{gas}=40$ K, and $T_{dust}=20$ K with the CMMC code for diffuse cloud is shown. The evolution of the neutrals and radicals are shown on the left whereas ions are shown on the right side. Derived abundances of all the species are found to be $<10^{-10}$ for such condition.}
\label{fig:diff_CMMC_ION_NEUTRAL}
\end{figure*}

\begin{figure}
\centering
\includegraphics[width=6cm, height=8cm, angle=270]{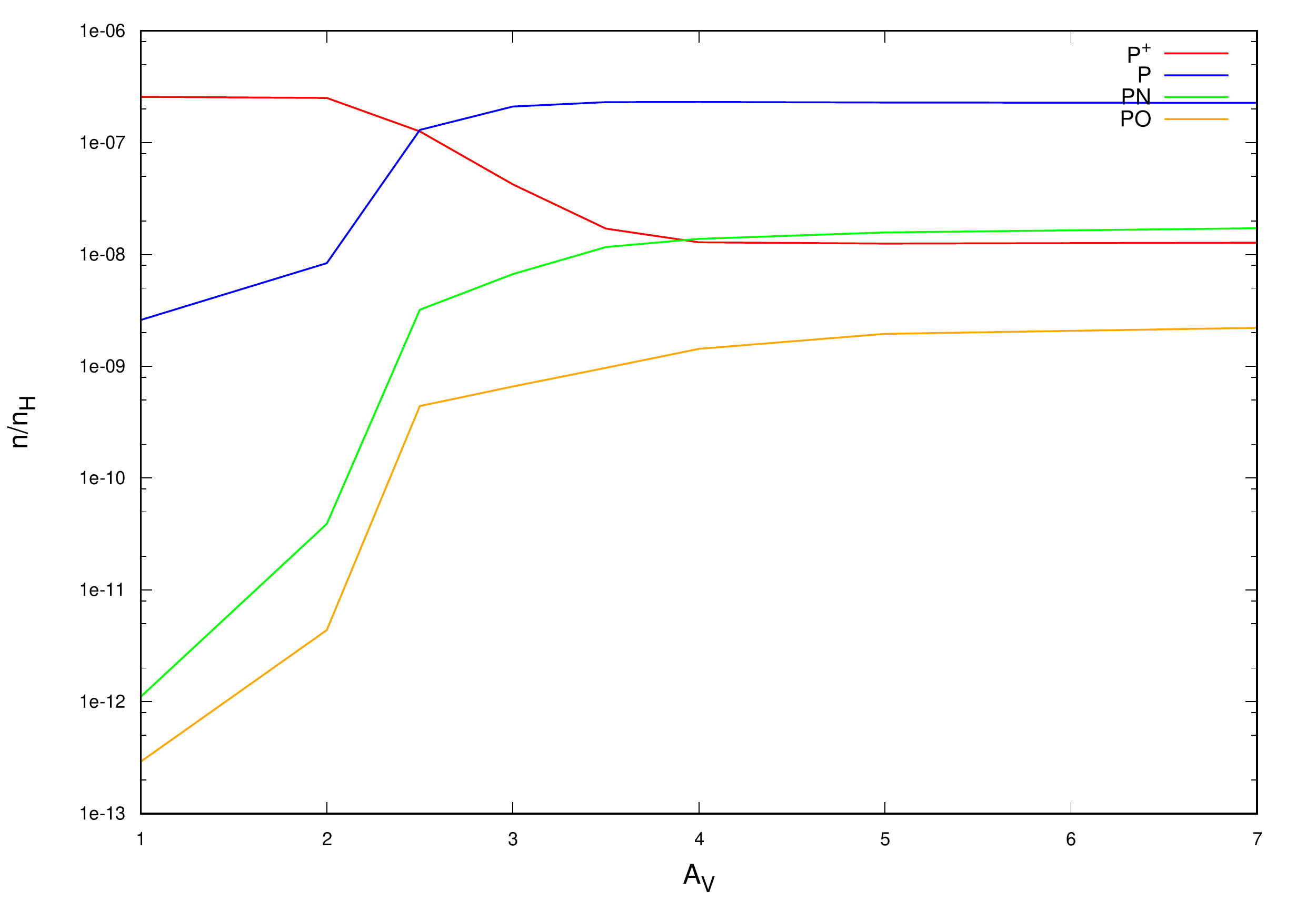}
\caption{The final abundance of P$^+$, P, PO, and PN are shown with A$_V$ with the CMMC code for diffuse cloud. The conversion of P$^+$ to P takes place in the range $A_V=2-3$ mag. Beyond A$_V=4$ mag no changes in the abundances of P$^+$ and P are obtained. The abundance ratio of PO and PN is found to be always $<1$.}
\label{fig:pconv}
\end{figure}

We consider the initial elemental abundances for the diffuse cloud model as in \cite{chan20} (see Table \ref{tab:diff_cloud_CMMC}). The dust to gas ratio of $0.01$ is considered in our model. We consider a photodesorption rate of $3 \times 10^{-3}$ molecules per incident UV photon for all the molecules. \cite{ober07} experimentally derived this rate from the laboratory measurement of CO ice. A non-thermal desorption constant $a=0.01$ and a cosmic ray ionization rate of $1.7 \times 10^{-16}$ s$^{-1}$ are considered. We keep the gas and dust temperatures constant at $40$ K and $20$ K, respectively.

Figure \ref{fig:diff_CMMC} represents the results obtained with the CMMC code for the diffuse cloud region. The solid curve in the Figure represents the case when we consider $n_H=100$ cm$^{-3}$ and the dotted curve represents the case when we consider $n_H=500$ cm$^{-3}$. Figure \ref{fig:diff_CMMC} depicts that in the ranges of $A_V=2-3$ mag, and n$_H=100-500$ cm$^{-3}$, we have a better agreement with the observed abundance.

Based on the obtained results, Figure \ref{fig:diff_CMMC_ION_NEUTRAL} shows the chemical evolution of most of the P-bearing species considered in this study.
In our case, we did not have a high abundance of PH$_3$ under this situation. It is because of the inclusion of the destruction pathways of PH$_3$ by H and OH in our network. The dashed green curve shows the PH$_3$ when we do not consider its destruction by H and OH in gas and ice both the phases.
It is interesting to note that we have a couple of orders higher abundance of PH$_3$ in the absence of these destruction pathways. However, these destruction pathways are essential and need to be considered before constraining the PH$_3$ abundance.
The right panel of Figure \ref{fig:diff_CMMC_ION_NEUTRAL} depicts that the abundance of P$^+$ remains constant for this case. Figure \ref{fig:pconv} shows the final abundance of P and P$^+$ with a variation of $A_V$. It shows that P$^+$ to P conversion is possible between $A_V=2-3$ mag and remains invariant beyond $A_V>4$ mag. For this case, we always get a higher peak abundance of PN compared to PO. At the end of the simulation, \cite{chan20} also obtained a comparatively higher abundance of PN than PO. With the \textsc{Cloudy} code, in Figure \ref{fig:diffuse_phosphorus_CLOUDY}, we also have obtained PO/PN $<1$.  The major difference between the CMMC model or the model used by \cite{chan20} and the \textsc{Cloudy} code is the consideration of the physical condition. The \textsc{Cloudy} code considers physical conditions more realistically. Figure \ref{fig:DIFF_HH2} (diffuse cloud results with \textsc{Cloudy}) shows a substantial temperature variation with $A_V$, whereas in the case of the CMMC model, we assume a constant  temperature ($T_{gas}=40$ K and $T_{ice}=20$ K) for all $A_V$. In the CMMC model, dust to gas ratio of $\sim 0.01$ is considered, whereas in the diffuse cloud model using \textsc{Cloudy}, it gives $6.594 \times 10^{-3}$.

\begin{figure*}
\centering
\includegraphics[width=7cm, height=7cm, angle=270]{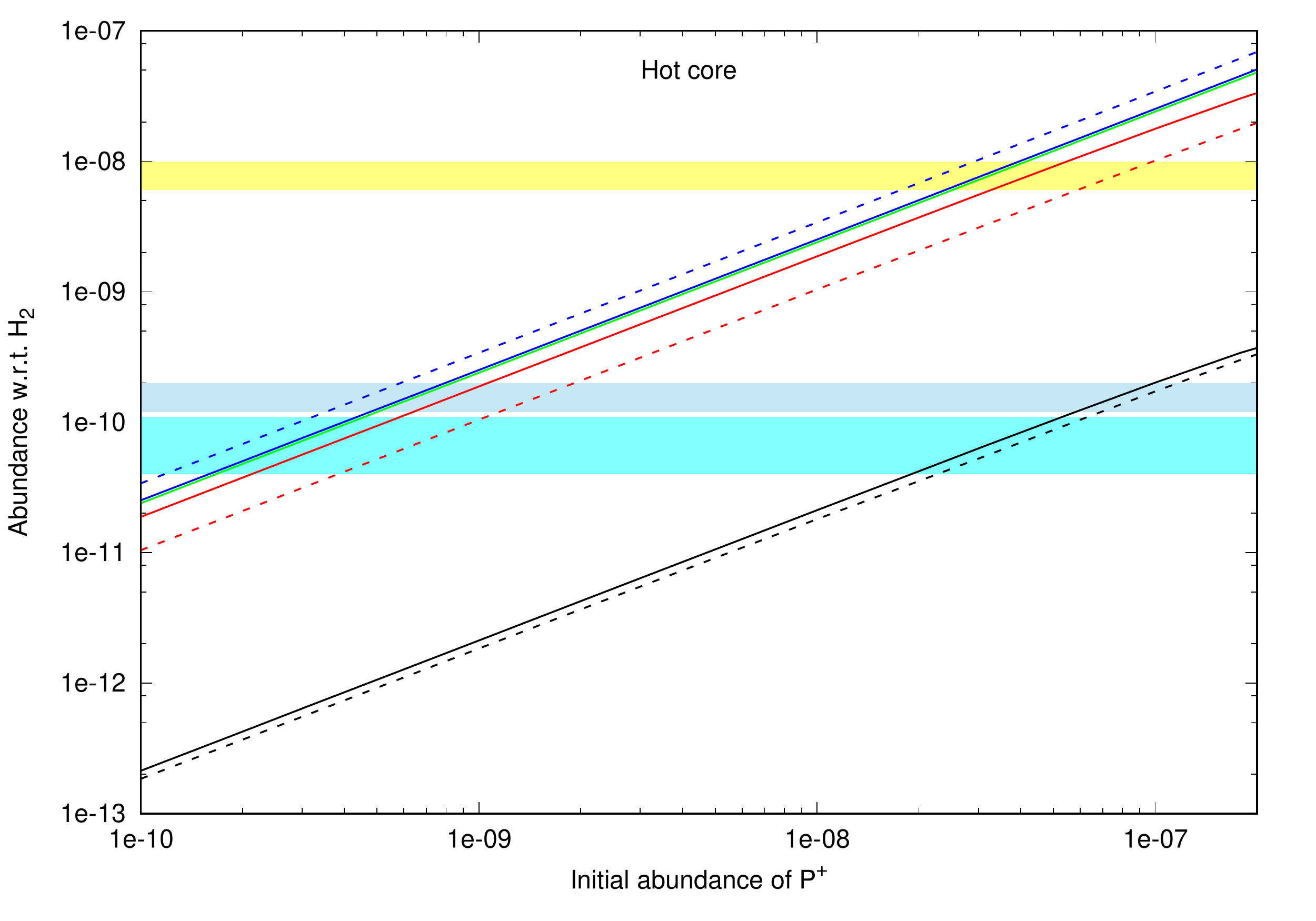}
\includegraphics[width=7cm, height=9.5cm, angle=270]{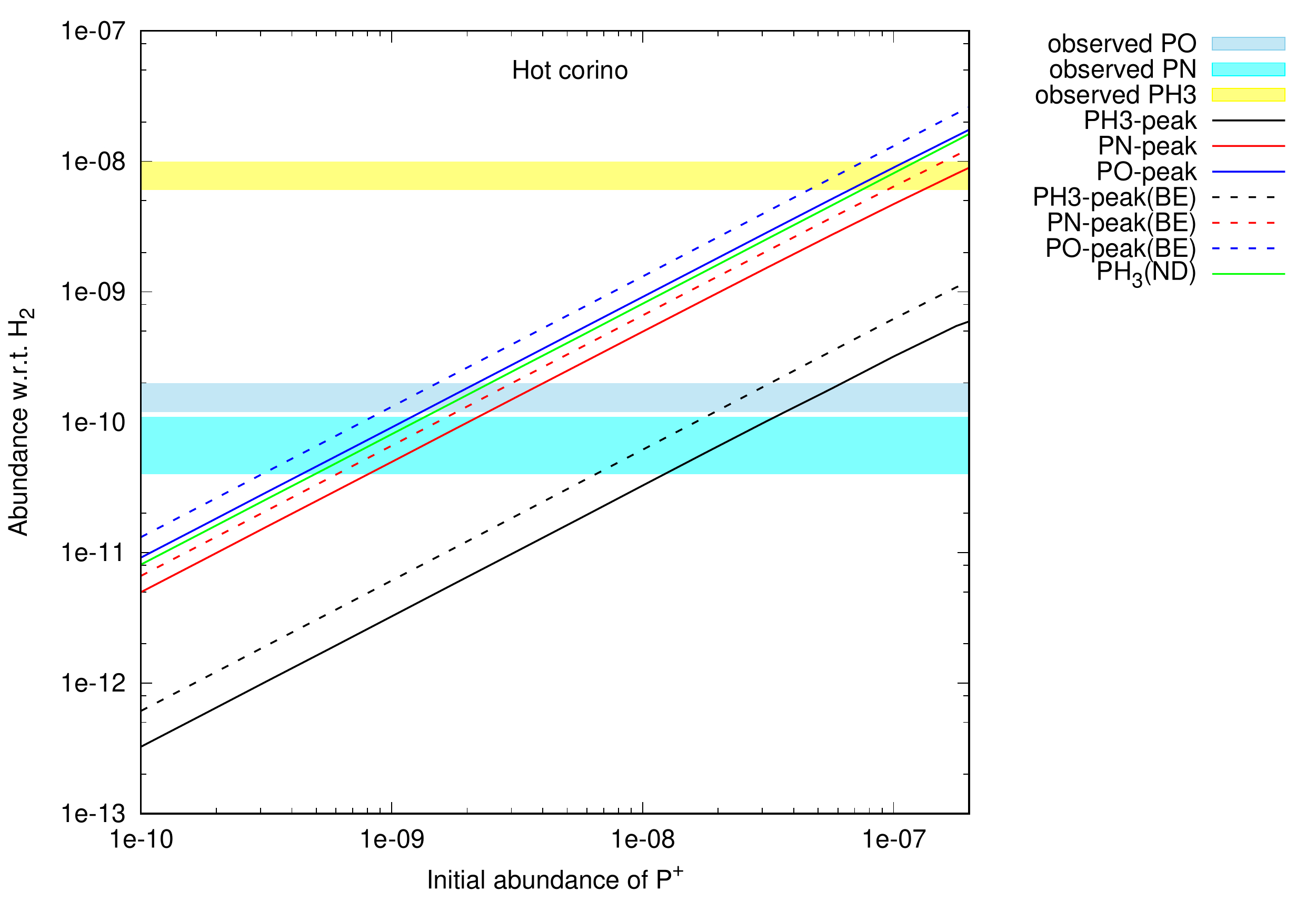}
 \caption{The observed abundances of PO and PN in the high mass star-forming region W51 and W3(OH) \citep{rivi16} are shown along with our modeled peak abundances (taken beyond the isothermal phase) with hot core (left panel) and hot corino (right panel) cases of our CMMC model. 
The peak abundance variation of the PH$_3$ is also shown with the initial abundance of P$^+$. PH$_3$ is yet to be observed in hot core/corino. The obtained peak abundance of PH$_3$ is far below the observed limit of PH$_3$ in C-star envelope IRC +10216 \citep{agun08,agun14}. 
Peak abundances of PO, PN, and PH$_3$ are also shown with the dashed lines when the BE of the P-bearing species is considered from the tetramer configuration noted in Table \ref{tab:binding}.
The solid green curve shows the peak abundance of PH$_3$ in the absence of its destruction by H and OH. We obtained a significantly higher peak abundance of PH$_3$ with the lack of these destruction pathways.}
\label{fig:depletion}
\end{figure*}

\begin{deluxetable}{cccc}
\tablecaption{Initial elemental abundance for the hot core/corino model considered in the CMMC code \citep{wake08}. \label{tab:init_dense}}
\tablewidth{0pt}
\tabletypesize{\scriptsize} 
\tablehead{
\colhead{\bf Element} &\colhead{\bf Abundance} & \colhead{\bf Element} &\colhead{\bf Abundance}
}
\startdata
H$_2$     & $0.5$ & Fe$^+$       & $3.00 \times 10^{-9}$\\
He        & $1.40 \times 10^{-1}$ & Na$^+$       & $2.00 \times 10^{-9}$\\
N         & $2.14 \times 10^{-5}$ & Mg$^+$       & $7.00 \times 10^{-9}$\\
O         & $1.76 \times 10^{-4}$ & Cl$^+$       & $1.00 \times 10^{-9}$\\
C$^+$        & $7.30 \times 10^{-5}$ & P$^+$        & $2.00 \times 10^{-10}$\\
S$^+$        & $8.00 \times 10^{-8}$ & F$^+$        & $6.68 \times 10^{-9}$\\
Si$^+$       & $8.00 \times 10^{-9}$ & e$^-$        & $7.31 \times 10^{-5}$\\
\enddata
\end{deluxetable}

\subsubsection{Hot core / Hot corino model \label{HC}}
Here, we consider a 3-phase starless collapsing cloud model described in \cite{gora20a}. The first phase corresponds to an isothermal ($T_{dust}=T_{gas}=10$ K) collapsing phase where density can increase from $3 \times 10^3$ cm$^{-3}$ to $10^7$ cm$^{-3}$ and visual extinction parameter ($A_V$) can increase from $2$ to $200$. The second phase corresponds to a warm-up phase where temperature rises from 10 to 200 K keeping $A_V$ constant at its maximum value in $5 \times 10^4$ years. The last phase is a post-warm-up phase where density, temperature, and visual extinction are constant at their respective highest values and continue for $10^5$ years. Based on the time scale of the initial isothermal collapsing phase, we define the hot core and hot corino case. In the hot core case, we use an isothermal collapsing time scale $\sim 10^5$ years, whereas, for the hot corino case, a relatively longer timescale ($\sim 10^6$ years) is used. Thus, the total simulation time scale of $2.5 \times 10^5$ years and $1.15 \times 10^6$ years is considered for the hot core and hot corino cases, respectively.
Table \ref{tab:init_dense} shows the low metallic initial elemental abundance \citep{wake08}, which is considered here. 

\cite{rivi16} reported the observed molecular abundance with an uncertainty of $\sim (0.5-5)\times10^{-10}$ for PO and PN. The ratio between PO and PN varies in the range $1-5$. Figure \ref{fig:depletion} shows the variation of the peak abundance of PO and PN in the case of hot core and hot corino by considering various initial elemental P abundances. 
The peak abundances of PN and PO from our hot core (left panel) model shows a good match with the observation toward the massive star-forming region when an initial elemental P abundance (P$^+$ abundance) of 
$2 \times 10^{-10}- 2 \times 10^{-9}$ is considered.

The solid curve of Figure \ref{fig:depletion} represents the {\it BE} of the P-bearing species as it is in KIDA. 
The dashed curve represents our computed BE values (BE with water c-tetramer) of phosphorous (see the BE values in Table \ref{tab:binding}) and keeps all other {\it BE} values the same as in KIDA. We notice an increase in the peak abundances of PO and PN for the inclusion of our calculated {\it BEs} for the hot corino case. We have obtained that the peak abundance of PN increases, whereas it decreases for PO for the hot core case.
PO's peak abundance is always more prominent than the peak abundance (beyond the isothermal phase) of PN in both cases.
Variation of the PH$_3$ abundance is also shown, which is far below the derived upper limit for the C-rich envelop \cite{agun08,agun14}. 
The changes in {\it BE} reflects a significant increase in the gas phase abundance of PH$_3$ for the hot corino case, but
a marginal decrease in the peak abundance of PH$_3$ is obtained in the case of the hot core. 
Additionally, in Figure \ref{fig:depletion}, PH$_3$ abundance is shown with a solid-green curve (with the BE of KIDA) when its destruction by H and OH is not included. It offers a significantly higher abundance of PH$_3$. 

Unless otherwise stated, we always use a  non-thermal desorption factor of $a=0.01$. Additionally, we have tested with a reactive desorption factor of $0.023$ \citep{nguy20} for PH$_3$, which yields roughly $2.3$ times higher abundance of PH$_3$ in our case.

\begin{figure*}
\centering
\includegraphics[width=8cm, height=12cm, angle=270]{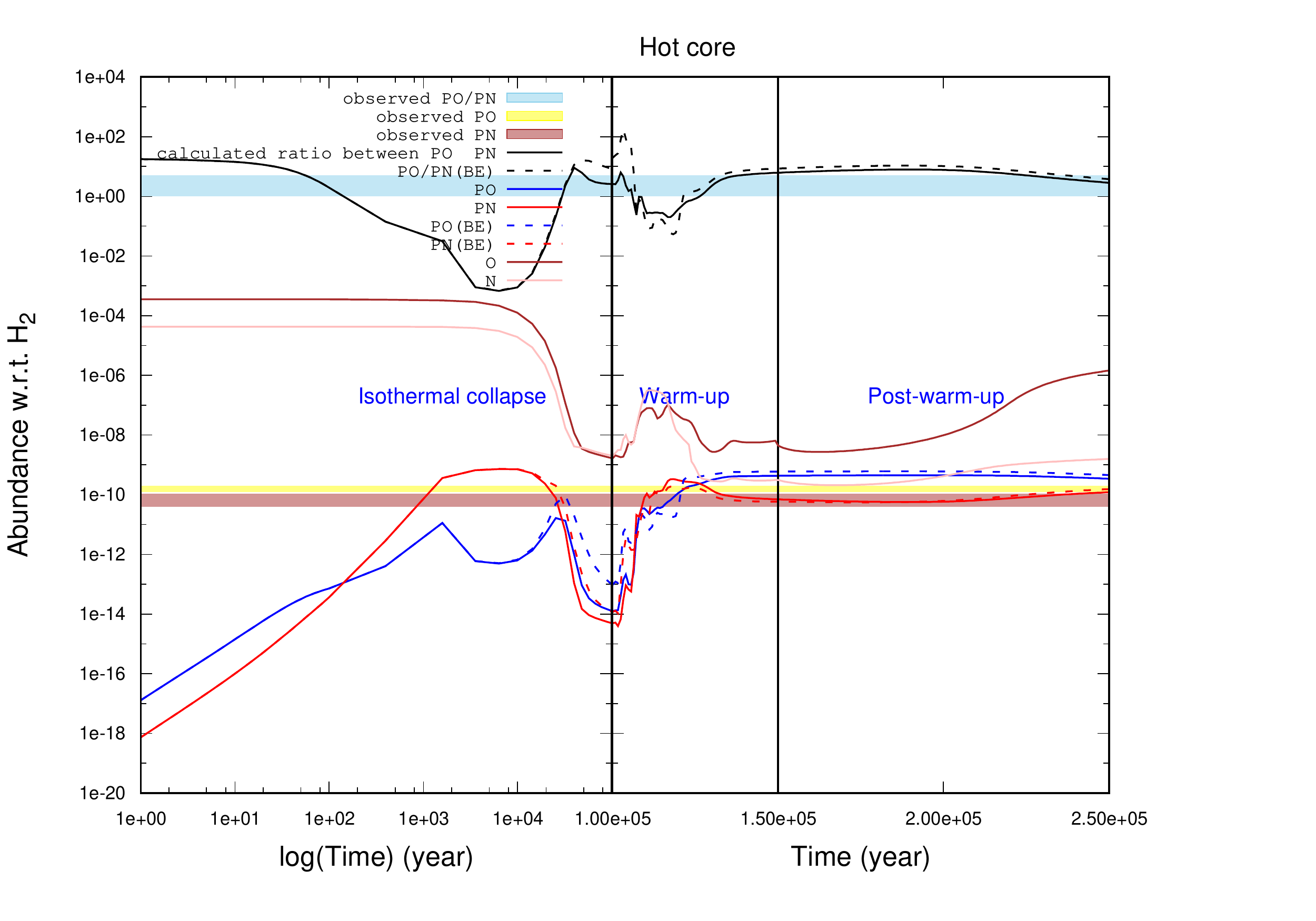}
\includegraphics[width=8cm, height=12cm, angle=270]{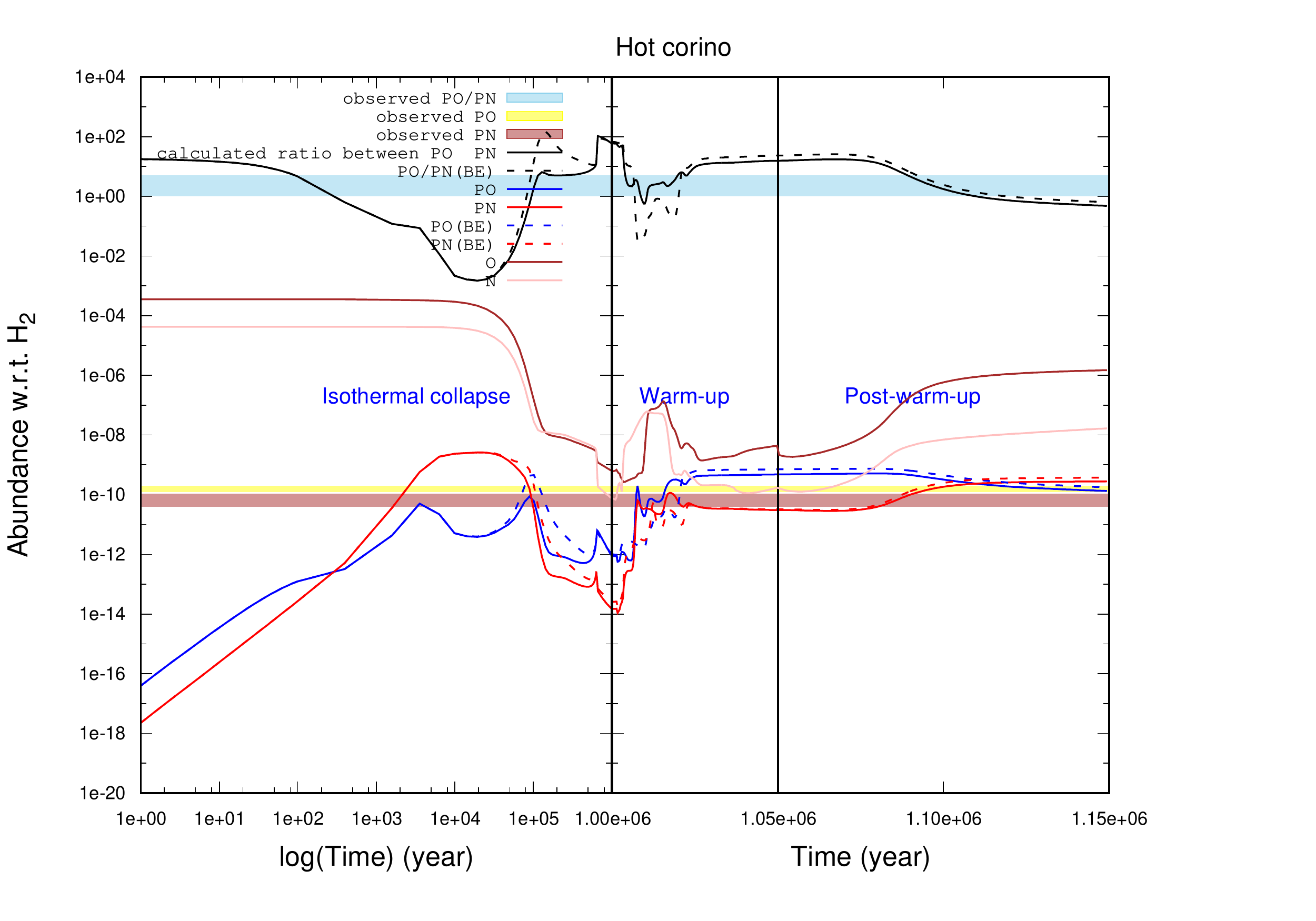}
\caption{Abundance variation of PO, PN, and PO/PN ratio {obtained from our CMMC model} by considering the initial elemental abundance of P$^+$ as $1.8 \times 10^{-9}$ for hot core and $5.6 \times 10^{-9}$ for the hot corino case. Solid lines represent the cases with the BEs from KIDA, and the dashed lines represent the values with the tetramer configuration of water. During the warm-up and post-warm-up phases, we have a good correlation with our results.}
\label{fig:PObyPN}
\end{figure*}

\begin{figure*}
\centering
\includegraphics[width=8cm, height=12cm, angle=270]{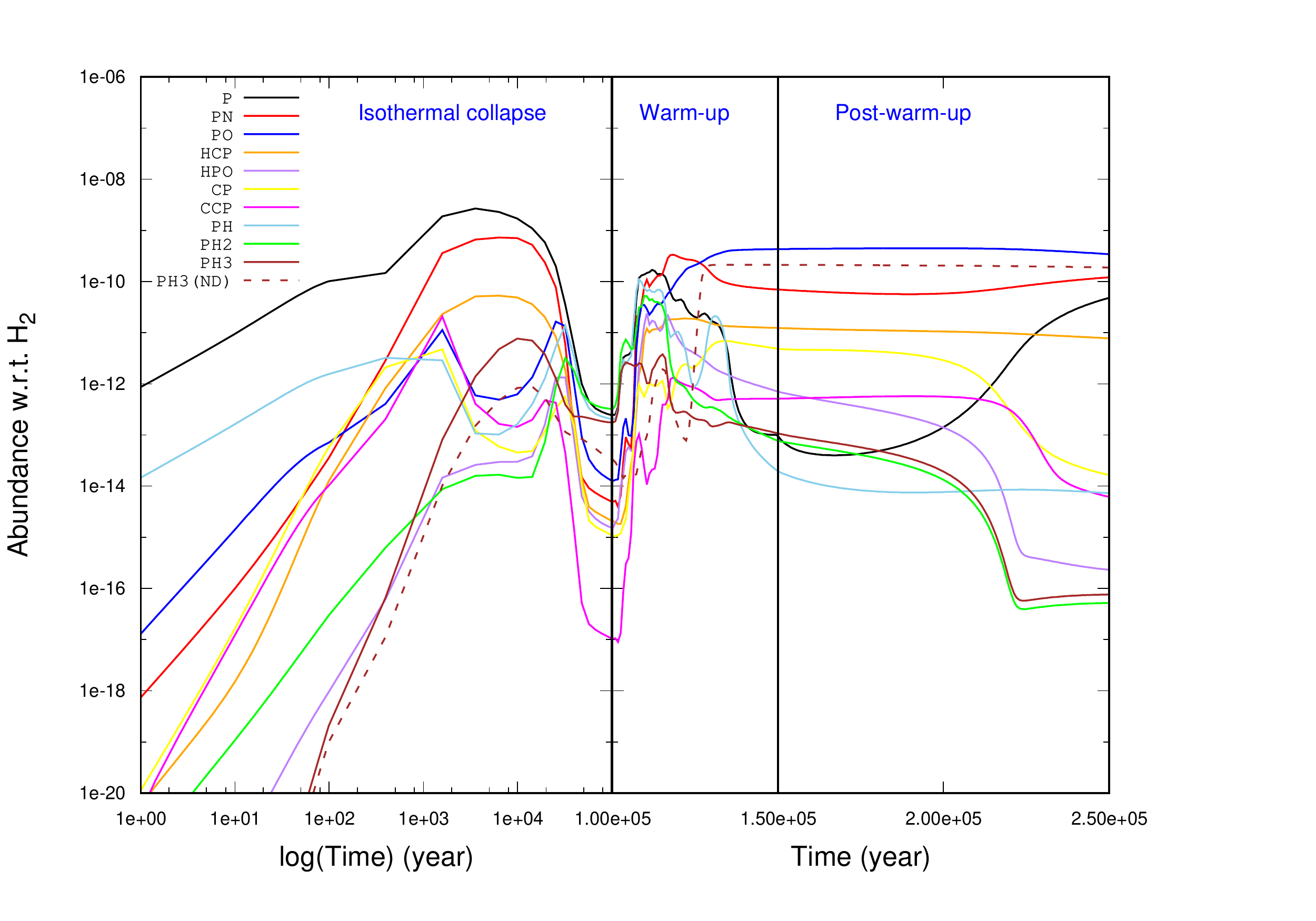}
\includegraphics[width=8cm, height=12cm, angle=270]{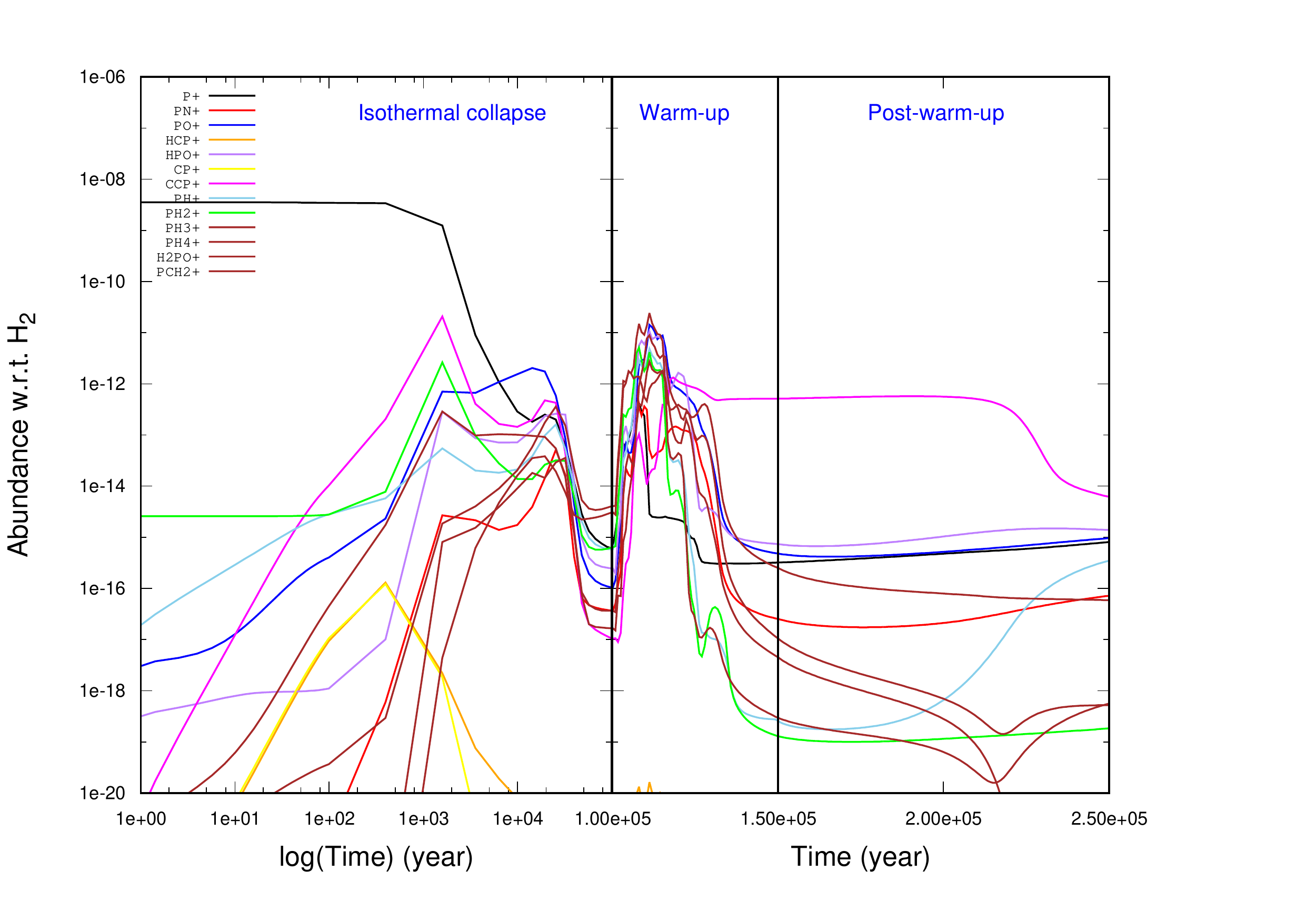}
\caption{Abundance variation of most of the P-bearing neutral and ionic species are shown from our CMMC model considering initial elemental abundance of P$^+$ as $1.8 \times 10^{-9}$ and for hot core region. In the absence of the destruction of PH$_3$ by H and OH, PH$_3$ is highly abundant.}
\label{fig:PObyPN-evo}
\end{figure*}

The solid lines of Figure \ref{fig:PObyPN} show the time evolution of the abundances of PO and PN and PO/PN for the hot core (upper panel) and hot corino (lower panel) case. The dashed lines represent the time evolution of the abundances of PO, PN, and the ratio PO/PN by considering the BEs of P-bearing species with the tetramer configuration of water. For a better understanding, we highlight the observed abundances and observed abundance ratio in the high mass star-forming region by \cite{rivi16}.
For the hot core case, we consider the initial P$^+$ abundance $\sim 1.8 \times 10^{-9}$, and for the hot corino case, a comparatively higher initial abundance of P$^+$ ($\sim 5.6 \times 10^{-9}$) is considered to match the observational result. 
We have obtained an exciting difference between the PO and PN abundances during the later stages of the simulation of the post-warm-up phase.
We have got PO/PN $> 1$ in the post-warm-up phase (latter stage of this phase) of the hot core, whereas it shows an opposite trend in the hot corino case.  The higher abundance of PN in the hot corino case happens due to the presence of a comparatively more elevated amount of atomic nitrogen (shown in Figure \ref{fig:PObyPN} with the pink curve) in this case.
This high abundance of atomic nitrogen undergoes $\rm{PO+ N \rightarrow PN + O}$ and yields PO/PN $<1$ in the hot corino case. Also, the abundance of atomic P increases due to the destruction of P-compounds.
Our modeling results for hot core (with PO/PN $>1$) and hot corino (with PO/PN $<1$) agree well with the modeling results presented by \cite{jime18}.
The peak abundance of Figure \ref{fig:PObyPN-evo} shows the time evolution of most of the P-bearing species with an initial P$^+$ abundance of $1.8 \times 10^{-9}$. It is interesting to note that at the end of the simulation time scale, mainly P is locked in the form of PO and PN. 

\begin{figure*}
\centering
\includegraphics[width=12cm, height=16cm, angle=270]{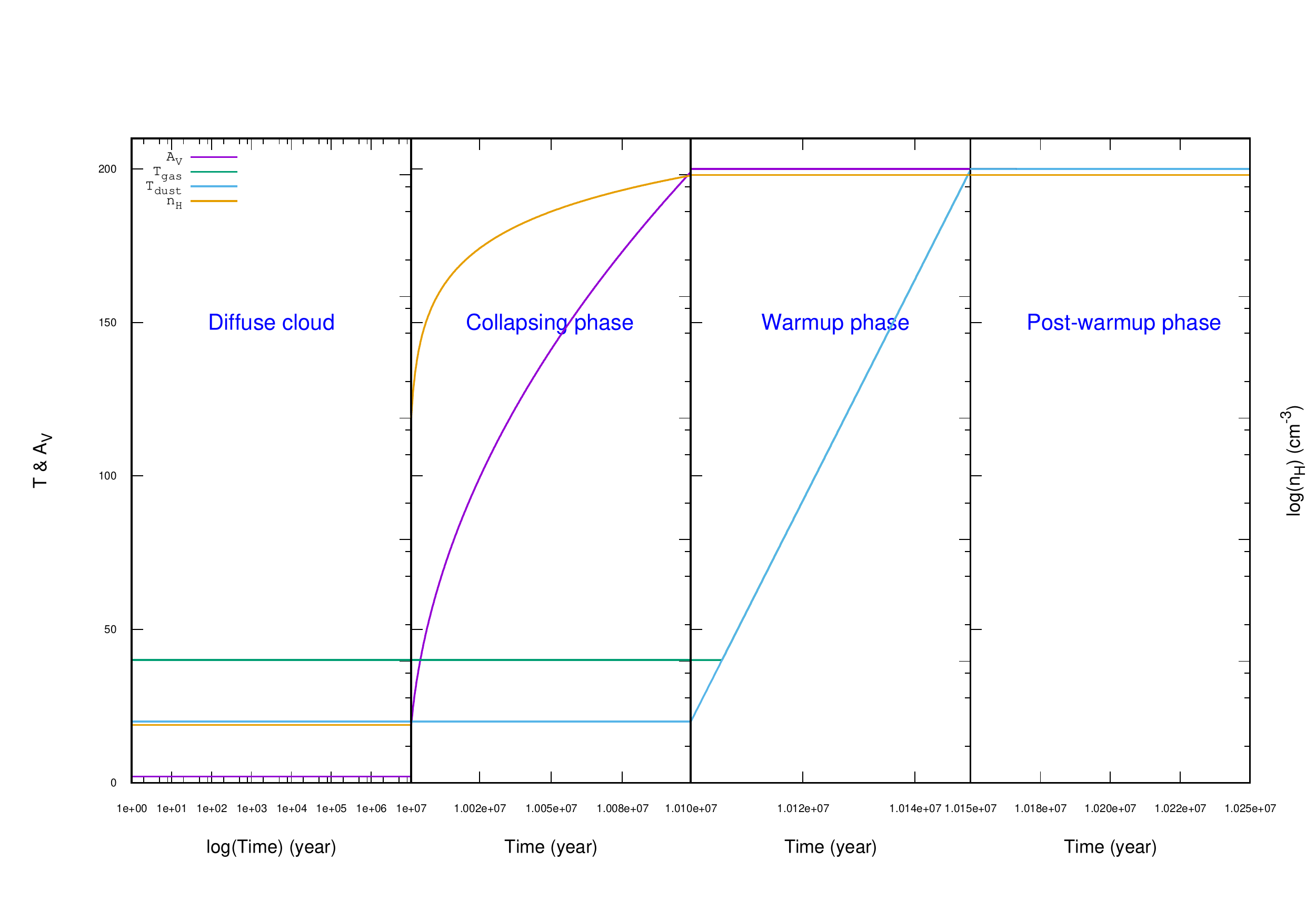}
\caption{Here, in our CMMC model, we consider four distinct phases for this simulation. In the first phase, the cloud remains in the diffuse stage for $10^7$ years. It starts to collapse in the next step, which continues for $10^5$ years. The collapsing phase is followed by a warm-up and post-warm-up phase, which continues for another $1.5 \times 10^5$ years.
\label{fig:diff_den}}
\end{figure*}

\begin{figure*}
\centering
\includegraphics[width=12cm, height=16cm, angle=270]{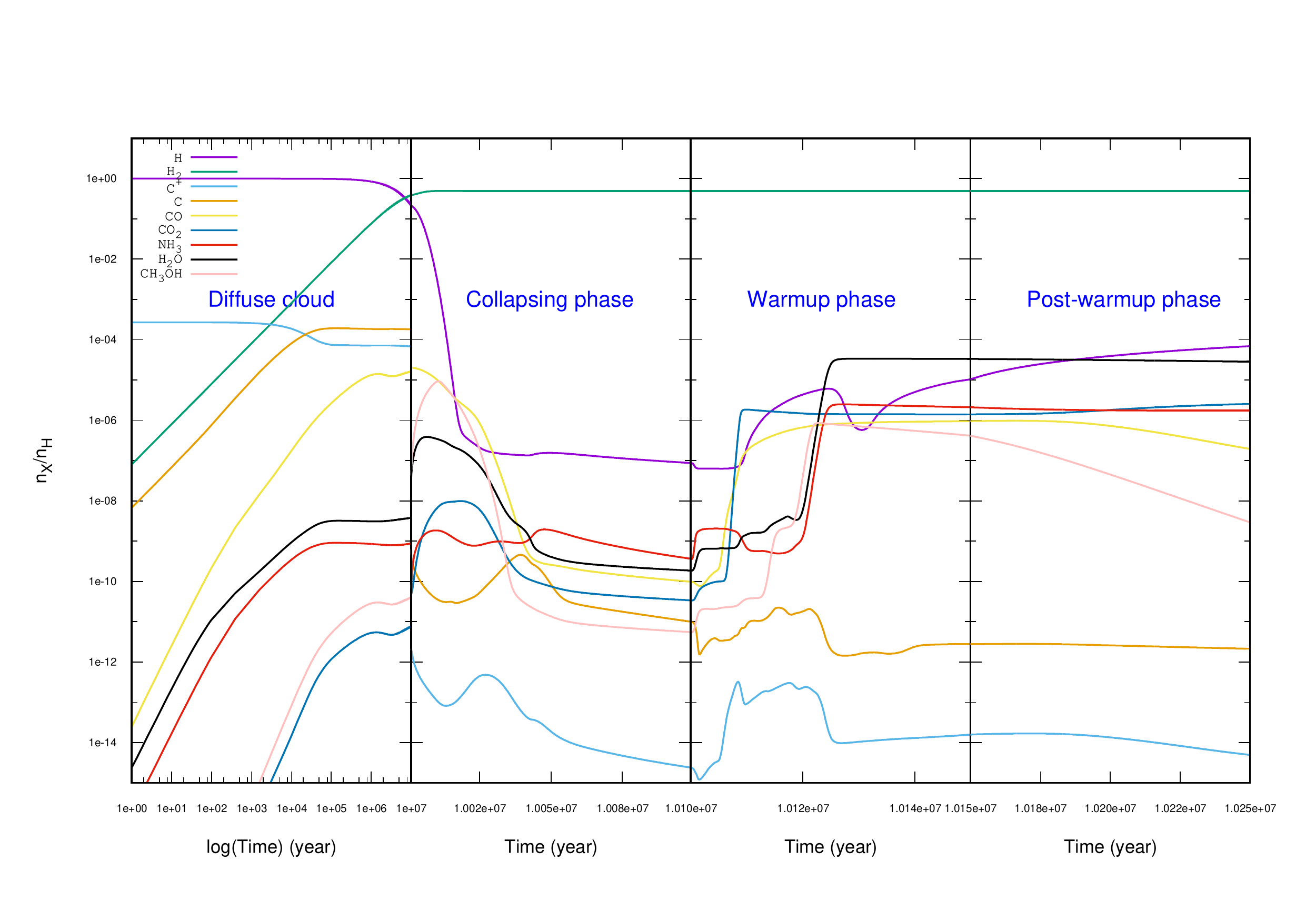}
\caption{Time evolution of H, H$_2$, C$^+$, C, CO, CO$_2$, NH$_3$, H$_2$O, and CH$_3$OH obtained from our CMMC model is shown. During the lifetime of the diffuse cloud H converts into H$_2$ and C$^+$ converts into C. During the collapsing phase, C atom is heavily depleted and converts into CO. 
\label{fig:mol_diff_den}}
\end{figure*}

\begin{figure*}
\centering
\includegraphics[width=12cm, height=16cm, angle=270]{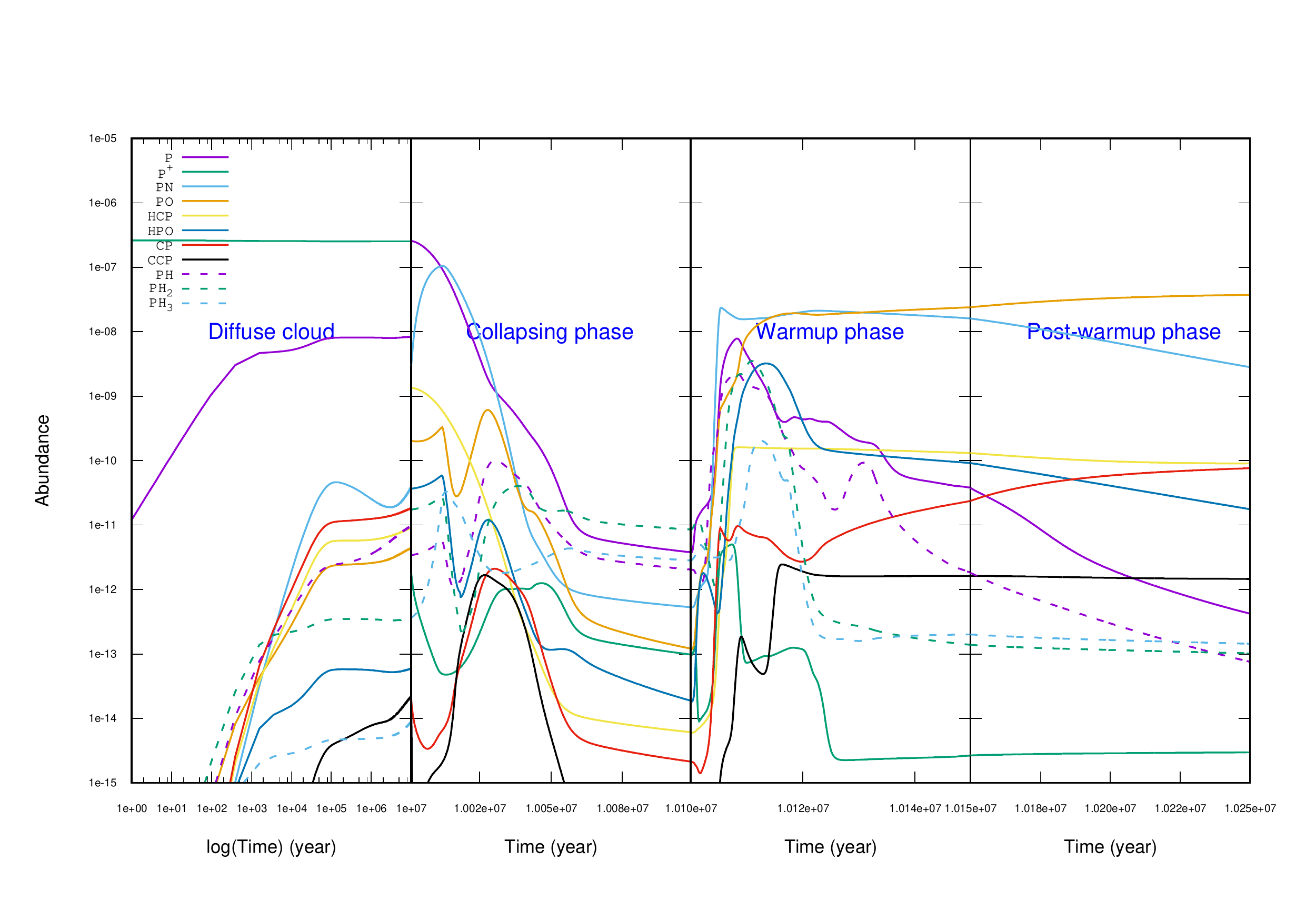}
\caption{Time evolution of the abundances (with respect to total hydrogen nuclei in all forms) of major P-bearing species from our CMMC model is shown.
\label{fig:P_diff_den}}
\end{figure*}

\subsubsection{Diffuse to Dense cloud model}
To avoid any ambiguity for considering the depletion factor, here we consider that the diffuse cloud converts into a molecular cloud after a sufficient time. The adopted physical condition for this model is shown in Figure \ref{fig:diff_den}.
 It depicts that the total simulation time is divided into four steps. The first step is the diffuse phase, and it continues for the initial $10^7$ years. The second step is the collapsing phase, whose span is for $10^5$ years. The warm-up phase starts after the collapsing phase, and it lasts for $5 \times 10^4$ years. Finally, the post-warm-up phase continues for another $10^5$ years. In total, the simulation time scale is $1.025 \times 10^7$ years.
Initial elemental abundance for this case was taken from \cite{chan20} (see Table \ref{tab:diff_cloud_CMMC}). All the BEs are used from the KIDA database.
Figure \ref{fig:mol_diff_den} shows the time evolution of the H, H$_2$, C$^+$, C, and CO. It is interesting to note that during the lifetime of the diffuse cloud ($\sim 10^7$ years), most of the H atoms convert into H$_2$. Ionized carbon is converted into neutral carbon. During the collapsing phase, the conversion of C to CO takes place. Once CO has formed, it starts to deplete on the grain and forms complex organic molecules (COMs). In the warm-up phase, these CO and its related molecules desorbed back to the gas phase. Figure \ref{fig:P_diff_den} shows the chemical evolution of the notable P-bearing species. Figure \ref{fig:P_diff_den} shows that in the warm-up phase, we have very high abundances of PN and PO. The abundance PH$_3$ is also significantly higher ($\sim 10^{-10}$ with respect to total H nuclei) comparable with \cite{jime18}. At the end of the simulation, we see that most of the P is locked in the form of PO and PN.

\begin{deluxetable*}{ccccccc}
\tablecaption{Peak abundances of the P-bearing species under various physical condition. The bracketed values in the CMMC model denote the case when the destruction of PH$_3$ by H and OH are ignored. \label{tab:abundances}}
\tablewidth{0pt}
\tabletypesize{\scriptsize} 
\tablehead{
\colhead{\bf Species} & \multicolumn{2}{c}{\textsc{Cloudy} output}&\multicolumn{4}{c}{CMMC output} \\
\colhead{ }& \colhead{Diffuse cloud ($n/n_H$)} & \colhead{PDR ($n/n_H$)} & \colhead{Diffuse cloud ($n/n_H$)}  & \colhead{Hot core $(n/n_{H_2}$)} & \colhead{Hot corino $(n/n_{H_2}$)} & \colhead{Diffuse-Dense ($n/n_{H_2}$)} \\
\colhead{ } &\colhead{($A_V=2$, $n_H=300$ cm$^{-3}$)} & \colhead{($A_V=5$, $T_{gas}=50$ K, $T_{ice}=20$ K} & \colhead{($A_V=2$, $T_{gas}=40$ K, $T_{ice}=20$ K} & \colhead{(initial P$^+ \sim 1.8 \times 10^{-9}$)} & \colhead{(initial P$^+ \sim 5.6 \times 10^{-9}$)} & \colhead{(initial P$^+ \sim 2.6 \times 10^{-7}$)} \\
\colhead{ } & \colhead{ } & \colhead{$n_H=10^{5.5}$ cm$^{-3}$)} & \colhead{$n_H=300$ cm$^{-3}$)} &\colhead{ } & \colhead{ }
}
\startdata
 PN&$3.33\times10^{-10}$& $4.02\times10^{-09}$ & $4.6 \times 10^{-11}$ & $1.7 \times 10^{-10} (2.9 \times 10^{-11})$ & $1.4 \times 10^{-10} (8.0 \times 10^{-11})$ & $1.04 \times 10^{-7}$ \\
 PO & $1.47\times10^{-10}$ & $2.60\times10^{-11}$ & $4.4 \times 10^{-12}$ & $2.3 \times 10^{-10} (4.9 \times 10^{-11})$ & $2.6 \times 10^{-10} (6.6 \times 10^{-11})$ & $3.7 \times 10^{-8}$ \\
PH & $1.13\times10^{-10}$ & $1.82\times10^{-11}$ & $9.6 \times 10^{-12}$ & $5.5 \times 10^{-11} (1.5 \times 10^{-11}$)& $6.5 \times 10^{-11} (8.9 \times 10^{-12})$ & $2.3 \times 10^{-9}$ \\
PH$_3$ & $6.30\times10^{-31}$& $1.16\times10^{-22}$ & $8.9 \times 10^{-15}$ & $1.9 \times 10^{-12} (2.1 \times 10^{-10})$ & $9.2 \times 10^{-12} (2.3 \times 10^{-10})$ & $2.0 \times 10^{-10}$ \\
CP & $2.03\times10^{-11}$ & $3.24\times10^{-12}$ & $1.8 \times 10^{-11}$ & $3.5 \times 10^{-12} (6.7 \times 10^{-13})$ & $3.1 \times 10^{-11} (3.2 \times 10^{-12})$ & $1.2 \times 10^{-10}$ \\
HCP & $1.38\times10^{-13}$ & $3.85\times10^{-14}$ & $9.1 \times 10^{-12}$ & $9.5 \times 10^{-12} (3.7 \times 10^{-12})$ & $3.5 \times 10^{-11} (4.3 \times 10^{-12})$ & $1.3 \times 10^{-9}$ \\
HPO & $4.43\times10^{-17}$ & $1.28\times10^{-19}$ & $5.9 \times 10^{-14}$ & $1.2 \times 10^{-11} (1.6 \times 10^{-12})$ & $1.5 \times 10^{-11} (9.8 \times 10^{-13})$ & $3.2 \times 10^{-9}$ \\
\enddata
\end{deluxetable*}

\begin{deluxetable*}{ccccccccc}
\tablecaption{\bf Summary of the obtained abundance ratio between PO and PN in different astrophysical environment. \label{tab:PO/PN}}
\tablewidth{0pt}
\tabletypesize{\scriptsize} 
\tablehead{
\colhead{\bf References} & \colhead{\bf Type of study} &\colhead{\bf PDR} & \colhead{\bf Diffuse cloud} &\colhead{\bf Hot core} & \colhead{\bf Hot corino} & \colhead{\bf Shock} & \colhead{\bf Comet} & \colhead{\bf Other}
}
\startdata
This work & Modeling & $<<1$ & $<1$         & $\ge 1^l$& $\ge 1^m$, $\le 1^n$ & \nodata & \nodata & \nodata \\
\cite{bern21} & Observation & \nodata & \nodata & \nodata & \nodata & \nodata & \nodata & 2.7$^a$ \\
\cite{rivi20}    & Observation & \nodata  & \nodata  & \nodata & \nodata & \nodata & $>10^b$ & 1.7$^c$ \\
\cite{chan20} & Modeling & & $<1$ & & & & & \\
\cite{berg19} & Observation & \nodata & \nodata & \nodata & \nodata & \nodata & \nodata & $(1-3)^d$ \\
\cite{rivi18} & Observation & \nodata & \nodata & \nodata & \nodata & \nodata & \nodata & $\sim1.5^e$ \\
\cite{jime18} & Modeling & $<<1$ & \nodata & $\geq 1$ & $<1$ & $\leq 1$ & \nodata & \nodata \\
\cite{lefl16} & Observation & \nodata & \nodata & \nodata & \nodata & $\approx3^f$ & \nodata & \nodata \\
\cite{rivi16} & Observation & \nodata & \nodata & $1.8^g$, $3^h$ & \nodata & \nodata & \nodata & \nodata \\
\cite{font16} & Observation & \nodata & \nodata & \nodata & \nodata & \nodata & \nodata & $<(1.3-4.5)^i$ \\
\cite{debe13} & Observation & \nodata & \nodata & \nodata & \nodata & \nodata & \nodata & $0.17-2^j$ \\
\cite{aota12} & Modeling  & \nodata & \nodata & \nodata & \nodata & $\sim(0.5-1.3)^{f,i}$ & \nodata & \nodata \\
\cite{tene07} & Observation & \nodata & \nodata & \nodata & \nodata & \nodata & \nodata & $2.2^k$ \\
\enddata
\tablecomments{
$^a$ Orion-KL,
$^b$ 67P/Churyumov-Gerasimenko \citep{altw16}. \\
$^c$ Average over multiple positions and velocity components toward AFGL 5142. \\
$^d$ Class I low-mass protostar B1-a. \\
$^e$ G+0.693-0.03, $^f$ L1157-B1, $^g$ W51 e1/e2, $^h$ W3(OH). \\
$^i$ Taking minimum and maximum values of upper limit of beam-averagd column densities of multiple sources. \\
$^j$ IK Tauri, $^k$ VY Canis Majoris.\\
$^l$ For the late warmup to post warmup phase of the hot core model.\\
$^m$ During the late warm-up to the initial post-warm-up of hot corino phase.\\
$^n$ During the late post warm-up phase.
}
\end{deluxetable*}

\subsection{An outline of our modeling results and a comparison with the earlier results}
Table \ref{tab:abundances} shows the peak abundances of the significant P-bearing species obtained with the two models (\textsc{Cloudy} and CMMC). However, due to more realistic physical conditions, the \textsc{Cloudy} code is preferred for the diffuse cloud region.
Both the diffuse cloud models (\textsc{Cloudy} and CMMC) show that PN's abundance is higher than that of PO.
 \cite{chan20} also predicted 
a higher abundance of PN ($4.8 \times 10^{-11}$) compared to PO  ($1.4 \times 10^{-11}$) in the diffuse cloud region. 
The abundances of PH$ _3$ dramatically vary between the two models; this is because \textsc{Cloudy} does not consider the grain surface reactions for the formation of PH$_3$, whereas the grain chemistry is adequately considered in CMMC. Moreover, the adopted physical condition differs between the two models. We have obtained a higher abundance of P-bearing species for our diffuse-dense model because of the consideration of the initially high abundance of P$^+$ ($\sim 2.6 \times 10^{-7}$).
In general, we have obtained a nice trend with the peak abundance ratio between PO and PN. We find PO/PN ratio $<1$ for the diffuse cloud region, $<<1$ for the PDR region, and $>1$ for the hot core/cornio (late warm-up phase)  region. In the late post-warm-up phase, we have PO/PN $>1$ for hot core and $<1$ for the hot corino. We have noticed that the reaction
$\rm{PO+N \rightarrow PN + O}$ is mainly responsible for controlling this ratio.

 The results of our diffuse cloud modeling with the \textsc{Cloudy} and CMMC code is in line with that obtained by \cite{chan20}. The significant difference between our diffuse cloud model with the \textsc{Cloudy} code and the model presented in \cite{chan20} is in consideration of realistic physical conditions instead of just considering the constant temperature. In contrast, the diffuse cloud model with CMMC code adopted similar physical parameters considered in \cite{chan20}. Since \textsc{Cloudy} was not equipped with adequate surface chemistry treatment, we did not have a notable abundance of PH$_3$. Our CMMC results show lower PH$_3$ because of its destruction by H and OH.
 
 \cite{char94} studied the chemistry of P in hot molecular cores. They considered that the gas phase P-chemistry in the hot core starts from PH$_3$ release from the interstellar grains. In their case, PH$_3$ was gradually destroyed and transformed into P, PO, and PN. They noticed that other P-bearing species formation time scale is longer than those of the hot core. We consider the in situ formation of PH$_3$ via chemical reaction on interstellar grains. Like \cite{char94}, in Figure \ref{fig:PObyPN-evo}, we also notice that at the end of the warm-up phase, the abundances of most of the P-bearing species remain in the form of P, PO, and PN.
 
 \cite{rivi16} found that PO and PN are chemically associated and formed during the cold collapse phase by the gas phase reactions. At the end of the collapsing phase, these two species were deposited to the interstellar grain. These again desorb back to the gas phase during the warm-up phase when the temperature is around 35 K. Our CMMC model shown in Figure \ref{fig:PObyPN} shows the similar behavior of PO and PN. In the isothermal collapsing phase, the gas-phase abundance of these two species offers a higher abundance, while at the end of this phase, these two are depleted to the grain. Once the temperature has become $>35$ K in the warm-up phase, it desorbed back to the gas phase.
 
\cite{jime18} constructed their model to explain the abundances of P-bearing species in a wide range of astrophysical conditions.
They constructed a short-lived and long-lived chemical model depending on the time of the collapse. In the short-lived collapse, they stopped their calculations when the gas density reaches its maximum value. In the long-lived collapse, they considered some additional time after getting the final density. They noticed that the P and PN are the most abundant phosphorous-bearing species in the collapsing phase. Their maximum abundance varies in the range $(5-10) \times 10^{-10}$. From our hot core model, in the collapsing phase (see Figure \ref{fig:PObyPN-evo}), we also have obtained that P and PN remain the most abundant P-bearing species, and their peak abundance varies in the range $(7-27) \times 10^{-10}$. 

\cite{jime16} carried out a high sensitivity observation toward the L1544 pre-stellar core. They were not able to identify the PN transitions. However, they predicted an upper limit of $\sim 4.6 \times 10^{-13}$ for the abundance of PN. The first phase of Figure \ref{fig:PObyPN-evo} represents the isothermal collapsing phase of a hot core. At the end of the collapsing phase, we have obtained a PN abundance of $\sim 5.4 \times 10^{-15}$, which is consistent with this upper limit.
\cite{font16} identified PN in various dense cores, which are in different evolutionary stages (starless, with a protostellar object, and with ultracompact H II region) of the intermediate and high-mass star formation. They obtained all the transitions of PN where the temperature is $<100$ K, and linewidths are $<5$ km s$^{-1}$, suggesting that the origin of PN in the quiescent and cold region. Because of the lack of data of the thermal dust continuum emission (at the millimeter/submillimeter regime for all these sources), PN's abundances were not derived. \cite{mini18} identified a few transitions of PN toward some of these sources (a sample of nine massive dense cores in different evolutionary stages). They had calculated the H$_2$ total column densities of the sources and derived the abundances of PN. 
For the slightly warmer region ($25-30$ K), \cite{font16} and \cite{mini18} obtained $10^{-11}$ and $5 \times 10^{-12}$, respectively. Our Figure \ref{fig:PObyPN-evo} depicts that when the temperature is $\sim 35$ K in the warm-up phase, we have a little higher PN abundance $\sim 2.08 \times 10^{-11}$ with respect to H$_2$. 
The main reason behind this difference is that. \cite{mini18} found the excitation temperature of PN $\sim 5-30$ K. Since the average density ($\sim10^{4-5}$ cm$^{-3}$) of their targeted regions are below the critical density ($10^{5-6}$ cm$^{-3}$) of the PN, they suggested a sub thermal excitation of PN. The total hydrogen density can reach as high as $10^7$ cm$^{-3}$ in our isothermal phase.
In the warm-up phase, we have considered the same density. So in our case, we have an average density, which is greater than the critical density of PN. So, a direct comparison between our model and the observation of \cite{mini18} and \cite{font16} would not be appropriate. Here, we have referred to these observations to infer the enhancement of the PN abundance with the increase in temperature from $10$ K to $35$ K only.

After releasing to the gas phase, PH$_3$ can destroy rapidly \citep{jime18}. The gas-phase formation of PH$_3$ can continue by the reactions R144 and R145 of Table \ref{tab:reaction_path} (in the Appendix). These two reactions were considered in \cite{jime18} and found to contribute marginally. Here, we have some additional destruction reactions of PH$_3$ by H (grain phase reaction R4 and gas-phase reaction R149 of Table \ref{tab:reaction_path}) and OH (grain phase reaction R5 and gas-phase reaction R161 of Table \ref{tab:reaction_path}) which yields a much lower PH$_3$ in the gas phase. 

In Table \ref{tab:PO/PN}, a summary of PO/PN abundance ratio is listed in the different astrophysical environments, along with a comparison with the earlier literature \citep{bern21,rivi20,chan20,berg19,jime18,rivi18,rivi16,lefl16,font16,debe13,aota12,tene07}.
Our modeling results agree well with the observed \citep{rivi16} and modeled \citep{jime18,chan20} PO/PN ratios.

\section{1D-RATRAN radiative transfer model \label{sec:RATRAN}}
Here, we employ a 1D radiative transfer model \citep{hoge00} to look into the transitions from the notable P-bearing species in the diffuse cloud region and around the more evolved stage, the hot core/corino region.

\begin{figure} 
\centering
\includegraphics[width=8cm, height=6cm]{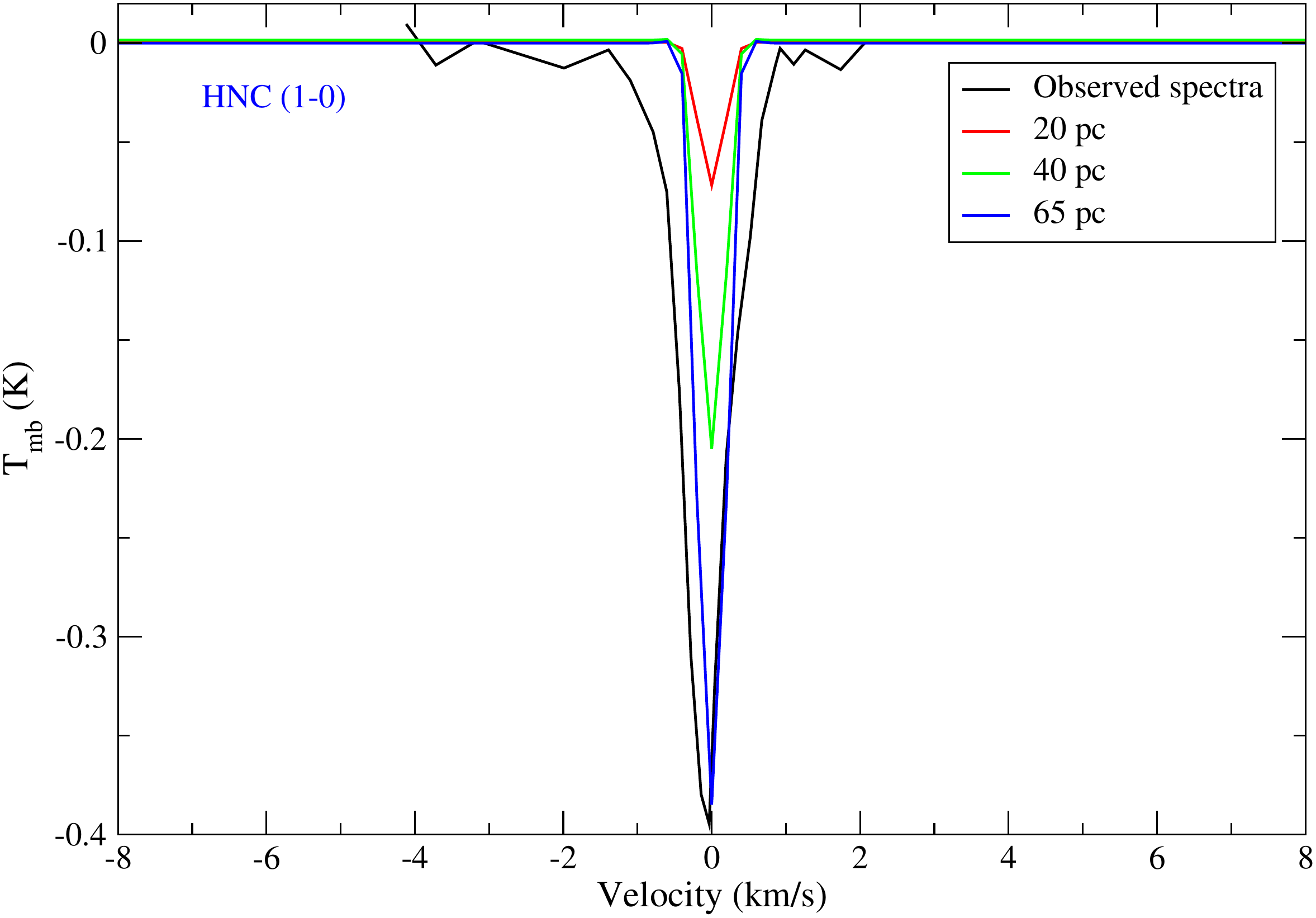}
\caption{Modeled $1 \rightarrow 0$ transitions of HNC by considering the different sizes of the diffuse cloud in the RATRAN model. It is interesting to note that the intensity of the absorption increases with the increase in the size of the cloud. Abundance of $2\times10^{-10}$ and Doppler parameter of 0.2 km/s are used.}
\label{fig:cloud_size}
\end{figure}

\begin{figure*} 
\centering
\includegraphics[width=8cm, height=6cm]{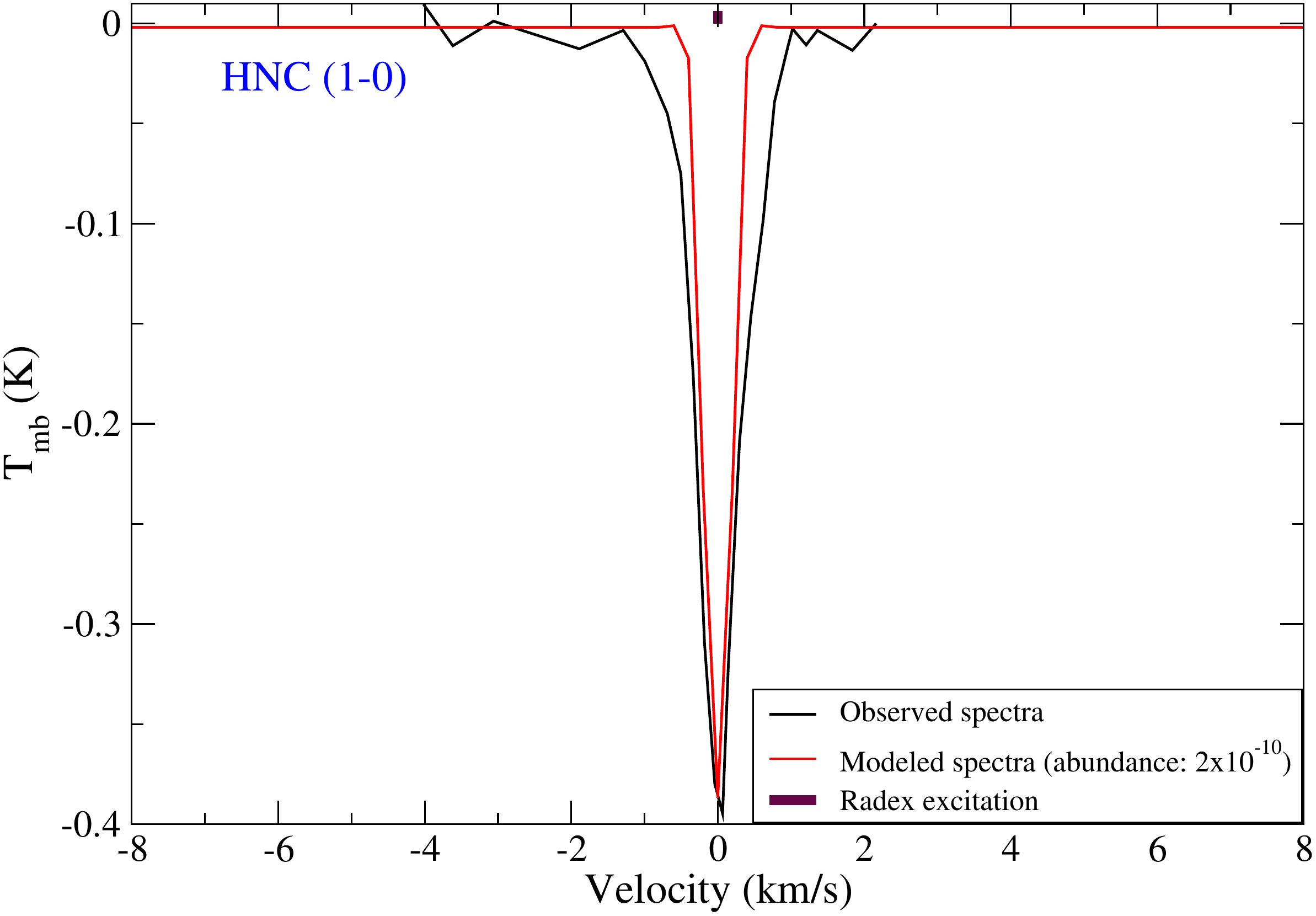}
\includegraphics[width=8cm, height=6cm]{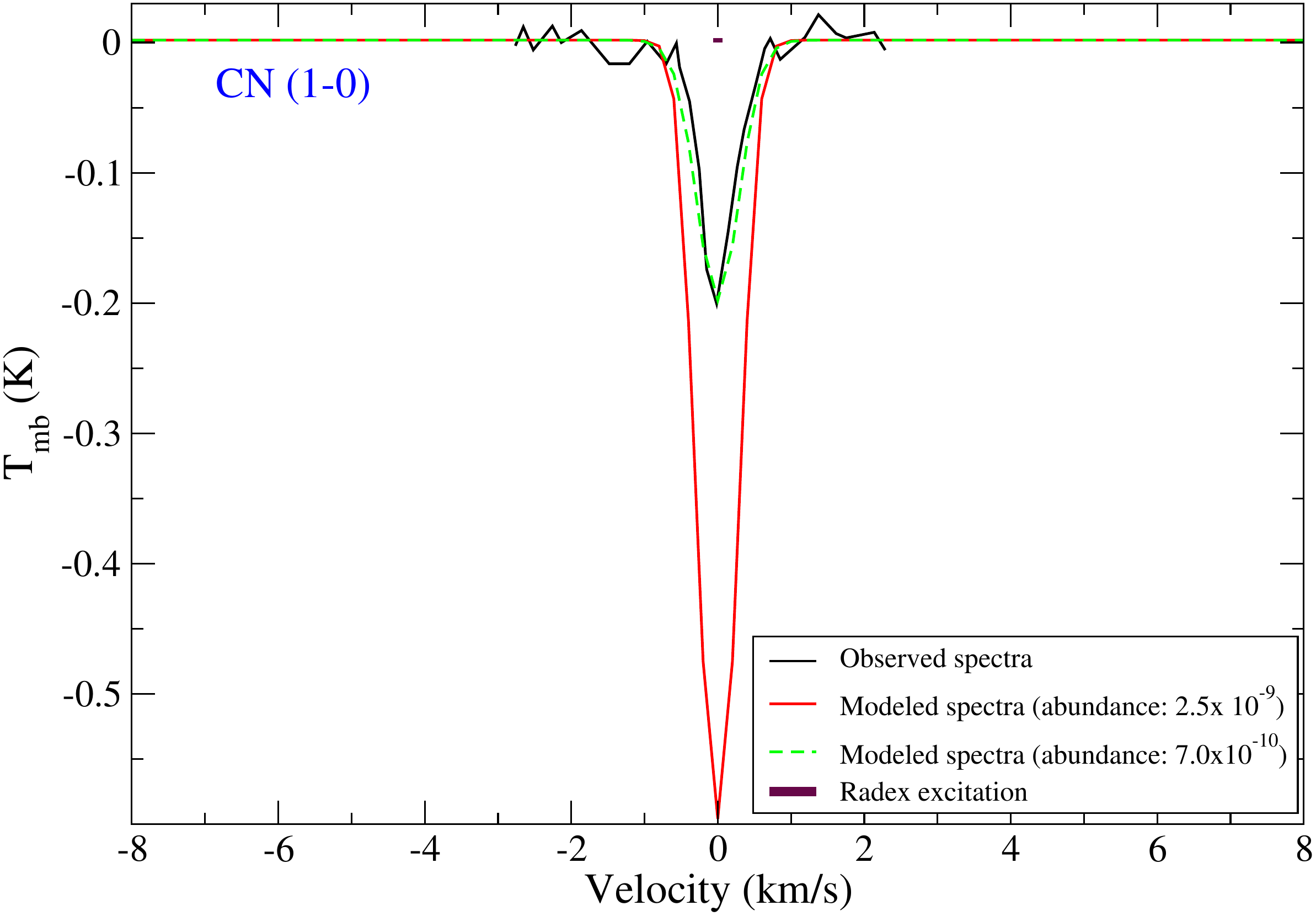}
\includegraphics[width=8cm, height=6cm]{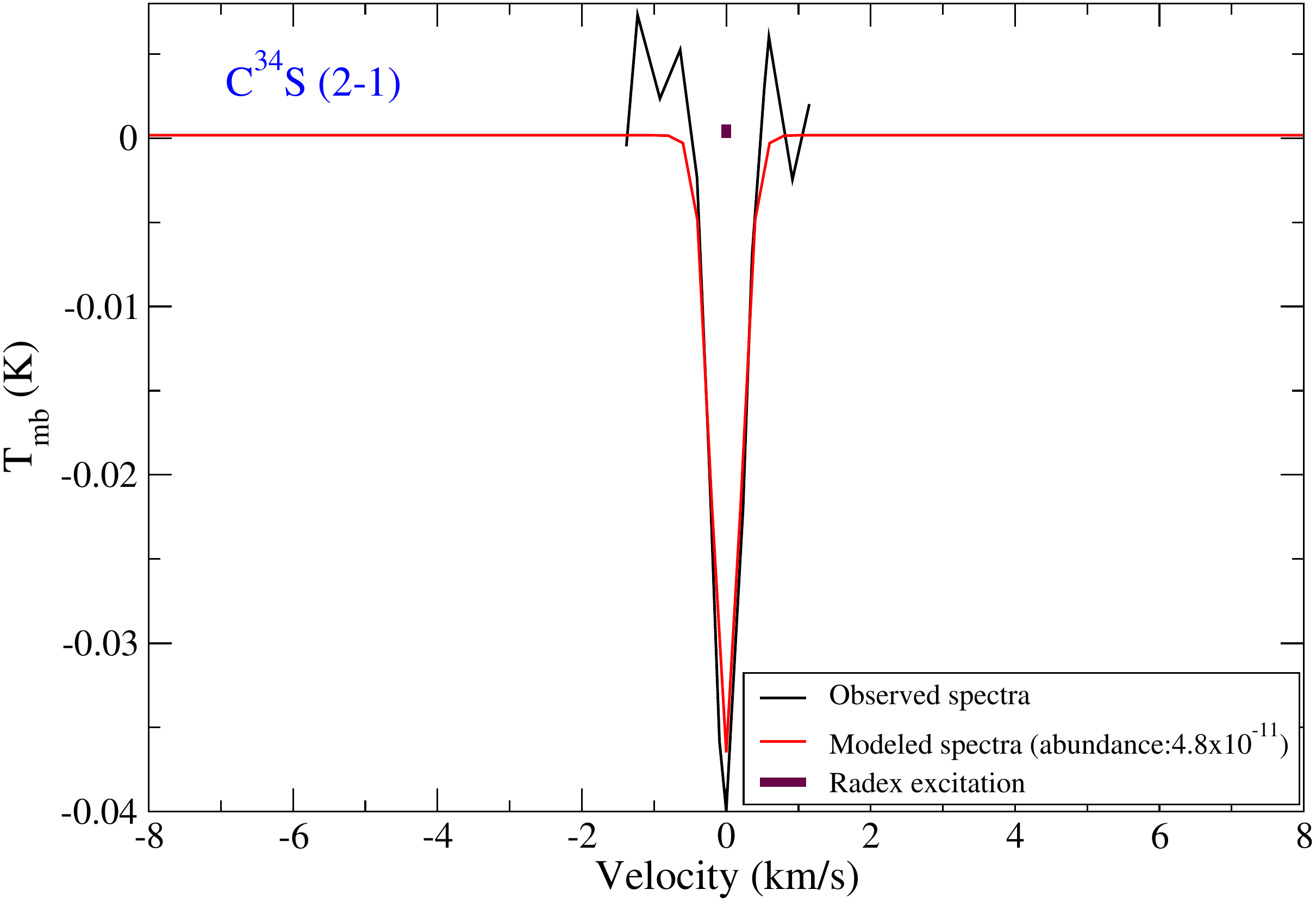}
\includegraphics[width=8cm, height=6cm]{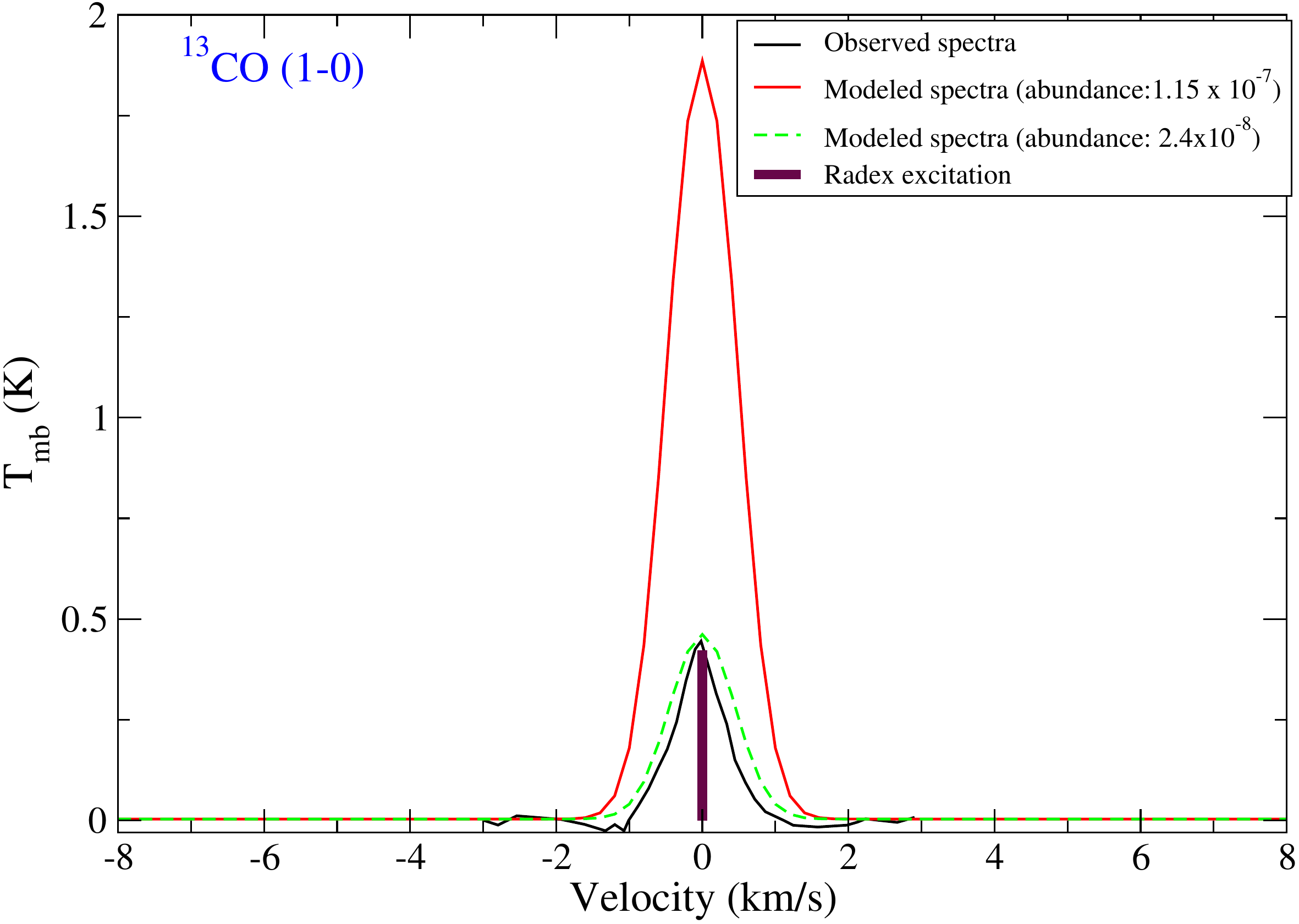}
\caption{Modeled line profile (with the RATRAN code) of HNC (1-0), CN (1-0), C$^{34}$S (2-1), and $^{13}$CO (1-0) for the diffuse cloud region. The black curve represents the observed line profile, whereas the red curve shows the modeled line profiles. Here, we consider H and H$_2$ as the collision partners having number density $100$ and $400$ cm$^{-3}$ for H and H$_2$ respectively. A temperature of $70$ K is considered for both gas and ice phases. The maroon vertical line shows the excitation with RADEX.
\label{fig:diffuse_observed}}
\end{figure*}

\begin{figure*} 
\centering
\includegraphics[width=15cm, height=10cm]{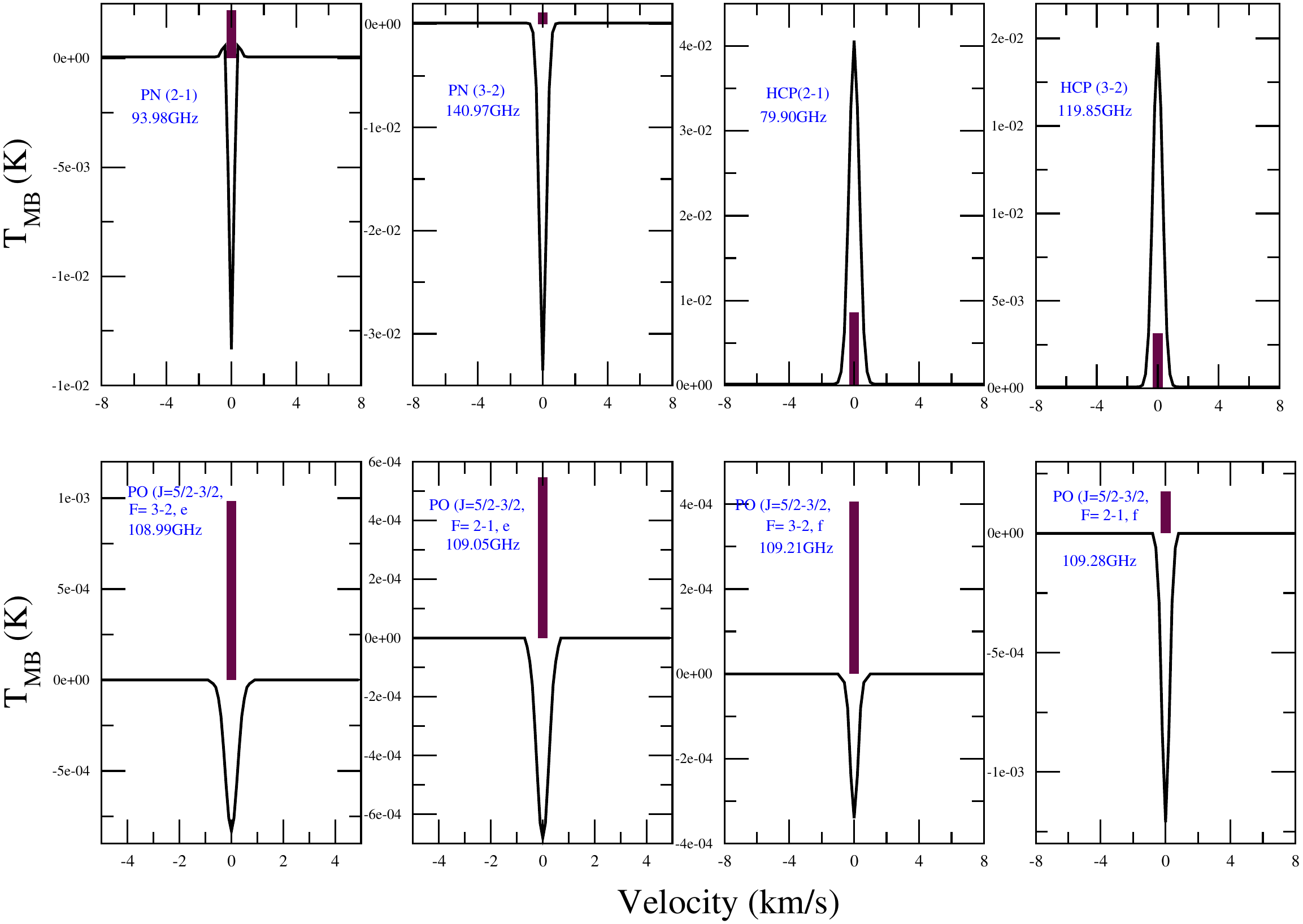}
\caption{Modeled line profile (with the RATRAN code) of PN, HCP, PO which can be observed in the diffuse cloud region. The black curve represent the modeled spectra whereas the maroon line represents the line excitation using RADEX. Here, we consider H and H$_2$ as the collision partners having number density $100$ and $400$ cm$^{-3}$ for H and H$_2$ respectively. A temperature of $70$ K is considered for both gas and ice phases.
\label{fig:diffuse_proposed}}
\end{figure*}

\subsection{Diffuse cloud model}
Here we model the galactic diffuse cloud region present in the line of sight toward the strong continuum source B0355+508 \citep{chan20}. We divide the entire cloud into $23$ spherical shells. Since around the diffuse cloud region, hardly a few mono-layers of ice could survive, we consider a dust absorption coefficient with the bare grain (MRN model with no ice mantle) with a coagulation time of $10^8$ years \citep[see Table 1 of][]{osse94}. We have used the dust emissivity ($\kappa$) by considering a power-law emissivity model.\begin{equation}
    \kappa=\kappa_0(\nu/\nu_0)^\beta,
\end{equation}
where $\kappa_0$=1520 cm$^2$/gm and $\nu_0$=1.62$\times$10$^{13}$ Hz \citep{osse94}. The power-law index $\beta$ is considered $\sim 1.0$. We have run several cases varying the number density of the collision partners (atomic and molecular hydrogen), the kinetic temperature of the gas, abundance profile, the Doppler broadening parameter, size of the cloud, etc. The diffuse cloud model discussed in the earlier section shows that at the end of the simulation ($\sim 10^7$ years), a significant portion of hydrogen would convert into H$_2$. Keeping this in mind, we consider $\rm{n_H=100}$ cm$^{-3}$ and $\rm{n_{H_2}=400}$ cm$^{-3}$ for this simulation. 

In Figure \ref{fig:cloud_size}, we show the absorption profile of the $1 \rightarrow 0$ transition of HNC as obtained from the 1D RATRAN model. Here, we consider the different sizes of the cloud to represent the absorption observed by \cite{chan20}. We notice that with the increase in the size of the cloud, the absorption is getting stronger. For the 65 pc size, we obtain a good match with the observed intensity. We use an abundance of $2 \times 10^{-10}$ and the Doppler parameter of $\sim 0.2$ km/s for HNC. In Figure \ref{fig:diffuse_observed}, the extracted observed spectra  of HNC(1-0), CN (1-0), C$^{34}$S (2-1) and $^{13}$CO (1-0) \citep{chan20} in black, along with the modeled spectra in red, are shown.
It is interesting to note that with the chosen parameter, we can successfully explain the absorption features of HNC, CN, and C$^{34}$S and emission feature of $^{13}$CO. 

In contrast to the modeling result of \cite{chan20} and the modeling results discussed in the earlier sections above, a combination of higher gas temperature ($\sim 70$ K) and density is required to explain the observed line profiles. We consider the gas and dust temperature to be equal in our calculation. Here, we use the collisional data file in the LAMDA database for HNC, CN, $^{13}$CO with H$_2$. 
No collisional rates with H were available. For simplicity, we consider it to be the same as it is with H$_2$. 
We further have tested using the scaled rate for H \citep{scho05}, no significant difference in synthetic spectra is observed.
Since the collisional data file of C$^{34}$S was not available, we consider it similar to the C$^{32}$S from the LAMDA database.  We obtain the best-fitted Doppler broadening parameter of 0.2, 0.4, 0.25, and 0.592 km/s for HNC, CN, C$^{34}$S, and $^{13}$CO, respectively, are consistent with the FWHM values obtained by \cite{chan20}. It is to be noted that our results can simultaneously explain the absorption features of HNC, CN, C$^{34}$S, and the emission feature of $^{13}$CO. 

Here, we convolve the synthetic spectra with the perfect beam size obtained for IRAM-30m observation.
We obtain a best-fitted abundance of  $ \sim 2 \times 10^{-10}$ for HNC. \cite{chan20} obtained an abundance of HNC $\sim 2.1 \times 10^{-10}$ from the modeling. 
\cite{chan20} calculated $^{32}$S/$^{34}$S isotopic ratio for the $-17$ km/s cloud is $18.7$, and $^{12}$CO/$^{13}$CO isotopic ratio is $16.7$. Considering this fractionation ratio, the abundance ratio predicted by \cite{chan20} for HNC:CN:C$^{34}$S:$^{13}$CO  is $1:12.5:0.24:576.35$. 
Maintaining the same abundance ratio of \cite{chan20}, we have $2.5 \times 10^{-9}$, $4.8 \times 10^{-11}$, and $1.15 \times 10^{-7}$ for CN, C$^{34}$S, and $^{13}$CO, respectively. 
The modeled spectra with these abundances are shown in Figure \ref{fig:diffuse_observed} with the solid red lines. In the case of C$^{34}$S, we have an excellent match with this abundance. But for CN and $^{13}$CO, we have strong absorption and emission respectively with these abundances.
{ For CN, an abundance of $7 \times 10^{-10}$, and for $^{13}$CO, an abundance of  $2.4 \times 10^{-8}$ 
shows an excellent match (see the green dashed lines).}

In Figure \ref{fig:diffuse_proposed}, the line profiles of the P-bearing species PN, HCP, and PO in the diffuse cloud region are shown. Using the same diffuse cloud model and considering abundances obtained from the CMMC model, i.e., $4.6 \times 10^{-11}$ for PN, $9.1\times10^{-12}$ for HCP, and $4.4\times10^{-12}$ for PO, we generate the synthetic spectra for these three P-bearing molecules. For this purpose, {we use the Doppler parameter to be $0.3$ km/s for PN, HCP, and PO, respectively, from \cite{chan20} which is consistent with the low FWHM values observed for other molecules in the diffuse cloud}. The collisional data for HCP and PN are taken from Basecol\footnote{\url{(https://basecol.vamdc.eu/index.html)}} database. The collisional rate of PO 
was taken from the  LAMDA\footnote{\url{(https://home.strw.leidenuniv.nl/~moldata/)}} database. The collisional rate with the H atom is considered to be the same as it is with H$_2$. For the IRAM-30m telescope, the beam convolution is considered using the beam sizes mentioned in the Appendix Section \ref{sec:append_b} (see Tables \ref{table:PN_observation}, \ref{table:PO_observation}, \ref{table:HCP_observation}, \ref{table:PH3_observation}). Interestingly, for PN and PO, we have obtained the spectra in absorption, whereas for the HCP, we have obtained it in emission. To further check the excitation condition for the molecules observed and the phosphorus-bearing molecules targeted, we employ the RADEX code \citep{vand07}. For the RADEX calculation, we consider the column densities for HNC, CN, C$^{34}$S, $^{13}$CO as 0.69$\times$10$^{12}$, 0.87$\times$10$^{13}$, 1.64$\times$10$^{11}$ and 3.98$\times$10$^{14}$ cm$^{-2}$ and FWHM 0.73, 0.54, 0.43, 0.82 km/s respectively from table 4 of \cite{chan20} for the diffuse cloud at an offset velocity $\simeq$ -17 km/s with respect to the source. We consider the input kinetic temperature 40 K and number density of collision partner H as 300 cm$^{-3}$ and a very low H$_{2}$ density 10 cm$^{-3}$. A standard background of $2.73$ K is considered. In Figures \ref{fig:diffuse_observed} and \ref{fig:diffuse_proposed}, the line excitation obtained with the RADEX is shown with the maroon vertical line at the 0 km/s. 
Interestingly, in Figure \ref{fig:diffuse_observed}, the excitation for HNC, CN, and C$^{34}$S (for which absorption was observed) is very weak for the previously observed molecules. In contrast, the excitation with the RADEX completely matches with the observed spectra for $^{13}$CO, observed in emission. 
Under the optically thin approximation and considering a very crude assumption, for the two-level system, the critical density can be represented by the ratio between the Einstein A coefficient in s$^{-1}$  and collisional rates in cm$^3$ s$^{-1}$ \citep{shir15}. In Table \ref{tab:critical}, we have noted down the Einstein A coefficient and collisional rate and critical density of these transitions. It is clear from the table that, as expected, we have obtained the transitions of HCP and $^{13}$CO in emission (critical density low) and the rest are in absorption (critical density high) with the RATRAN code (see Figures \ref{fig:diffuse_observed} and \ref{fig:diffuse_proposed}).

\begin{deluxetable*}{cccccc}
\tablecaption{Critical density of some transitions under the optically thin approximation. \label{tab:critical}}
\tablewidth{0pt}
\tabletypesize{\scriptsize} 
\tablehead{
\colhead{\bf Species} &\colhead{\bf Transitions} &\colhead{\bf Frequency}&\colhead{\bf Einstein coefficient} &\colhead{\bf Collision rate}&\colhead{\bf Critical density}\\
&&\colhead{\bf (GHz)}&\colhead{\bf (s$^{-1})$}&\colhead{\bf at $\sim 10$ K (cm$^{3} s^{-1}$)}&\colhead{\bf (cm$^{-3}$)}
}
\startdata
CN&$N=1-0,\ J=1/2-1/2,\ F=3/2-1/2$&113.169&$1.182\times10^{-5}$&$9.10\times10^{-12}$&$1.299\times10^{6}$\\
HNC&$J=1-0$&90.664&$2.69\times10^{-5}$&$9.71\times10^{-11}$&$2.770\times10^{5}$\\
C$^{34}$S&$J = 2 - 1$ &96.413&$1.60\times10^{-5}$&$5.06\times10^{-11}$&$3.162\times10^{5}$\\
$^{13}$CO&$J = 1 - 0$ &110.201&$6.294\times10^{-8}$&$3.302\times10^{-11}$&$1.906\times10^{3}$\\
\hline
HCP&$J = 2 - 1$ &79.903&$3.61\times10^{-7}$&$6.884\times10^{-11}$&$5.244\times10^{3}$\\
HCP&$J = 3 - 2$ &119.854&$1.31\times10^{-6}$&$7.33\times10^{-11}$&$1.786\times10^{4}$\\
PN& $J = 2 - 1$ &93.979&$2.92\times10^{-5}$&$4.538\times10^{-11}$&$6.534\times10^{5}$\\
PN& $J = 3 - 2$ &140.968&$1.05\times10^{-4}$&$5.218\times10^{-11}$&$2.012\times10^{6}$\\
PO&$J = 5/2 - 3/2,\ \Omega = 1/2,\ F = 3 - 2,\ e$ &108.998&$2.132\times10^{-5}$&$5.10\times10^{-12}$&$4.179\times10^{6}$\\
PO&$J = 5/2 - 3/2,\ \Omega = 1/2,\ F = 2 - 1,\ e$ &109.045&$1.92\times10^{-5}$&$1.93\times10^{-11}$&$9.952\times10^{5}$\\
PO&$J = 5/2 - 3/2,\ \Omega = 1/2,\ F = 3 - 2,\ f$ &109.206&$2.143\times10^{-5}$&$1.161\times10^{-10}$&$1.846\times10^{5}$\\
PO&$J = 5/2 - 3/2,\ \Omega = 1/2,\ F = 2 - 1,\ f$ &109.281&$1.93\times10^{-5}$&$6.30\times10^{-12}$&$3.062\times10^{6}$\\
\enddata
\end{deluxetable*}

In Figure \ref{fig:diffuse_proposed}, line excitation of the P-bearing molecules are shown. The excitations obtained from the RADEX is well matched for HCP, which is in emission. For PN, we have seen emissions. For PO, we have obtained a strong emission with the RADEX, whereas the results obtained with the RATRAN code show these in absorption.

\subsection{Hot core/corino model}
Here, we execute a RATRAN radiative transfer model to explain the anticipated line profiles of the P-bearing molecules in the hot core/corino region.
\cite{rolf10} fitted the observed HCN profiles toward Sgr B2(M). They provided the best-suited spatially varying parameters (density, temperature, and velocity) for this case. Interestingly, they found that in-fall dominates in the outer parts, whereas multiple outflows were expelling matter from the inner region. Both density and temperature are almost linearly increasing toward the center of the hot core region.
Using this physical parameter as input of the RATRAN model, we generate the synthetic spectra for different P-bearing molecules (PN, PO, HCP, and PH$_3$) for our hot core model.
For the hot corino model, we also consider the same physical condition for simplicity (line profiles by considering more realistic physical requirements for the hot corino region are also reported in section \ref{sec:append_c}).
For the first time, \cite{rivi16} identified PO in the two star-forming regions W51 and W3(OH) in emission using the IRAM-30m telescope. Previously PO was detected in the envelope of the evolved stars but not in the star-forming regions. \cite{rivi16} also recognized some transitions of PN in emission. They explained the observed column densities of PN and PO using a chemical model. We use the abundances of PO, PN, HCP, and PH$_3$ for the hot core/corino model from  Table \ref{tab:abundances}.
Following \cite{rolf10}, here, we assume a bare dust grain \citep{osse94} due to the high temperature of the source.

The massive hot core Sgr B2(M) is situated at a distance of $\simeq$ 7.8 kpc \citep{reid09}. 
 \cite{rivi16} obtained FWHM for PO and PN as 7.0 and 8.2 km/s, respectively. 
 Based on these choices, here, we use line-broadening parameter (i.e., the Doppler parameter) for PO and PN as 4.2 and 4.9 km/s, respectively. For HCP, we consider it the same as PN (i.e., 4.9 km/s). For PH$_3$, we consider it 1.9 km/s. We consider the collisional data file for PN and HCP with H$_2$ from the BASECOL database \citep{dube13}. No collisional data files were available for PH$_3$. Looking at the structural similarity, we consider the H$_2$ collisional rate of NH$_3$ in place of PH$_3$. The same numerical values are considered for the hot corino case.

We calculate the beam size for IRAM-30m\footnote{\url{http://www.iram.es/IRAMES/mainWiki/Iram30mEfficiencies}} and for SOFIA-GREAT\footnote{\url{https://www.sofia.usra.edu/science/proposing-and-observing/observers-handbook-cycle-7/7-great/72-planning-observations}}. Estimated telescopic parameters (main beam temperature, beam size, integration time, and visibility) for the diffuse cloud and hot core/corino region are noted in Tables \ref{table:PN_observation}, \ref{table:PO_observation}, \ref{table:HCP_observation}, and \ref{table:PH3_observation} for PN, PO, HCP, and PH$_3$ respectively. Only the potentially detectable transitions in the hot core/corino region are noted in the tables. The predicted line profiles of these transitions after performing beam convolution are shown in Figures \ref{fig:PN}, \ref{fig:PO}, \ref{fig:HCP}, and \ref{fig:PH3} of the Appendix Section \ref{sec:append_b}. Interestingly, we have obtained ``inverse P-Cygni type'' spectral profiles for some transitions of PH$_3$ (see Figure \ref{fig:PH3}). This unique type of velocity profile represents both in-fall and outflow are present in this source.
Theoretically, \cite{rolf10} predicted the existence of accelerating in-fall having a density power-law index of  $\sim 1.5$ to support a spherically symmetric constant mass accretion rate in Sgr B2(M). However, they were also unable to obtain the observational evidence of accelerating infall. In Appendix Section \ref{sec:append_c}, we have shown the obtained line profiles when physical condition related to IRAS4A was considered as a representative of a hot corino case. We notice that the physical input parameters are very much susceptible to the derived line profiles. We do not obtain the inverse P-cygni profile while we use the physical parameters of IRAS4A. But it is out of scope for this work to comment elaborately on such concerns here.  

\begin{deluxetable*}{ccccccccc}
\tablecaption{Experimental and calculated infrared data for PH$_3$. \label{tab:IR}}
\tablewidth{0pt}
\tabletypesize{\scriptsize} 
\tablehead{
\colhead{Assignment} & \multicolumn{5}{c}{Frequency [cm$^{-1}$]} & \multicolumn{3}{c}{Absorption Coefficients [cm molecule$^{-1}$]} \\
\colhead{ } &\colhead{ Experimental$^a$} &\multicolumn{2}{c}{Calculated$^b$} & \multicolumn{2}{c}{Calculated$^b$ $\times$ 0.967} & \colhead{Experimental$^a$} & \multicolumn{2}{c}{Calculated$^b$} \\
\colhead{ } &\colhead{ } &\colhead{Harmonic} & \colhead{Anharmonic} & \colhead{Harmonic} & \colhead{Anharmonic} & \colhead{ } & \colhead{Harmonic} & \colhead{Anharmonic}
}
\startdata
\multicolumn{9}{c}{Fundamental Bands} \\
\hline
$\nu_2$ (bending) & 982 & 1007.267 & 990.267 & 974.027 & 957.588 & $0.51\times10^{-18}$ & $5.40\times10^{-18}$ & $5.11\times10^{-18}$ \\
$\nu_4$ (scissoring) & 1096 & 1134.859 & 1110.02 & 1097.409 & 1073.389 & $0.71\times10^{-18}$ & $2.78\times10^{-18}$ & $2.55\times10^{-18}$\\
scissoring & & 1134.967 & 1103.681 & 1097.513 & 1067.260 & & $2.78\times10^{-18}$ & $2.58\times10^{-18}$\\
$\nu_1$ (symmetric stretching) & 2303 & 2396.887 & 2296.619 & 2317.790 & 2220.831 & $2.4\times10^{-18}$ & $7.82\times10^{-18}$ & $7.86\times10^{-18}$ \\
$\nu_3$ (asymmetric stretching) & 2316 & 2401.893 & 2303.377 & 2322.630 & 2227.366 & $8.4\times10^{-18}$ & $1.38\times10^{-17}$ & $1.47\times10^{-17}$ \\
asymmetric stretching & & 2405.153 & 2283.634 & 2325.783 & 2208.274 & & $1.34\times10^{-17}$ & $1.45\times10^{-17}$ \\
\hline
\multicolumn{9}{c}{Overtones} \\
\hline
2$\nu_4$ & 2195 & & 2213.024 & & 2139.994 & & & $2.98\times10^{-19}$ \\
3$\nu_2$ & 2905 & & & & & & & \\
2$\nu_1$ & 4536 & & 4540.981 & & 4391.129 & & & $1.35\times10^{-22}$ \\
\hline
\multicolumn{9}{c}{Combination Bands} \\
\hline
$\nu_2+\nu_4$ & 2067, 2083 & & 2093.256 & & 2024.178 & & & $2.86\times10^{-20}$ \\
$\nu_1/\nu_3+\nu_L$ & 2376, 2426, 2461 & & & & & & & \\
$\nu_1+\nu_2$ & 3288 & & 3280.701 & & 3172.438 & & & $6.50\times10^{-22}$ \\
$\nu_1+\nu_4$ & 3392 & & 3391.715 & & 3279.788 & & & $7.12\times10^{-20}$ \\
$\nu_3+\nu_4$ & 3405 & & 3392.344 & & 3280.397 & & & $5.67\times10^{-20}$ \\
$\nu_1+\nu_3$ & 4621 & & 4523.242 & & 4373.975 & & & $1.35\times10^{-21}$ \\
\enddata
\tablecomments{
$^a$ \cite{turn15}, \\
$^b$ This work.}
\end{deluxetable*}

\begin{figure*}
\centering
\includegraphics[width=12cm, height=8cm]{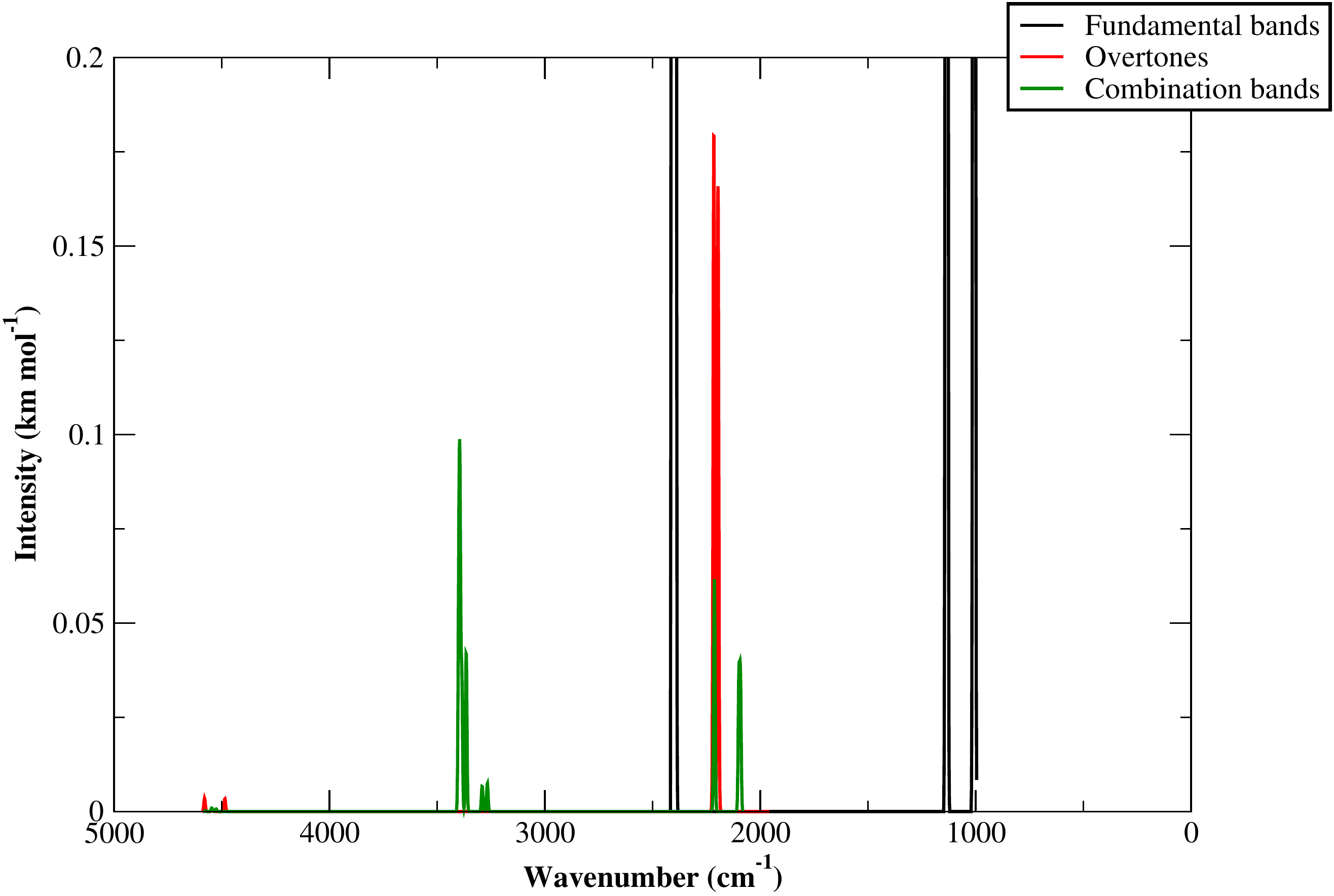}
\caption{Ice phase IR absorption spectra of PH$_3$ including the harmonic fundamental bands and anharmonic overtones and combination bands.}
\label{fig:IR-PH3}
\end{figure*}

\begin{figure*}
\centering
\includegraphics[width=\textwidth]{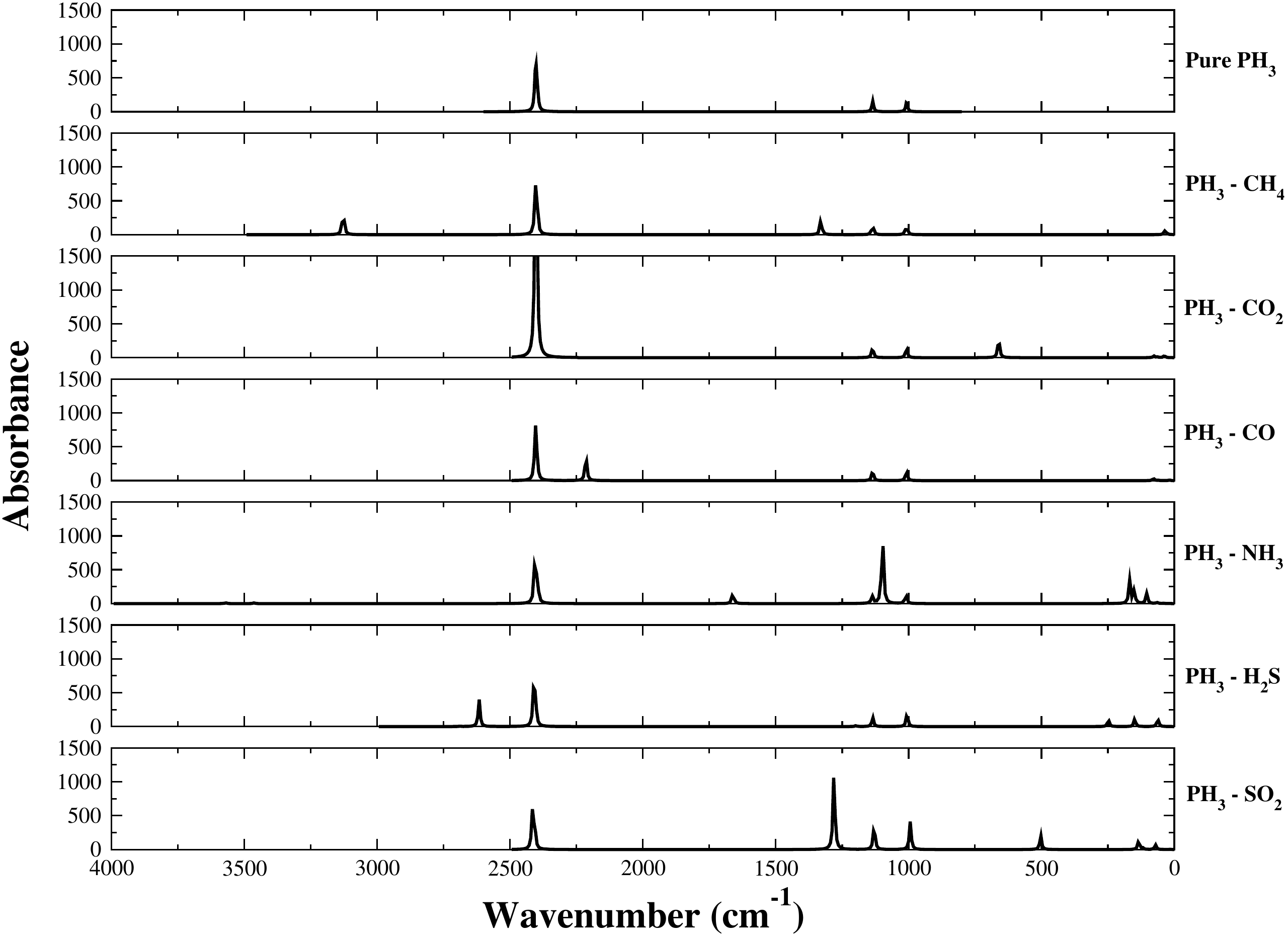}
\caption{Comparison of ice-phase absorption spectra of pure PH$_3$ and mixture of PH$_3$ with other volatiles (CH$_4$, CO$_2$, CO, NH$_3$, H$_2$S, and SO$_2$) considering the harmonic fundamental bands.}
\label{fig:IR-imp}
\end{figure*}

\begin{figure*}
\centering
\includegraphics[width=\textwidth]{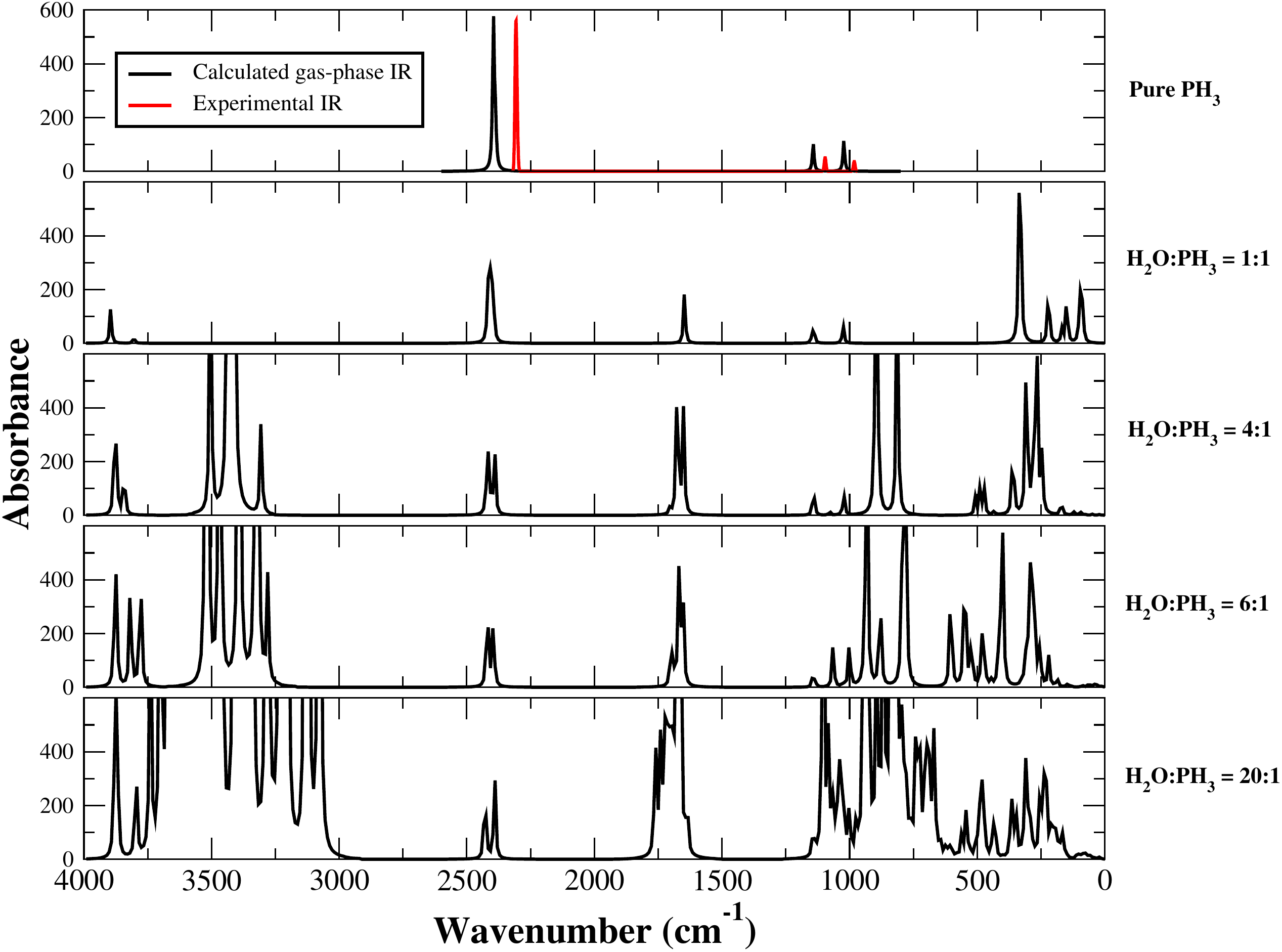}
\caption{Gas-phase IR absorption spectra of PH$_3$ with increasing concentration of H$_2$O.}
\label{fig:IR-H2O}
\end{figure*}

\section{Infrared spectra of PH$_3$} \label{sec:IR_spectra}
Phosphine is an important reservoir of interstellar phosphorous \citep{char94}. Since it is of particular interest to astronomers, the IR vibrational spectra analysis of PH$_3$ could be helpful to the community. 
The vibrational spectra of PH$_3$ would be beneficial for future astronomical observations with the James Web Space Telescope (JWST). To precisely estimate the frequency and interpret the intensities,  it is necessary to go beyond the harmonic approximation. The anharmonic calculations not only show a significant deviation from the harmonic calculations. Another advantage of the anharmonic analysis is that overtones and combination bands can also be analyzed with this. When anharmonicity is considered, their intensity vanishes at the harmonic level. 
The Density Functional Theory (DFT) is employed to investigate the IR feature of PH$_3$. Within the DFT approach, the standard B3LYP functional \citep{beck93} has been used in conjunction with the 6-311G(d,p) basis set. All DFT computations have been performed employing the \textsc{Gaussian} suite of programs for quantum chemistry \citep{fris13}.
Figure \ref{fig:IR-PH3} shows the calculated infrared feature of the ice phase PH$_3$. 
To mimic the ice features, the PH$_3$ molecule is embedded in a continuum solvation field to represent local effects. The integral equation formalism (IEF) variant of the polarizable continuum model (PCM) as a default self-consistent reaction field (SCRF) method is employed with water as a solvent \citep{canc97,toma05}. The implicit solvation
model places the molecule of interest inside a cavity in a continuous homogeneous dielectric medium representing the solvent.
The fundamental modes of vibration, along with the overtones and combination bands, are shown in Figure \ref{fig:IR-PH3}. Table \ref{tab:IR} also notes down the wavenumbers (in cm$^{-1}$) and corresponding absorption coefficients (IR intensities in cm molecule$^{-1}$) of the fundamental bands, overtones, and combination bands.

A comparison between our computed spectra and that obtained experimentally by \cite{turn15} is shown in Table \ref{tab:IR}. 
The computed vibrational frequencies are often scaled to resemble the experimental results. This scaling factor varies with the choice of the basis sets and implemented method. The NIST database\footnote{\url{https://cccbdb.nist.gov/vibscalejust.asp}} noted some of the scaling factors, which are helpful to compare with the experimental values. 
Based on our method and the chosen basis set, we use a vibrational scaling factor of $\sim 0.967$. \cite{puzz14a} demonstrated that the best state-of-the-art theoretical estimates have an accuracy of about $\rm{5 \ cm^{-1} - 10 \ cm^{-1} \ and \ 10 \ cm^{-1} - 20 \ cm^{-1}}$ for fundamental
and non-fundamental transitions, respectively. This accuracy allows for reliable simulations of IR
spectra supporting astronomical observations.
The harmonic frequencies presented in Table \ref{tab:IR} agree with the experimental values noted in \cite{turn15} within a limit of about $\rm{1 \ cm^{-1} - 15 \ cm^{-1}}$ after scaling. Our calculated values are in excellent agreement with the experimental values for the overtones and combination bands even without scaling.

Figure \ref{fig:IR-imp} shows the IR feature of PH$_3$ in the presence of various impurities. H$_2$O molecules would cover a significant portion of the interstellar ice mantles in the dense interstellar region. Some other species, such as CO, CO$_2$, CH$_4$, NH$_3$, etc. \citep{boog15,gora20c} can also constitute a sizeable portion of the grain mantle. These molecules can influence the band profile of PH$_3$. We notice that the stretching of PH$_3$ around 2400 cm$^{-1}$ is getting more robust in the presence of  CO$_2$. H$_2$S and SO$_2$ are among the main components of Venusian atmospheres \citep{sous20,grea20}. It is important to note that a fundamental mode of SO$_2$ coincides with the bending-scissoring modes of PH$_3$ around $\sim (1000-1100)$ cm$^{-1}$, which can blend the PH$_3$ transitions. Since H$_2$O is the major component of the interstellar dense cloud region, we check the effect of increasing concentration of H$_2$O on PH$_3$ separately in Figure \ref{fig:IR-H2O}. We notice that with increasing H$_2$O concentration, PH$_3$ fundamental bands are highly affected.

\section{Conclusions} \label{sec:Conclusion}
In this paper, we have computed the abundance of the P-bearing species under various interstellar circumstances. We draw the following major conclusions.

\begin{itemize}

 \item From this work, we found that the abundance of PH$_3$ is low in the diffuse cloud region and PDR. However, in the hot core region, a detectable amount of PH$_3$ could be produced.

 \item There appears to be a clear trend between the abundances of PO and PN.  The PO/PN ratio is mainly governed by the reaction $\rm{PO + N \rightarrow PN + O}$. Since we were having a high abundance of atomic N in the diffuse clouds and PDRs ($\sim 6.8 \times 10^{-5}$) compared to hot cores ($\sim 10^{-6}$), we found PO/PN $<1$ for the diffuse cloud region, $<<1$ for the PDR region. For the late warm-up to the post-warm-up region of the hot core, we obtained PO/PN $\ge 1$.  For the hot corino case, we noticed PO/PN $\ge 1$ for the late warm-up to the initial post-warm-up phase, whereas it is $\le 1$ for the late post-warm-up phase.

 \item The {\it BE} of PH$_3$ with water is found to be lower as compared to that of NH$_3$. In \cite{das18}, the {\it BE} of NH$_3$ with a c-hexamer configuration of water was found to be 5163 K. In the current paper,
the same for PH$_3$ is found to be only $2395$ K.

  \item The abundance of PH$_3$ could be significantly affected by the destruction of H and OH. Without the inclusion of these, any inference on the abundance of PH$_3$ in astrophysical environments would be inaccurate.

 \item The radiative transfer model for the diffuse cloud region can successfully  explain the observed line profiles of HNC, CN, C$^{34}$S in absorption, and one transition of $^{13}$CO in emission. Some of the line profiles  of the P-bearing species are estimated and proposed for future observation in the hot core region.  An inverse P-Cygni profile of PH$_3$ is expected in the hot core region.

 \item A good agreement between the calculated IR wavenumbers of PH$_3$ and the experimental feature of \cite{turn15} was seen.

 \item The stretching and bending-scissoring modes of  PH$_3$ could be affected by the CO$_2$ and SO$_2$, respectively, in the ice.
 
 \item PH$_3$ fundamental bands are highly affected with increasing H$_2$O concentration in the gas phase.

\end{itemize}

\acknowledgments
M.S. gratefully acknowledges DST, the Government of India for providing financial assistance through DST-INSPIRE Fellowship [IF160109] scheme. S.S. acknowledges UGC, New Delhi, India, for providing a fellowship. B.B. acknowledges DST-INSPIRE Fellowship [IF170046] for providing financial support. S.K.M. acknowledges CSIR fellowship (Ref no. 18/06/2017(i) EU-V). P.G. acknowledges support from a Chalmers Cosmic Origins postdoctoral fellowship. A.P. acknowledges financial support from DST SERB EMR grant, 2017 (SERB-EMR/2016/005266), Banaras Hindu University, Varanasi and The Inter-University Centre for Astronomy and Astrophysics, Pune for associateship.
A.D. acknowledges ISRO respond project (Grant number ISRO/RES/2/402/16-17) for partial financial support. This work was supported by the Japan-India Science Cooperative Program between JSPS and DST, Grant number JPJSBP120207703. This research was possible due to a Grant-In-Aid from the Higher Education Department of the Government of West Bengal.

\software{\textsc{Gaussian} 09 \citep{fris13}, \textsc{Cloudy} 17.02 \citep{ferl17}, RATRAN \citep{hoge00}, ATRAN \citep{lord92}, RADEX \citep{vand07}.}

\clearpage
\appendix

\restartappendixnumbering
\section{Complete Phosphorus Chemistry Network} \label{sec:append_a}
All the gas and ice phase pathways considered in this work are shown in Table \ref{tab:reaction_path} with proper references.

\startlongtable
\begin{deluxetable*}{ l c c c c c}
\tablecaption{Reaction pathways \label{tab:reaction_path}}
\tablewidth{0pt}
\tabletypesize{\scriptsize} 
\tablehead{
\colhead{\bf Reaction}   & \colhead{\bf Reactions}                               & \multicolumn{3}{c}{\bf Rate coefficient} & \colhead{\bf Reference} \\
\colhead{\bf Number (Type)}          & \colhead{ }                                        & \colhead{$\alpha$}    & \colhead{$\beta$} & \colhead{$\gamma$} & \colhead{ }
}
\startdata
\multicolumn{6}{c}{Gas-phase pathways} \\
\hline
 R1 (IN) & $\rm{C^+ + P \rightarrow P^+ + C}$ & $1.0\times10^{-09}$ & 0.0 & 0.0 & \cite{mcel13} \\
 R2 (IN) & $\rm{H^+ + P \rightarrow P^+ + H}$ & $1.0\times10^{-09}$ & 0.0 & 0.0 & \cite{mcel13} \\ 
 R3 (IN) & $\rm{He^+ + P \rightarrow P^+ + He}$ & $1.0\times10^{-09}$ & 0.0 & 0.0 & \cite{mcel13} \\
 R4 (IN) & $\rm{NH_3 + P^+ \rightarrow P + NH_3^+}$ & $3.08\times10^{-10}$ & -0.5 & 0.0 & \cite{mcel13} \\
 R5 (IN) & $\rm{Si + P^+ \rightarrow P + Si^+}$ & $1.0\times10^{-09}$ & 0.0 & 0.0 & \cite{mcel13} \\
 R6 (CR) & $\rm{P + CRPHOT \rightarrow P^+ + e^-}$ & $1.30\times10^{-17}$ & 0.0 & 750 & \cite{mcel13} \\
 R7 (PH) & $\rm{P + h\nu \rightarrow P^+ + e^-}$ & $1.0\times10^{-09}$ & 0.0 & 2.7 & \cite{mcel13} \\
 R8 (ER) & $\rm{P^+ + e^- \rightarrow P + h\nu}$ & $3.41\times10^{-12}$ & -0.65 & 0.0 & \cite{mcel13} \\
 R9 (NR) & $\rm{P + CN \rightarrow PN + C}$ & $3.0\times10^{-10}$ & - & - & \cite{jime18} \\
 R10 (NR) & $\rm{N + PH \rightarrow PN + H}$ & $5.0\times10^{-11}$ & 0.0 & 0.0 & \cite{smit04} \\
 R11 (NR) & $\rm{N + PO \rightarrow PN + O}$ & $3.0\times10^{-11}$ & $-0.6$ & 0.0 & \cite{smit04} \\
 R12 (NR) & $\rm{N + PO \rightarrow P + NO}$ & $2.55\times10^{-12}$ & 0.0 & 0.0 & \cite{mill87} \\
 R13 (NR) & $\rm{N + CP \rightarrow PN + C}$ & $3.0\times10^{-10}$ & - & - & \cite{jime18} \\
 R14 (NR) & $\rm{C + PN^+ \rightarrow PN + C^+}$ & $1.0\times10^{-09}$ & 0.0 & 0.0 & \cite{mcel13} \\
 R15 (DR) & $\rm{PN^+ + e^- \rightarrow P + N}$ & $1.8\times10^{-7}$ & $-0.5$ & 0.0 & \cite{mcel13} \\
 R16 (DR) & $\rm{HPN^+ + e^- \rightarrow PN + H}$ & $1.0\times10^{-7}$ & $-0.5$ & 0.0 & \cite{mill91} \\
 R17 (DR) & $\rm{HPN^+ + e^- \rightarrow P + NH}$ & $1.0\times10^{-7}$ & $-0.5$ & 0.0 & \cite{mcel13} \\
 R18 (DR) & $\rm{PNH_2^+ + e^- \rightarrow PN + H_2}$ & $1.5\times10^{-7}$ & $-0.5$ & 0.0 & \cite{mcel13} \\
 R19 (DR) & $\rm{PNH_3^+ + e^- \rightarrow PN + H_2}$ & $1.5\times10^{-7}$ & $-0.5$ & 0.0 & \cite{mcel13} \\
 R20 (CR) & $\rm{PN + CRPHOT \rightarrow P + N}$ & $1.30\times10^{-17}$ & 0.0 & 250 & \cite{mcel13} \\
 R21 (IN) & $\rm{H^+ + PN \rightarrow PN^+ + H}$ & $1.0\times10^{-9}$ & $-0.5$ & 0.0 & \cite{mill91} \\
 R22 (IN) & $\rm{He^+ + PN \rightarrow P^+ + N + He}$ & $1.0\times10^{-9}$ & $-0.5$ & 0.0 & \cite{mill91} \\
 R23 (IN) & $\rm{H_3^+ + PN \rightarrow HPN^+ + H_2}$ & $1.0\times10^{-9}$ & $-0.5$ & 0.0 & \cite{mill91} \\
 R24 (IN) & $\rm{H_3O^+ + PN \rightarrow HPN^+ + H_2O}$ & $1.0\times10^{-9}$ & $-0.5$ & 0.0 & \cite{mill91} \\
 R25 (IN) & $\rm{HCO^+ + PN \rightarrow HPN^+ + CO}$ & $1.0\times10^{-9}$ & $-0.5$ & 0.0 & \cite{mill91} \\
 R26 (NR) & $\rm{N + PN \rightarrow P + N_2}$ & $1.0\times10^{-18}$ & 0.0 & 0.0 & \cite{mill87} \\
 R27 (PH) & $\rm{PN + h\nu \rightarrow P + N}$ & $5.0\times10^{-12}$ & 0.0 & 3.0 & \cite{mcel13} \\
 R28 (CR) & $\rm{HPO + CRPHOT \rightarrow PO + H}$ & $1.30\times10^{-17}$ & 0.0 & 750 & \cite{mcel13} \\
 R29  (DR) & $\rm{HPO^+ + e^- \rightarrow PO + H}$ & $1.50\times10^{-7}$ & $-0.5$ & 0.0 & \cite{thor84} \\
 R30 (DR) & $\rm{HPO^+ + e^- \rightarrow O + PH}$ & $1.50\times10^{-7}$ & $-0.5$ & 0.0 & \cite{mill91} \\
 R31 (DR) & $\rm{HPO^+ + e^- \rightarrow P + O + H}$ & $1.00\times10^{-7}$ & $-0.5$ & 0.0 & \cite{mcel13} \\
 R32 (DR) & $\rm{PO^+ + e^- \rightarrow P + O}$ & $1.80\times10^{-7}$ & $-0.5$ & 0.0 & \cite{mcel13} \\
 R33 (IN) & $\rm{H_2O + HPO^+ \rightarrow PO + H_3O^+}$ & $3.40\times10^{-10}$ & $-0.5$ & 0.0 & \cite{mcel13} \\
 R34 (IN) & $\rm{H_2O + P^+ \rightarrow PO^+ + H_2}$ & $5.50\times10^{-11}$ & $-0.5$ & 0.0 & \cite{mcel13} \\
 R35 (NR) & $\rm{O + HPO \rightarrow PO + OH}$ & $3.80\times10^{-11}$ & 0.0 & 0.0 & \cite{smit04} \\
 R36 (NR) & $\rm{O + PH_2 \rightarrow PO + H_2}$ & $4.0\times10^{-11}$ & 0.0 & 0.0 & \cite{mill87} \\
 R37 (NR) & $\rm{O + PH_2 \rightarrow H + HPO}$ & $8.0\times10^{-11}$ & 0.0 & 0.0 & \cite{smit04} \\
 R38 (NR) & $\rm{O + PH \rightarrow PO + H}$ & $1.0\times10^{-10}$ & 0.0 & 0.0 & \cite{smit04} \\
 R39 (NR) & $\rm{P + OH \rightarrow PO + H}$ & $6.1\times10^{-11}$ & $-0.23$ & 14.9 & \cite{jime18} \\
 R40 (NN) & $\rm{O_2 + P \rightarrow O + PO}$ & $1.0\times10^{-13}$ & 0.0 & 0.0 & \cite{mcel13} \\
 R41 (NN) & $\rm{O_2 + PH_2^+ \rightarrow PO^+ + H_2O}$ & $7.8\times10^{-11}$ & 0.0 & 0.0 & \cite{mcel13} \\
 R42 (IN) & $\rm{P^+ + CO_2 \rightarrow PO^+ + CO}$ & $4.60\times10^{-10}$ & 0.0 & 0.0 & \cite{mcel13} \\
 R43 (IN) & $\rm{P^+ + O_2 \rightarrow PO^+ + O}$ & $5.60\times10^{-10}$ & 0.0 & 0.0 & \cite{mcel13} \\
 R44 (IN) & $\rm{P^+ + OCS \rightarrow PO^+ + CS}$ & $4.18\times10^{-10}$ & 0.0 & 0.0 & \cite{mcel13} \\
 R45 (PH) & $\rm{HPO + h\nu \rightarrow PO + H}$ & $1.70\times10^{-10}$ & 0.0 & $5.3$ & \cite{mcel13} \\
 R46 (CR) & $\rm{PO + CRPHOT \rightarrow P + O}$ & $1.30\times10^{-17}$ & 0.0 & 250 & \cite{mcel13} \\
 R47 (IN) & $\rm{C^+ + PO \rightarrow PO^+ + C}$ & $1.0\times10^{-9}$ & $-0.5$ & 0.0 & \cite{thor84} \\
 R48 (IN) & $\rm{H^+ + PO \rightarrow PO^+ + H}$ & $1.0\times10^{-9}$ & $-0.5$ & 0.0 & \cite{thor84} \\ 
 R49 (IN) & $\rm{H_3^+ + PO \rightarrow HPO^+ + H_2}$ & $1.0\times10^{-9}$ & $-0.5$ & 0.0 & \cite{thor84} \\
 R50 (IN) & $\rm{HCO^+ + PO \rightarrow HPO^+ + CO}$ & $1.0\times10^{-9}$ & $-0.5$ & 0.0 & \cite{thor84} \\ 
 R51 (IN) & $\rm{He^+ + PO \rightarrow P^+ + O + He}$ & $1.0\times10^{-9}$ & $-0.5$ & 0.0 & \cite{thor84} \\
 R52 (PH) & $\rm{PO + h\nu \rightarrow P + O}$ & $3.0\times10^{-10}$ & 0.0 & 2.0 & \cite{mcel13} \\
 R53 (DR) & $\rm{PCH_2^+ + e^- \rightarrow HCP + H}$ & $1.50\times10^{-7}$ & $-0.5$ & 0.0 & \cite{mill91} \\
 R54 (CR) & $\rm{HCP + CRPHOT \rightarrow CP + H}$ & $1.30\times10^{-17}$ & 0.0 & 750 & \cite{mcel13} \\
 R55 (IN) & $\rm{C^+ + HCP \rightarrow CCP^+ + H}$ & $5.0\times10^{-10}$ & 0.0 & 0.0 & \cite{mill91} \\
 R56 (IN) & $\rm{C^+ + HCP \rightarrow HCP^+ + C}$ & $5.0\times10^{-10}$ & 0.0 & 0.0 & \cite{mill91} \\
 R57 (IN) & $\rm{C + HCP^+ \rightarrow CCP^+ + H}$ & $2.0\times10^{-10}$ & 0.0 & 0.0 & \cite{mcel13} \\
 R58 (IN) & $\rm{H^+ + HCP \rightarrow HCP^+ + H}$ & $1.0\times10^{-9}$ & 0.0 & 0.0 & \cite{mill91} \\
 R59 (IN) & $\rm{H_3^+ + HCP \rightarrow PCH_2^+ + H_2}$ & $1.0\times10^{-9}$ & 0.0 & 0.0 & \cite{mill91} \\
 R60 (IN) & $\rm{H_3O^+ + HCP \rightarrow PCH_2^+ + H_2O}$ & $1.0\times10^{-9}$ & 0.0 & 0.0 & \cite{mill91} \\
 R61 (IN) & $\rm{HCO^+ + HCP \rightarrow PCH_2^+ + CO}$ & $1.0\times10^{-9}$ & 0.0 & 0.0 & \cite{adam90} \\
 R62 (IN) & $\rm{He^+ + HCP \rightarrow P^+ + CH + He}$ & $5.0\times10^{-10}$ & 0.0 & 0.0 & \cite{mill91} \\
 R63 (IN) & $\rm{He^+ + HCP \rightarrow PH + C^+ + He}$ & $5.0\times10^{-10}$ & 0.0 & 0.0 & \cite{mill91} \\
 R64 (NR) & $\rm{O + HCP \rightarrow PH + CO}$ & $3.61\times10^{-13}$ & 2.1 & 3080.0 & \cite{smit04} \\
 R65 (PH) & $\rm{HCP + h\nu \rightarrow CP + H}$ & $5.48\times10^{-10}$ & 0.0 & 2.0 & \cite{mcel13} \\
 R66 (NR) & $\rm{CCP + O \rightarrow CO + CP}$ & $6.0\times10^{-12}$ & 0.0 & 0.0 & \cite{smit04} \\
 R67 (IN) & $\rm{CCP + He^+ \rightarrow He + CP + C^+}$ & $5.0\times10^{-10}$ & $-0.5$ & 0.0 & \cite{mill91} \\
 R68 (IN) & $\rm{CCP + He^+ \rightarrow He + C_2 + P^+}$ & $5.0\times10^{-10}$ & $-0.5$ & 0.0 & \cite{mill91} \\
 R69 (DR) & $\rm{PCH_4^+ + e^- \rightarrow CP + C_3H}$ & $7.5\times10^{-8}$ & $-0.5$ & 0.0 & \cite{mcel13} \\
 R70 (DR) & $\rm{CCP^+ + e^- \rightarrow C + CP}$ & $1.5\times10^{-7}$ & $-0.5$ & 0.0 & \cite{mill91} \\
 R71 (DR) & $\rm{CCP^+ + e^- \rightarrow P + C_2}$ & $1.5\times10^{-7}$ & $-0.5$ & 0.0 & \cite{mill91} \\
 R72 (DR) & $\rm{PCH_2^+ + e^- \rightarrow CP + H_2}$ & $1.50\times10^{-7}$ & $-0.5$ & 0.0 & \cite{mill91} \\
 R73 (DR) & $\rm{HCP^+ + e^- \rightarrow CP + H}$ & $1.50\times10^{-7}$ & $-0.5$ & 0.0 & \cite{mill91} \\
 R74 (DR) & $\rm{HCP^+ + e^- \rightarrow P + CH}$ & $1.50\times10^{-7}$ & $-0.5$ & 0.0 & \cite{mill91} \\
 R75 (DR) & $\rm{CP^+ + e^- \rightarrow P + C}$ & $1.00\times10^{-7}$ & $-0.5$ & 0.0 & \cite{mcel13} \\ 
 R76 (NR) & $\rm{C + PH \rightarrow CP + H}$ & $7.50\times10^{-11}$ & 0.0 & 0.0 & \cite{smit04} \\
 R77 (CR) & $\rm{CP + CRPHOT \rightarrow C + P}$ & $1.30\times10^{-17}$ & 0.0 & 250 & \cite{mcel13} \\
 R78 (IN) & $\rm{C^+ + CP \rightarrow CP^+ + C}$ & $1.0\times10^{-9}$ & 0.0 & 0.0 & \cite{mill91} \\
 R79 (IN) & $\rm{H^+ + CP \rightarrow CP^+ + H}$ &$1.0\times10^{-9}$ & 0.0 & 0.0 & \cite{mill91} \\
 R80 (IN) & $\rm{H_3^+ + CP \rightarrow HCP^+ + H_2}$ & $1.0\times10^{-9}$ & 0.0 & 0.0 & \cite{mill91} \\
 R81 (IN) & $\rm{H_2 + CP^+ \rightarrow HCP^+ + H}$ & $1.0\times10^{-9}$ & 0.0 & 0.0 & \cite{mcel13} \\
 R82 (IN) & $\rm{H_3O^+ + CP \rightarrow HCP^+ + H_2O}$ & $1.0\times10^{-9}$ & 0.0 & 0.0 & \cite{mill91} \\
 R83 (IN) & $\rm{HCO^+ + CP \rightarrow HCP^+ + CO}$ & $1.0\times10^{-9}$ & 0.0 & 0.0 & \cite{adam90} \\
 R84 (IN) & $\rm{He^+ + CP \rightarrow P^+ + C + He}$ & $5.0\times10^{-10}$ & 0.0 & 0.0 & \cite{mill91} \\
 R85 (IN) & $\rm{He^+ + CP \rightarrow P + C^+ + He}$ & $5.0\times10^{-10}$ & 0.0 & 0.0 & \cite{mill91} \\
 R86 (NR) & $\rm{O + CP \rightarrow P + CO}$ & $4.0\times10^{-11}$ & 0.0 & 0.0 & \cite{mill91} \\
 R87 (IN) & $\rm{O + CP^+ \rightarrow P^+ + CO}$ & $2.0\times10^{-10}$ & 0.0 & 0.0 & \cite{mcel13} \\
 R88 (PH) & $\rm{CP + h\nu \rightarrow C + P}$ & $1.0\times10^{-9}$ & 0.0 & 2.8 & \cite{mcel13} \\
 R89 (PH) & $\rm{C + P \rightarrow CP + h\nu}$ & $1.41\times10^{-18}$ & 0.03 & 55.0 & \cite{mcel13} \\
 R90 (DR) & $\rm{PC_2H_3^+ + e^- \rightarrow PH + C_2H_2}$ & $1.00\times10^{-7}$ & $-0.50$ & 0.0 & \cite{mcel13} \\ 
 R91 (DR) & $\rm{PNH_2^+ + e^- \rightarrow PH + NH}$ & $1.50\times10^{-7}$ & $-0.50$ & 0.0 & \cite{mcel13} \\ 
 R92 (DR) & $\rm{PNH_3^+ + e^- \rightarrow PH + NH_2}$ & $1.50\times10^{-7}$ & $-0.50$ & 0.0 & \cite{mcel13} \\
 R93 (DR) & $\rm{PNH_2^+ + e^- \rightarrow P + NH_2}$ & $1.50\times10^{-7}$ & $-0.50$ & 0.0 & \cite{mcel13} \\
 R94 (DR) & $\rm{PNH_3^+ + e^- \rightarrow P + NH_3}$ & $3.0\times10^{-7}$ & $-0.50$ & 0.0 & \cite{mcel13} \\
 R95 (DR) & $\rm{PH_2^+ + e^- \rightarrow PH + H}$ & $9.38\times10^{-8}$ & $-0.64$ & 0.0 & \cite{mcel13} \\
 R96 (DR) & $\rm{PH_3^+ + e^- \rightarrow PH + H_2}$ & $1.5\times10^{-7}$ & $-0.5$ & 0.0 & \cite{mill91} \\
 R97 (DR) & $\rm{HPN^+ + e^- \rightarrow PH + N}$ & $1.0\times10^{-7}$ & $-0.5$ & 0.0 & \cite{mill91} \\
 R98 (DR) & $\rm{H_2PO^+ + e^- \rightarrow PH + OH}$ & $1.5\times10^{-7}$ & $-0.5$ & 0.0 & \cite{mcel13} \\
 R99 (DR) & $\rm{H_2PO^+ + e^- \rightarrow HPO + H}$ & $1.5\times10^{-7}$ & $-0.5$ & 0.0 & \cite{mcel13} \\
 R100 (IN) & $\rm{H_2O + PH_2^+ \rightarrow PH + H_3O^+}$ & $1.62\times10^{-10}$ & $-0.5$ & 0.0 & \cite{adam90} \\
 R101 (IN) & $\rm{NH_3 + PH^+ \rightarrow P + NH_4^+}$ & $3.99\times10^{-10}$ & $-0.5$ & 0.0 & \cite{mcel13} \\
 R102 (IN) & $\rm{NH_3 + PH^+ \rightarrow PNH_2^+ + H_2}$ & $5.88\times10^{-10}$ & $-0.5$ & 0.0 & \cite{mcel13} \\
 R103 (IN) & $\rm{NH_3 + PH^+ \rightarrow PNH_3^+ + H}$ & $1.11\times10^{-9}$ & $-0.5$ & 0.0 & \cite{mcel13} \\
 R104 (IN) & $\rm{NH_3 + P^+ \rightarrow PNH_2^+ + H}$ & $2.67\times10^{-10}$ & $-0.5$ & 0.0 & \cite{mcel13} \\
 R105 (IN) & $\rm{NH_3 + PH_2^+ \rightarrow PH + NH_4^+}$ & $3.8\times10^{-10}$ & $-0.5$ & 0.0 & \cite{adam90} \\
 R106 (IN) & $\rm{NH_3 + PH_2^+ \rightarrow PNH_3^+ + H_2}$ & $1.62\times10^{-9}$ & $-0.5$ & 0.0 & \cite{mcel13} \\
 R107 (NR) & $\rm{O + PH_2 \rightarrow PH + OH}$ & $2.0\times10^{-11}$ & 0.0 & 0.0 & \cite{smit04} \\
 R108 (NR) & $\rm{C + HPO \rightarrow PH + CO}$ & $4.0\times10^{-11}$ & 0.0 & 0.0 & \cite{mill91} \\
 R109 (CR) & $\rm{PH + CRPHOT \rightarrow P + H}$ & $1.3\times10^{-17}$ & 0.0 & 250 & \cite{mcel13} \\
 R110 (IN) & $\rm{C^+ + PH \rightarrow PH^+ + C}$ & $1.0\times10^{-9}$ & 0.0 & 0.0 & \cite{mill91} \\
 R111 (IN) & $\rm{H^+ + PH \rightarrow PH^+ + H}$ & $1.0\times10^{-9}$ & 0.0 & 0.0 & \cite{mill91} \\
 R112 (IN) & $\rm{H_3^+ + P \rightarrow PH^+ + H_2}$ & $1.0\times10^{-9}$ & 0.0 & 0.0 & \cite{adam90} \\
 R113 (IN) & $\rm{H_3^+ + PH \rightarrow PH_2^+ + H_2}$ & $2.0\times10^{-9}$ & 0.0 & 0.0 & \cite{mill91} \\
 R114 (IN) & $\rm{HCO^+ + P \rightarrow PH^+ + CO}$ & $1.0\times10^{-9}$ & 0.0 & 0.0 & \cite{adam90} \\
 R115 (IN) & $\rm{O + HCP^+ \rightarrow PH^+ + CO}$ & $2.0\times10^{-10}$ & 0.0 & 0.0 & \cite{mill91} \\
 R116 (IN) & $\rm{O + HPO^+ \rightarrow PH^+ + O_2}$ & $2.0\times10^{-10}$ & 0.0 & 0.0 & \cite{thor84} \\
 R117 (IN) & $\rm{HCO^+ + PH \rightarrow PH_2^+ + CO}$ & $1.0\times10^{-9}$ & 0.0 & 0.0 & \cite{mill91} \\
 R118 (IN) & $\rm{He^+ + PH \rightarrow P^+ + He + H}$ & $1.0\times10^{-9}$ & 0.0 & 0.0 & \cite{mill91} \\
 R119 (IN) & $\rm{He^+ + HPO \rightarrow PH^+ + O + He}$ & $5.0\times10^{-10}$ & $-0.5$ & 0.0 & \cite{mill91} \\
 R120 (IN) & $\rm{He^+ + HPO \rightarrow PO^+ + H + He}$ & $5.0\times10^{-10}$ & $-0.5$ & 0.0 & \cite{mcel13} \\
 R121 (NR) & $\rm{H + PH \rightarrow P + H_2}$ & $1.50\times10^{-10}$ & 0.0 & 416 & \cite{kaye83} \\
 R122 (IN) & $\rm{O + PH^+ \rightarrow PO^+ + H}$ & $1.0\times10^{-9}$ & 0.0 & 0.0 & \cite{thor84} \\
 R123 (IN) & $\rm{HCN + PH^+ \rightarrow HCNH^+ + P}$ & $3.06\times10^{-10}$ & $-0.5$ & 0.0 & \cite{mcel13} \\
 R124 (IN) & $\rm{O + PN^+ \rightarrow PO^+ + N}$ & $2.0\times10^{-10}$ & 0.0 & 0.0 & \cite{mcel13} \\
 R125 (IN) & $\rm{OH + P^+ \rightarrow PO^+ + H}$ & $5.0\times10^{-10}$ & $-0.5$ & 0.0 & \cite{mcel13} \\
 R126 (IN) & $\rm{PH^+ + O_2 \rightarrow PO^+ + OH}$ & $5.4\times10^{-10}$ & 0.0 & 0.0 & \cite{adam90} \\
 R127 (DR) & $\rm{PH^+ + e^- \rightarrow P + H}$ & $1.0\times10^{-7}$ & $-0.5$ & 0.0 & \cite{thor84} \\
 R128 (PH) & $\rm{PH + h\nu \rightarrow P + H}$ & $4.0\times10^{-10}$ & 0.0 & 1.5 & \cite{mcel13} \\
 R129 (DR) & $\rm{PH_3^+ + e^- \rightarrow H + PH_2}$ & $1.5\times10^{-7}$ & $-0.5$ & 0.0 & \cite{mcel13} \\
 R130 (DR) & $\rm{PH_2^+ + e^- \rightarrow P + H_2}$ & $5.36\times10^{-8}$ & $-0.64$ & 0.0 & \cite{mcel13} \\
 R131 (DR) & $\rm{PH_2^+ + e^- \rightarrow P + H + H}$ & $5.23\times10^{-7}$ & $-0.64$ & 0.0 & \cite{mcel13} \\
 R132 (IN) & $\rm{NH_3 + PH_3^+ \rightarrow PH_2 + NH_4^+}$ & $2.3\times10^{-9}$ & $-0.5$ & 0.0 & \cite{adam90} \\
 R133 (DR) & $\rm{PH_4^+ + e^- \rightarrow PH_2 + H_2}$ & $1.50\times10^{-7}$ & $-0.5$ & 0.0 & \cite{char94} \\
 R134 (CR) & $\rm{PH_2 + CRPHOT \rightarrow PH + H}$ & $1.3\times10^{-17}$ & 0.0 & 750.0 & \cite{mcel13} \\
 R135 (IN) & $\rm{H^+ + PH_2 \rightarrow PH_2^+ + H}$ & $1.0\times10^{-9}$ & 0.0 & 0.0 & \cite{mill91} \\
 R136 (NR) & $\rm{H + PH_2 \rightarrow PH + H_2}$ & $6.20\times10^{-11}$ & 0.0 & 318 & \cite{kaye83} \\
 R137 (IN) & $\rm{H_3^+ + PH_2 \rightarrow PH_3^+ + H_2}$ & $2.0\times10^{-9}$ & 0.0 & 0.0 & \cite{mill91} \\
 R138 (IN) & $\rm{HCO^+ + PH_2 \rightarrow PH_3^+ + CO}$ & $1.0\times10^{-9}$ & 0.0 & 0.0 & \cite{mill91} \\
 R139 (IN) & $\rm{He^+ + PH_2 \rightarrow P^+ + He + H_2}$ & $1.0\times10^{-9}$ & 0.0 & 0.0 & \cite{mill91} \\
 R140 (IN) & $\rm{PH_2^+ + O_2 \rightarrow PO^+ + H_2O}$ & $7.8\times10^{-11}$ & 0.0 & 0.0 & \cite{adam90} \\
 R141 (IN) & $\rm{H_2 + P^+ \rightarrow PH_2^+ + h\nu}$ & $7.5\times10^{-18}$ & $-1.3$ & 0.0 & \cite{adam90} \\
 R142 (PH) & $\rm{PH_2 + h\nu \rightarrow PH_2^+ + e^-}$ & $1.73\times10^{-10}$ & 0.0 & 2.6 & \cite{mcel13} \\
 R143 (PH) & $\rm{PH_2 + h\nu \rightarrow PH + H}$ & $2.11\times10^{-10}$ & 0.0 & 1.5 & \cite{mcel13} \\
 R144 (DR) & $\rm{PH_4^+ + e^- \rightarrow PH_3 + H}$ & $1.50\times10^{-7}$ & $-0.5$ & 0.0 & \cite{char94} \\
 R145 (IN) & $\rm{PH_4^+ + NH_3 \rightarrow PH_3 + NH_4^+}$ & $2.10\times10^{-9}$ & 0.0 & 0.0 & \cite{thor83} \\
 R146 (NR) & $\rm{PH_2 + H \rightarrow PH_3}$ & $3.7\times10^{-10}$ & 0.0 & 340 & \cite{kaye83} \\
 R147 (RA) & $\rm{H_2 + PH^+ \rightarrow PH_3^+ + h\nu}$ & $2.40\times10^{-17}$ & $-1.4$ & 0.0 & \cite{adam90} \\
 R148 (NR) & $\rm{H^+ + PH_3 \rightarrow PH_3^+ + H}$ & $7.22\times10^{-11}$ & 0.0 & 886 & \cite{sous20} \\
 R149 (NR) & $\rm{H + PH_3 \rightarrow PH_2 + H_2}$ & $7.22\times10^{-11}$ & 0.0 & 886 & \cite{sous20} \\
 R150 (RR) & $\rm{NH_2 + PH_3 \rightarrow PH_2 + NH_3}$ & $1.50\times10^{-12}$ & 0.0 & 928 & \cite{kaye83} \\
 R151 (IN) & $\rm{H^+ + PH_3 \rightarrow PH_3^+  + H}$ & $2.00\times10^{-9}$ & 0.0 & 0.0 & \cite{char94} \\
 R152 (IN) & $\rm{He^+ + PH_3 \rightarrow PH_2^+ + H + He}$ & $2.00\times10^{-9}$ & 0.0 & 0.0 & \cite{char94} \\
 R153 (IN) & $\rm{C^+ + PH_3 \rightarrow PH_3^+ + C}$ & $2.00\times10^{-9}$ & 0.0 & 0.0 & \cite{char94} \\
 R154 (IN) & $\rm{H_3^+ + PH_3 \rightarrow PH_4^+ + H_2}$ & $2.00\times10^{-9}$ & 0.0 & 0.0 & \cite{char94} \\
 R155 (IN) & $\rm{HCO^+ + PH_3 \rightarrow PH_4^+ + CO}$ & $2.00\times10^{-9}$ & 0.0 & 0.0 & \cite{char94} \\
 R156 (IN) & $\rm{H_3O^+ + PH_3 \rightarrow PH_4^+ + H_2O}$ & $2.00\times10^{-9}$ & 0.0 & 0.0 & \cite{char94} \\
 R157 (IN) & $\rm{PH_3^+ + PH_3 \rightarrow PH_4^+ + PH_2}$ & $1.10\times10^{-9}$ & 0.0 & 0.0 & \cite{smit89} \\
 R158 (PH) & $\rm{PH_3 + h\nu \rightarrow PH_2 + H}$ & $9.23\times10^{-10}$ & 0.0 & 2.1 & \cite{mcel13} [UMIST, following NH$_3$] \\
 R159 (PH) & $\rm{PH_3 + h\nu \rightarrow PH + H_2}$ & $2.76\times10^{-10}$ & 0.0 & 2.1 & \cite{mcel13} [UMIST, following NH$_3$] \\
 R160 (PH) & $\rm{PH_3 + h\nu \rightarrow PH_3^+ + e^-}$ & $2.80\times10^{-10}$ & 0.0 & 3.1 & \cite{mcel13} [UMIST, following NH$_3$] \\
 R161 (NR) & $\rm{PH_3 + OH \rightarrow H_2O + PH_2}$ & $2.71\times10^{-11}$ & 0.0 & 155 & \cite{sous20} \\
 R162 (DR) & $\rm{PC_2H_2^+ + e^- \rightarrow CCP + H_2}$ & $1.0\times10^{-7}$ & $-0.5$ & 0.0 & \cite{mcel13} \\
 R163 (DR) & $\rm{PC_2H_3^+ + e^- \rightarrow CCP + H_2 + H}$ & $1.0\times10^{-7}$ & $-0.5$ & 0.0 & \cite{mcel13} \\
 R164 (DR) & $\rm{PC_2H_4^+ + e^- \rightarrow CCP + H_2 + H_2}$ & $1.0\times10^{-7}$ & $-0.5$ & 0.0 & \cite{mcel13} \\
 R165 (DR) & $\rm{PC_3H^+ + e^- \rightarrow CCP + CH}$ & $1.0\times10^{-7}$ & $-0.5$ & 0.0 & \cite{mcel13} \\
 R166 (DR) & $\rm{PC_4H^+ + e^- \rightarrow CCP + C_2H}$ & $7.5\times10^{-8}$ & $-0.5$ & 0.0 & \cite{mcel13} \\
 R167 (CR) & $\rm{CCP + CRPHOT \rightarrow C_2 + P}$ & $1.3\times10^{-17}$ & 0.0 & 375.0 & \cite{mcel13} \\
 R168 (CR) & $\rm{CCP + CRPHOT \rightarrow CP + C}$ & $1.3\times10^{-17}$ & 0.0 & 375.0 & \cite{mcel13} \\
 R169 (IN) & $\rm{C^+ + CCP \rightarrow CCP^+ + C}$ & $5.0\times10^{-10}$ & $-0.5$ & 0.0 & \cite{mcel13} \\
 R170 (IN) & $\rm{C^+ + CCP \rightarrow CP^+ + C_2}$ & $5.0\times10^{-10}$ & $-0.5$ & 0.0 & \cite{mcel13} \\
 R171 (IN) & $\rm{H^+ + CCP \rightarrow CCP^+ + H}$ & $1.0\times10^{-9}$ & $-0.5$ & 0.0 & \cite{mcel13} \\
 R172 (PH) & $\rm{CCP + h\nu \rightarrow C_2 + P}$ & $1.0\times10^{-10}$ & 0.0 & 1.7 & \cite{mcel13} \\
 R173 (PH) & $\rm{CCP + h\nu \rightarrow CP + C}$ & $1.0\times10^{-9}$ & 0.0 & 1.7 & \cite{mcel13} \\
 R174 (IN) & $\rm{C^+ + HPO \rightarrow HPO^+ + C}$ & $1.0\times10^{-9}$ & $-0.5$ & 0.0 & \cite{mcel13} \\
 R175 (IN) & $\rm{H^+ + HPO \rightarrow HPO^+ + H}$ & $1.0\times10^{-9}$ & $-0.5$ & 0.0 & \cite{mcel13} \\
 R176 (IN) & $\rm{H_2O + P^+ \rightarrow HPO^+ + H}$ & $4.95\times10^{-10}$ & $-0.5$ & 0.0 & \cite{mcel13} \\
 R177 (IN) & $\rm{H_2O + PH^+ \rightarrow HPO^+ + H_2}$ & $7.44\times10^{-10}$ & $-0.5$ & 0.0 & \cite{mcel13} \\
 R178 (IN) & $\rm{H_3O^+ + P \rightarrow HPO^+ + H_2}$ & $1.00\times10^{-9}$ & 0.0 & 0.0 & \cite{mcel13} \\
 R179 (IN) & $\rm{H_2O + PH^+ \rightarrow H_2PO^+ + H}$ & $2.04\times10^{-10}$ & $-0.5$ & 0.0 & \cite{mcel13} \\
 R180 (IN) & $\rm{H_2O + PH^+ \rightarrow P + H_3O^+}$ & $2.52\times10^{-10}$ & $-0.5$ & 0.0 & \cite{mcel13} \\
 R181 (IN) & $\rm{H_2O + PH_2^+ \rightarrow H_2PO^+ + H_2}$ & $3.28\times10^{-10}$ & $-0.5$ & 0.0 & \cite{mcel13} \\
 R182 (IN) & $\rm{H_3^+ + HPO \rightarrow H_2PO^+ + H_2}$ & $1.00\times10^{-9}$ & $-0.5$ & 0.0 & \cite{mcel13} \\
 R183 (IN) & $\rm{H_3O^+ + HPO \rightarrow H_2PO^+ + H_2O}$ & $1.00\times10^{-9}$ & $-0.5$ & 0.0 & \cite{mcel13} \\
 R184 (IN) & $\rm{HCO^+ + HPO \rightarrow H_2PO^+ + CO}$ & $1.00\times10^{-9}$ & $-0.5$ & 0.0 & \cite{mcel13} \\
 R185 (IN) & $\rm{CH_3^+ + P \rightarrow PCH_2^+ + H}$ & $1.0\times10^{-9}$ & 0.0 & 0.0 & \cite{mcel13} \\
 R186 (IN) & $\rm{CH_4 + P^+ \rightarrow PCH_2^+ + H_2}$ & $9.6\times10^{-10}$ & 0.0 & 0.0 & \cite{mcel13} \\
 R187 (IN) & $\rm{H_2 + HCP^+ \rightarrow PCH_2^+ + H}$ & $1.00\times10^{-9}$ & 0.0 & 0.0 & \cite{mcel13} \\
 R188 (DR) & $\rm{PCH_3^+ + e^- \rightarrow CP + H_2 + H}$ & $1.5\times10^{-7}$ & $-0.5$ & 0.0 & \cite{mcel13} \\
 R189 (DR) & $\rm{PCH_3^+ + e^- \rightarrow HCP + H_2}$ & $1.5\times10^{-7}$ & $-0.5$ & 0.0 & \cite{mcel13} \\
 R190 (DR) & $\rm{PCH_3^+ + e^- \rightarrow P + CH_3}$ & $3.0\times10^{-7}$ & $-0.5$ & 0.0 & \cite{mcel13} \\
 R191 (IN) & $\rm{CH_4 + PH^+ \rightarrow PCH_3^+ + H_2}$ & $1.05\times10^{-9}$ & 0.0 & 0.0 & \cite{mcel13} \\
 R192 (IN) & $\rm{H^+ + CH_2PH \rightarrow PCH_3^+ + H}$ & $1.00\times10^{-9}$ & 0.0 & 0.0 & \cite{mcel13} \\
 R193 (DR) & $\rm{PCH_4^+ + e^- \rightarrow CH_2PH + H_2 + H}$ & $1.5\times10^{-7}$ & $-0.5$ & 0.0 & \cite{mcel13} \\
 R194 (DR) & $\rm{PCH_4^+ + e^- \rightarrow HCP + H_2 + H}$ & $1.5\times10^{-7}$ & $-0.5$ & 0.0 & \cite{mcel13} \\
 R195 (DR) & $\rm{PCH_4^+ + e^- \rightarrow P + CH_4}$ & $3.0\times10^{-7}$ & $-0.5$ & 0.0 & \cite{mcel13} \\
 R196 (IN) & $\rm{CH_4 + PH^+ \rightarrow PCH_4^+ + H}$ & $5.5\times10^{-11}$ & 0.0 & 0.0 & \cite{mcel13} \\
 R197 (IN) & $\rm{CH_4 + PH_2^+ \rightarrow PCH_4^+ + H_2}$ & $1.1\times10^{-9}$ & 0.0 & 0.0 & \cite{mcel13} \\
 R198 (IN) & $\rm{H_3^+ + CH_2PH \rightarrow PCH_4^+ + H_2}$ & $1.00\times10^{-9}$ & 0.0 & 0.0 & \cite{mcel13} \\
 R199 (IN) & $\rm{H_3O^+ + CH_2PH \rightarrow PCH_4^+ + H_2O}$ & $1.00\times10^{-9}$ & 0.0 & 0.0 & \cite{mcel13} \\
 R200 (IN) & $\rm{HCO^+ + CH_2PH \rightarrow PCH_4^+ + CO}$ & $1.00\times10^{-9}$ & 0.0 & 0.0 & \cite{mcel13} \\
 R201 (IN) & $\rm{H^+ + HC_2P \rightarrow HC_2P^+ + H}$ & $1.00\times10^{-9}$ & 0.0 & 0.0 & \cite{mcel13} \\
 R202 (CR) & $\rm{HC_2P + CRPHOT \rightarrow CCP + H}$ & $1.3\times10^{-17}$ & 0.0 & 750.0 & \cite{mcel13} \\
 R203 (DR) & $\rm{PC_2H_2^+ + e^- \rightarrow HC_2P + H}$ & $1.0\times10^{-7}$ & $-0.5$ & 0.0 & \cite{mcel13} \\
 R204 (DR) & $\rm{PC_2H_3^+ + e^- \rightarrow HC_2P + H_2}$ & $1.0\times10^{-7}$ & $-0.5$ & 0.0 & \cite{mcel13} \\
 R205 (DR) & $\rm{PC_2H_4^+ + e^- \rightarrow HC_2P + H_2 + H}$ & $1.0\times10^{-7}$ & $-0.5$ & 0.0 & \cite{mcel13} \\
 R206 (DR) & $\rm{PC_3H^+ + e^- \rightarrow HC_2P + C}$ & $1.0\times10^{-7}$ & $-0.5$ & 0.0 & \cite{mcel13} \\
 R207 (IN) & $\rm{C^+ + HC_2P \rightarrow CCP^+ + CH}$ & $5.00\times10^{-10}$ & 0.0 & 0.0 & \cite{mcel13} \\
 R208 (IN) & $\rm{C^+ + HC_2P \rightarrow CP^+ + C_2H}$ & $5.00\times10^{-10}$ & 0.0 & 0.0 & \cite{mcel13} \\
 R209 (IN) & $\rm{H_3^+ + HC_2P \rightarrow PC_2H_2^+ + H_2}$ & $1.00\times10^{-9}$ & 0.0 & 0.0 & \cite{mcel13} \\
 R210 (IN) & $\rm{H_3O^+ + HC_2P \rightarrow PC_2H_2^+ + H_2O}$ & $1.00\times10^{-9}$ & 0.0 & 0.0 & \cite{mcel13} \\
 R211 (IN) & $\rm{HCO^+ + HC_2P \rightarrow PC_2H_2^+ + CO}$ & $1.00\times10^{-9}$ & 0.0 & 0.0 & \cite{mcel13} \\
 R212 (IN) & $\rm{He^+ + HC_2P \rightarrow CP^+ + CH + He}$ & $5.00\times10^{-10}$ & 0.0 & 0.0 & \cite{mcel13} \\
 R213 (IN) & $\rm{He^+ + HC_2P \rightarrow CP + CH^+ + He}$ & $5.00\times10^{-10}$ & 0.0 & 0.0 & \cite{mcel13} \\
 R214 (NN) & $\rm{O + HC_2P \rightarrow HCP + CO}$ & $4.00\times10^{-11}$ & 0.0 & 0.0 & \cite{mcel13} \\
 R215 (PH) & $\rm{HC_2P + h\nu \rightarrow CCP + H}$ & $5.48\times10^{-10}$ & 0.0 & 2.0 & \cite{mcel13} \\
 R216 (IN) & $\rm{H^+ + C_4P \rightarrow C_4P^+ + H}$ & $1.00\times10^{-9}$ & $-0.5$ & 0.0 & \cite{mcel13} \\
 R217 (CR) & $\rm{C_4P + CRPHOT \rightarrow C_3P + C}$ & $1.3\times10^{-17}$ & 0.0 & 750.0 & \cite{mcel13} \\
 R218 (DR) & $\rm{PC_4H^+ + e^- \rightarrow C_4P + H}$ & $7.50\times10^{-8}$ & $-0.5$ & 0.0 & \cite{mcel13} \\
 R219 (IN) & $\rm{H_3^+ + C_4P \rightarrow PC_4H^+ + H_2}$ & $1.00\times10^{-9}$ & $-0.5$ & 0.0 & \cite{mcel13} \\
 R220 (IN) & $\rm{H_3O^+ + C_4P \rightarrow PC_4H^+ + H_2O}$ & $1.00\times10^{-9}$ & $-0.5$ & 0.0 & \cite{mcel13} \\
 R221 (IN) & $\rm{HCO^+ + C_4P \rightarrow PC_4H^+ + CO}$ & $1.00\times10^{-9}$ & $-0.5$ & 0.0 & \cite{mcel13} \\
 R222 (IN) & $\rm{He^+ + C_4P \rightarrow CCP^+ + C_2 + He}$ & $5.00\times10^{-10}$ & $-0.5$ & 0.0 & \cite{mcel13} \\
 R223 (IN) & $\rm{He^+ + C_4P \rightarrow CCP + C_2^+ + He}$ & $5.00\times10^{-10}$ & $-0.5$ & 0.0 & \cite{mcel13} \\
 R224 (NN) & $\rm{O + C_4P \rightarrow C_3P + CO}$ & $1.00\times10^{-11}$ & 0.0 & 0.0 & \cite{mcel13} \\
 R225 (PH) & $\rm{C_4P + h\nu \rightarrow C_3 + CP}$ & $5.48\times10^{-10}$ & 0.0 & 1.7 & \cite{mcel13} \\
 R226 (CR) & $\rm{C_3P + CRPHOT \rightarrow CCP + C}$ & $1.3\times10^{-17}$ & 0.0 & 750.0 & \cite{mcel13} \\
 R227 (DR) & $\rm{C_4P^+ + e^- \rightarrow C_3P + C}$ & $1.50\times10^{-7}$ & $-0.5$ & 0.0 & \cite{mcel13} \\
 R228 (DR) & $\rm{PC_3H^+ + e^- \rightarrow C_3P + H}$ & $1.00\times10^{-7}$ & $-0.5$ & 0.0 & \cite{mcel13} \\
 R229 (DR) & $\rm{PC_4H^+ + e^- \rightarrow C_3P + CH}$ & $7.50\times10^{-8}$ & $-0.5$ & 0.0 & \cite{mcel13} \\
 R230 (IN) & $\rm{H_3^+ + C_3P \rightarrow PC_3H^+ + H_2}$ & $1.00\times10^{-9}$ & $-0.5$ & 0.0 & \cite{mcel13} \\
 R231 (IN) & $\rm{H_3O^+ + C_3P \rightarrow PC_3H^+ + H_2O}$ & $1.00\times10^{-9}$ & $-0.5$ & 0.0 & \cite{mcel13} \\
 R232 (IN) & $\rm{HCO^+ + C_3P \rightarrow PC_3H^+ + CO}$ & $1.00\times10^{-9}$ & $-0.5$ & 0.0 & \cite{mcel13} \\
 R233 (IN) & $\rm{He^+ + C_3P \rightarrow C_3^+ + P + He}$ & $5.00\times10^{-10}$ & $-0.5$ & 0.0 & \cite{mcel13} \\
 R234 (IN) & $\rm{He^+ + C_3P \rightarrow C_3 + P^+ + He}$ & $5.00\times10^{-10}$ & $-0.5$ & 0.0 & \cite{mcel13} \\
 R235 (NN) & $\rm{O + C_3P \rightarrow CCP + CO}$ & $4.00\times10^{-11}$ & 0.0 & 0.0 & \cite{mcel13} \\
 R236 (PH) & $\rm{C_3P + h\nu \rightarrow C_2 + CP}$ & $5.00\times10^{-10}$ & 0.0 & 1.8 & \cite{mcel13} \\
 R237 (CR) & $\rm{CH_2PH + CRPHOT \rightarrow HCP + H_2}$ & $1.3\times10^{-17}$ & 0.0 & 750.0 & \cite{mcel13} \\
 R238 (IN) & $\rm{C^+ + CH_2PH \rightarrow HC_2P^+ + H_2}$ & $1.00\times10^{-9}$ & 0.0 & 0.0 & \cite{mcel13} \\
 R239 (IN) & $\rm{He^+ + CH_2PH \rightarrow PH^+ + CH_2 + He}$ & $5.00\times10^{-10}$ & 0.0 & 0.0 & \cite{mcel13} \\
 R240 (IN) & $\rm{He^+ + CH_2PH \rightarrow PH + CH_2^+ + He}$ & $5.00\times10^{-10}$ & 0.0 & 0.0 & \cite{mcel13} \\
 R241 (NN) & $\rm{O + CH_2PH \rightarrow PH_2 + CO + H}$ & $4.00\times10^{-11}$ & 0.0 & 0.0 & \cite{mcel13} \\
 R242 (PH) & $\rm{CH_2PH + h\nu \rightarrow CH_2 + PH}$ & $9.54\times10^{-10}$ & 0.0 & 1.8 & \cite{mcel13} \\
 R243 (DR) & $\rm{PC_2H_2^+ + e^- \rightarrow P + C_2H_2}$ & $1.00\times10^{-7}$ & $-0.5$ & 0.0 & \cite{mcel13} \\
 R244 (IN) & $\rm{C_2H_2 + PH^+ \rightarrow PC_2H_2^+ + H}$ & $1.30\times10^{-9}$ & 0.0 & 0.0 & \cite{mcel13} \\
 R245 (IN) & $\rm{C_2H_2 + PH_2^+ \rightarrow PC_2H_2^+ + H_2}$ & $1.40\times10^{-9}$ & 0.0 & 0.0 & \cite{mcel13} \\
 R246 (IN) & $\rm{C_2H_2 + PH_3^+ \rightarrow PC_2H_3^+ + H_2}$ & $5.80\times10^{-10}$ & 0.0 & 0.0 & \cite{mcel13} \\
 \hline
\multicolumn{6}{c}{Ice-phase/grain-surface formation pathways of PH$_3$} \\
 \hline
 R1 & $\rm{gH + gP \rightarrow gPH}$ & - & - & - & \cite{chan20} \\
 R2 & $\rm{gH + gPH \rightarrow gPH_2}$ & - & - & - & \cite{chan20} \\
 R3 & $\rm{gH + gPH_2 \rightarrow gPH_3}$ & - & - & - & \cite{chan20} \\
 R4 & $\rm{gH + gPH_3 \rightarrow gPH_2 + gH_2}$ & - & - & - & This work, \cite{sous20} \\
 R5 & $\rm{gOH + gPH_3 \rightarrow gPH_2 + gH_2O}$ & - & - & - & This work, \cite{sous20} \\
 R6 & $\rm{gPH_3 \rightarrow PH_3}$ & - & - & - & \cite{chan20} \\
\enddata
\tablecomments{
CR refers to cosmic-rays, IN to ion-neutral reactions, NR to neutral-radical reactions, NN to neutral-neutral reactions, RR to radical-radical reactions, RA to radiative association reactions, ER to electronic recombination reactions for atomic ions, DR to dissociative recombination reactions for molecular ions, PH to photodissociation reactions, h$\nu$ to a photon. \\
Grain surface species are denoted by the letter ``g".}
\end{deluxetable*}

\clearpage
\restartappendixnumbering

\section{Estimated intensities from the radiative transfer model} \label{sec:append_b}
Here, we note the intensities of various P-bearing species in the diffuse and hot core regions. We also highlight the observability of these transitions with IRAM 30m, Herschel (not operational), and SOFIA.
We use the ATRAN program\footnote{\url{https://atran.arc.nasa.gov/cgi-bin/atran/atran.cgi}} to check the atmospheric transmission around these frequencies for the ground-based observation as well as space-based observations.\\

\begin{figure*}
\centering
\includegraphics[width=12cm, height=16cm, angle=270]{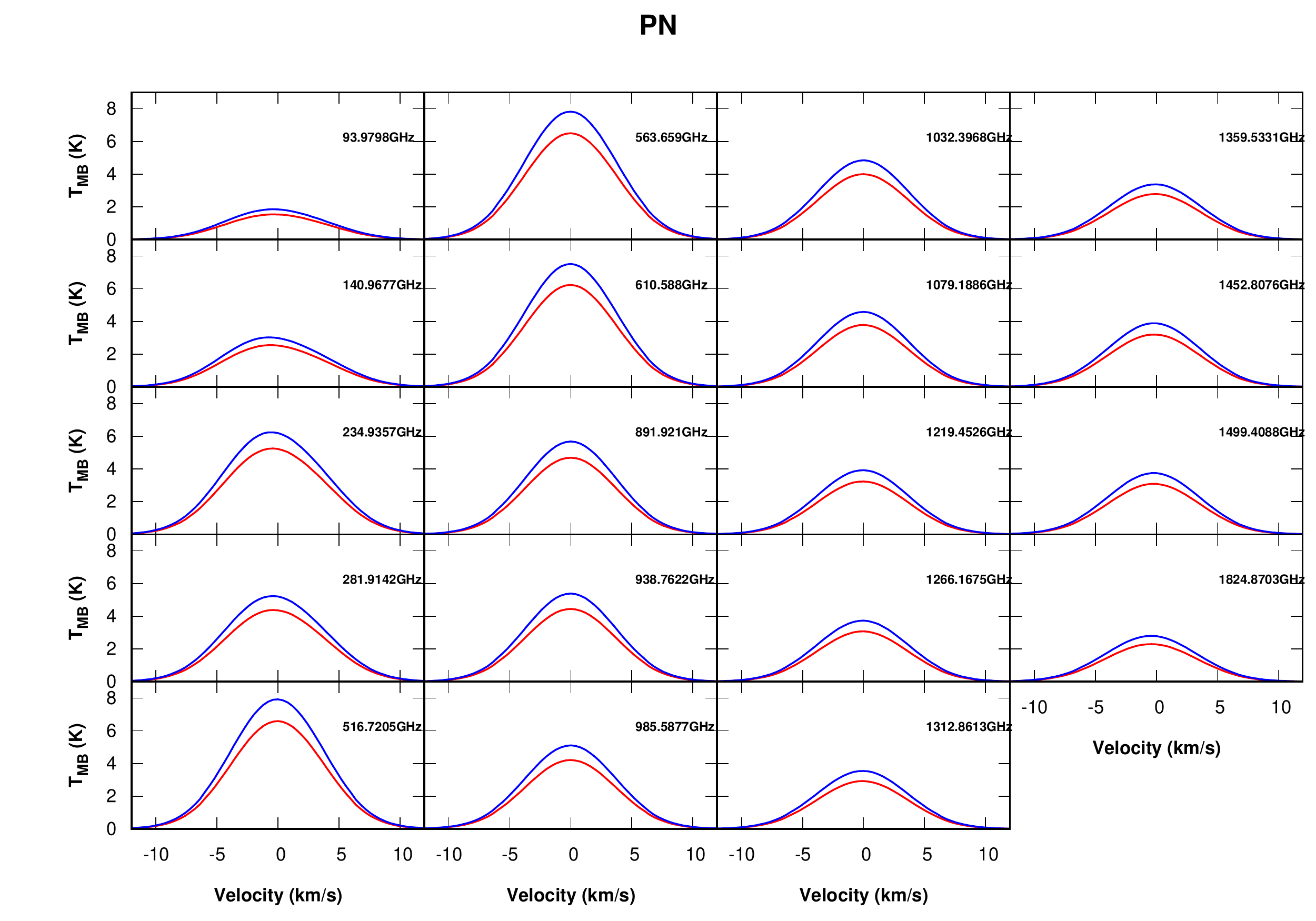}
\caption{Table \ref{table:PN_observation} shows possible transitions of PN, which could be observed with the IRAM-30m or SOFIA (GREAT) toward hot core/corino. The estimated line profile of these transitions obatined with the RATRAN model is shown here for hot core (blue) and hot corino (red) using the abundances from Table \ref{tab:abundances}.}
\label{fig:PN}
\end{figure*}

\begin{figure*}
\centering
\includegraphics[width=12cm, height=16cm, angle=270]{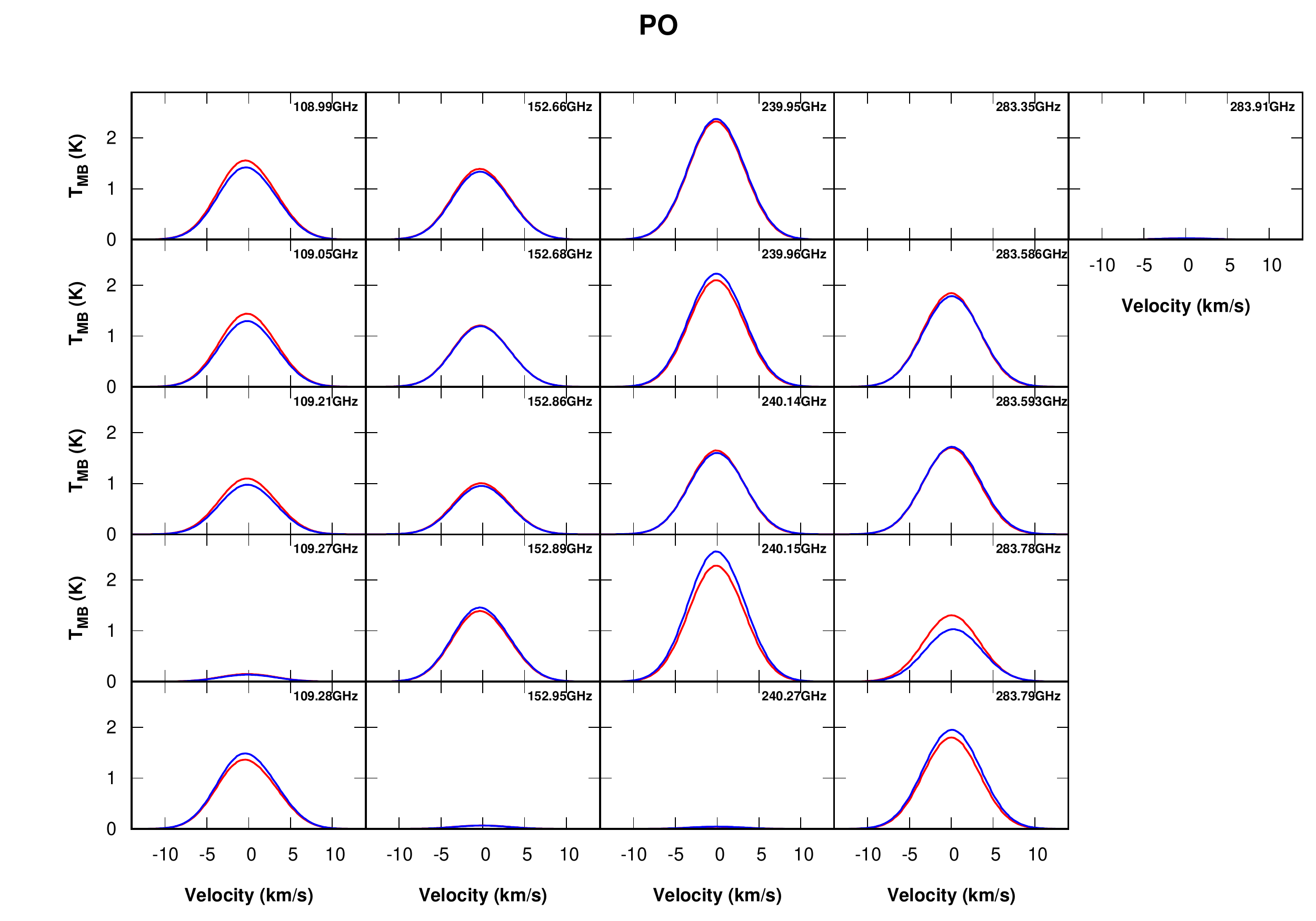}
\caption{Table \ref{table:PO_observation} shows the possible transitions of PO, which could be observed with the IRAM-30m toward Sgr-B2(M). The estimated line profile of these transitions with the RATRAN model is shown here for hot core (blue) and hot corino (red) using the abundances from Table \ref{tab:abundances}.
\label{fig:PO}}
\end{figure*}

\begin{figure*}
\centering
\includegraphics[width=10cm, height=14cm, angle=270]{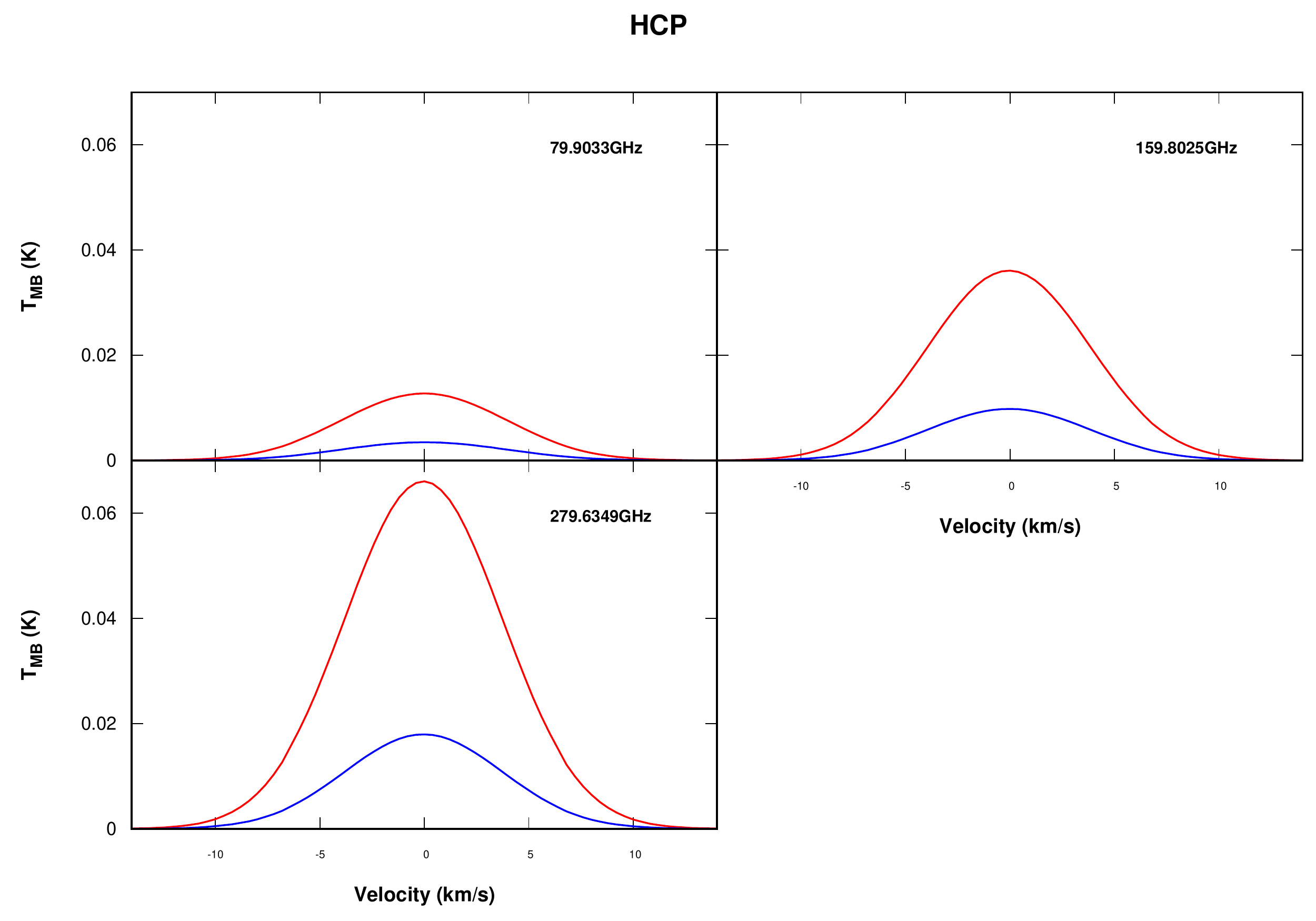}
\caption{Table \ref{table:HCP_observation} shows the possible transitions of HCP, which could be observed with the IRAM-30m toward Sgr-B2(M). The estimated line profile of these transitions is shown here for Hot core (blue) and Hot corino (red) using the abundances from Table \ref{tab:abundances}.}
\label{fig:HCP}
\end{figure*}

\begin{figure*}
\centering
\includegraphics[width=10cm, height=16cm, angle=270]{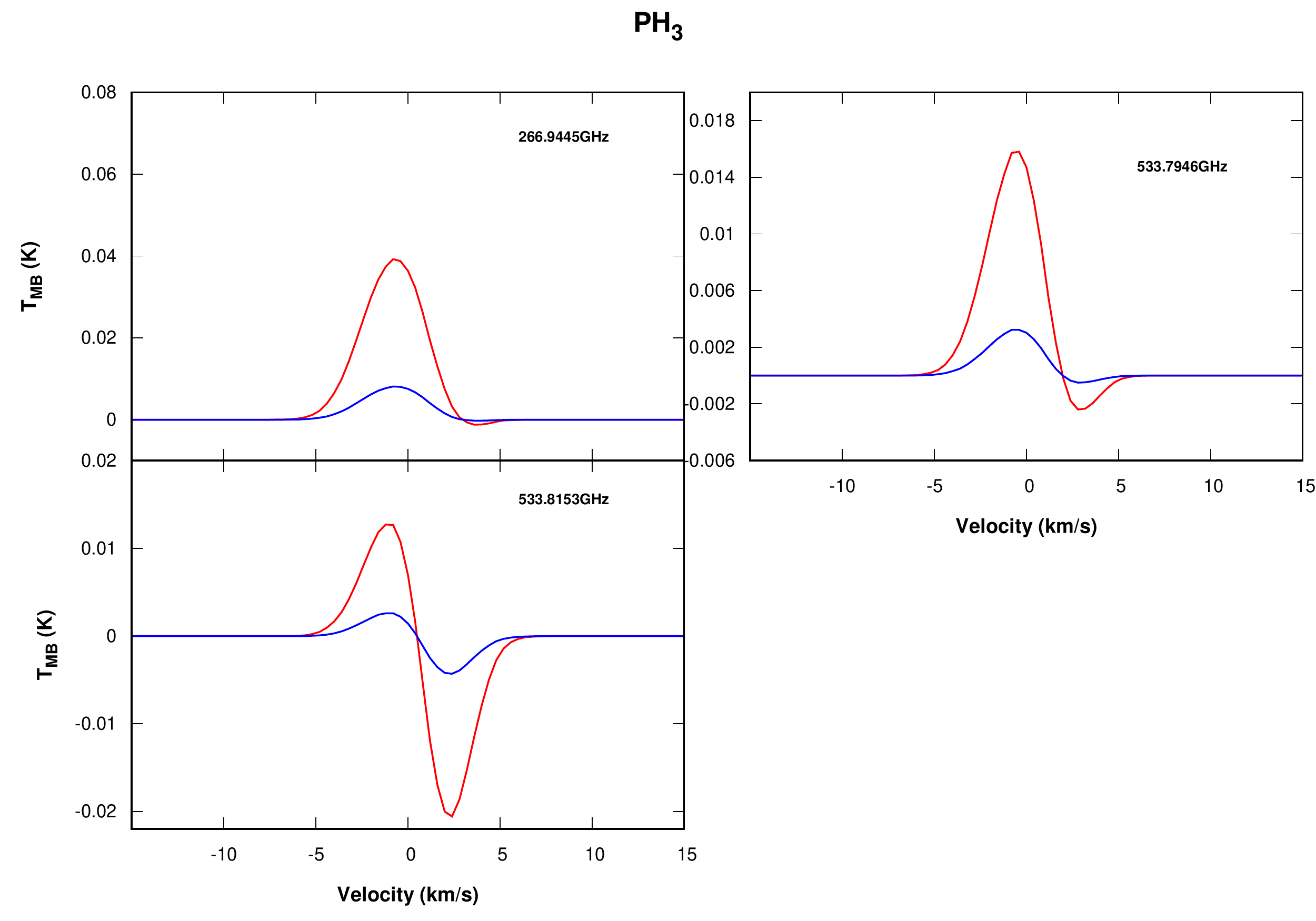}
\caption{Table \ref{table:PH3_observation} shows the possible transitions of PH$_3$, which could be observed with the IRAM-30m and SOFIA (GREAT) toward Sgr-B2(M). The estimated line profile of these transitions with the RATRAN model is shown here for hot core (blue) and hot corino (red) using the abundances from Table \ref{tab:abundances}.}
\label{fig:PH3}
\end{figure*}

\begin{deluxetable*}{ccccccccc}
 \caption{The telescopic parameters (IRAM and SOFIA) of some transitions of PN obtained toward a hot core/corino  or diffuse cloud regions obtained with the RATRAN model are shown. The footnote shows the adopted parameters for these calculations. \label{table:PN_observation}}
 \tablewidth{0pt}
 \tabletypesize{\scriptsize}
 \tablehead{\colhead{Frequency}&\colhead{Telescope}&\colhead{Beam size}&\colhead{Atmospheric}&\colhead{T$_{MB}$(K)}&\colhead{Integration time}&\colhead{T$_{MB}$(K)}&\colhead{Integration time}\\
 \colhead{(GHz)}&\colhead{}&\colhead{($^{''}$)}&\colhead{transmission$^a$}&\colhead{(Diffuse)}&\colhead{(Diffuse$^b$)}&\colhead{Hot core (Hot corino)}&\colhead{Hot core$^c$ (Hot corino$^c$)}}
 \startdata
 $^\star$93.9798 & IRAM-30m & 26.18 &0.7&$-0.0134$&4.7 hr&1.85 (1.54)&11.3 min (11.3 min)\\
 $^\star$140.9677&IRAM-30m & 17.45 &0.65&$-0.0337$&47.4 min&3.03 (2.55)&11.9 min (11.9 min)\\
 234.9357 & IRAM-30m & 10.47 &0.4&$-0.00216$&high&6.23 (5.24)&25.6 min (25.6 min)\\
 $^\star$281.9142 & IRAM-30m & 8.73 &0.35&0.0038&high&5.23 (4.38)&46.6 min (46.6 min)\\
 \hline
  516.7205&SOFIA-GREAT&18.2&0.97&0.0032&high&7.90 (6.58)&0.6sec (0.9 sec)\\
  563.6590&SOFIA-GREAT&18.2&0.72&0.0024&high&7.80 (6.48)&2.3 sec (3.3 sec)\\
  610.5880&SOFIA-GREAT&18.2&0.94&0.0019&high&7.51 (6.23)&0.8 sec (1.2 sec)\\
 891.9210&SOFIA-GREAT&18.2&0.99&0.0008&high&5.67 (4.68)&2.2 sec (3.2 sec)\\
 938.7622&SOFIA-GREAT&18.2&0.98&0.0008&high&5.37 (4.44)&2.4 sec (3.4 sec)\\
 985.5877&SOFIA-GREAT&18.2&0.17&0.0007&high&5.10 (4.20)&140.6 sec(207.4 sec)\\
 1032.3968&SOFIA-GREAT&18.2&0.7&0.0007&high&4.84 (4.00)&3.4 sec(5.0 sec)\\
 1079.1886&SOFIA-GREAT&18.2&0.81&0.0005&high&4.59 (3.78)&6.5 sec (9.6 sec)\\
 1219.4526&SOFIA-GREAT&18.2&0.71&0.0004&high&3.91 (3.23)&88.1 sec (129.1 sec)\\
 1266.1675&SOFIA-GREAT&18.2&0.95&0.0004&high&3.72 (3.07)&47.8 sec(70.2 sec)\\
 1312.8613&SOFIA-GREAT&18.2&0.88&0.0003&high&3.54 (2.92)&62.5 sec(91.8 sec)\\
 1359.5331&SOFIA-GREAT&18.2&0.96&0.0004&high&3.37 (2.78)&55.1 sec (81.0 sec)\\
 1452.8076&SOFIA-GREAT&14&0.96&0.0002&high&3.89 (3.20)&37.3 sec(55.2 sec)\\
 1499.4088&SOFIA-GREAT&14&0.98&0.0002&high&3.74 (3.08)&37.9 sec(55.9 sec)\\
 1824.8703&SOFIA-GREAT&14&0.93&0.0001&high&2.79 (2.29)&52.5 sec (77.9 sec)\\
  \enddata
 \tablecomments{$^a$ ATRAN (\url{https://atran.arc.nasa.gov/cgi-bin/atran/atran.cgi}) program is used for the calculation of the atmospheric transmission.\\ 
$^b$ For the diffuse cloud region, we use a frequency resolution of $1$ km/s and a signal to noise ratio $\geq$ 3 with IRAM-30m and a frequency resolution of 1 km/s and a signal to noise ratio $3$ with SOFIA.\\
$^c$ For the hot core/hot corino region, we use a frequency resolution of 1 km/s with a signal to noise ratio $10$ with SOFIA and a frequency resolution of 1 km/s and   a signal to noise ratio $\geq$ 3 with IRAM 30m.\\
$^\star$Observed by \cite{mini18}}
\end{deluxetable*}

 \begin{deluxetable*}{ccccccccc}
 \tablecaption{The telescopic parameters (IRAM and SOFIA) of some transitions of PO toward hot core/corino or diffuse cloud regions obtained with the RATRAN model. The footnote shows the adopted parameters for these calculations. \label{table:PO_observation}}
 \tablewidth{0pt}
 \tabletypesize{\scriptsize}
 \tablehead{\colhead{Frequency}&\colhead{Telescope}&\colhead{Beam size}& \colhead{Atmospheric} & \colhead{T$_{MB}$} & \colhead{Integration time$^b$} & \colhead{T$_{MB}$(K)} & \colhead{Integration time$^c$} \\
 \colhead{(GHz)}&\colhead{}&\colhead{($^{''}$)}&\colhead{transmission$^a$}&\colhead{(Diffuse)}&\colhead{(Diffuse)}&\colhead{Hot core (Hot corino)}&\colhead{Hot core (Hot corino)}} 
\startdata
108.7072&IRAM-30m&22.63&0.65&$-0.02793$&1.3 hr&\nodata (\nodata)& \nodata (\nodata)\\
 108.9984&IRAM-30m&22.57&0.65&$-8.32\times10^{-4}$&high&1.4164 (1.5525)& 12.6 min (12.6 min)\\
 109.0454&IRAM-30m&22.56&0.65&$-6.81\times10^{-4}$&high&1.295 (1.4385)& 12.6 min (12.6min)\\
 109.2062&IRAM-30m&22.53&0.64&$-3.41\times10^{-4}$&high&0.9740 (1.0954)& 12.8 min (12.8 min)\\
 109.2714&IRAM-30m&22.51&0.65&$-1.2\times 10^{-4}$&high&0.1358 (0.1490)&12.8 min (12.8 min)\\
 109.2812&IRAM-30m&22.51&0.64&$-1.21\times10^{-3}$&high&1.3576 (1.48)& 12.8 min (12.8 min)\\
 152.3891&IRAM-30m&16.14&0.6&$-0.0213$&2.3 hr&1.4532 (1.3870)& 12.5 min (12.5 min)\\
 $^\star$152.6570&IRAM-30m&16.11&0.6&$-4.11\times10^{-4}$&high&1.3359 (1.3907)&12.5 min (12.5 min)\\
 $^\star$152.6803&IRAM-30m&16.11&0.6&$-3.71\times10^{-4}$&high&1.1911 (1.2072)& 12.5 min (12.5 min)\\
 $^\star$152.8555&IRAM-30m&16.09&0.61&$-1.41\times10^{-4}$&high&0.9519 (1.0044)& 12.5 min (12.5 min)\\
 $^\star$152.8881&IRAM-30m&16.09&0.61&$-6.01\times10^{-4}$&high&1.4532 (1.3870)&12.5 min (12.5 min)\\
 152.9532&IRAM-30m&16.08&0.61&$-2.01\times10^{-5}$&high&0.0655 (0.0671)&12.5 min (12.5 min)\\
 239.7043&IRAM-30m&10.26&0.01&$-0.0392$&2.7 hr&\nodata (\nodata)&\nodata (\nodata)\\
 239.9490&IRAM-30m&10.25&0.02&$-1.3\times 10^{-4}$&high&2.3684 (2.3166)& 25.3min (25.3 min)\\
 239.9581&IRAM-30m&10.25&0.01&$-1.00\times10^{-4}$&high&2.2247 (2.0924)&25.3 min (25.3 min)\\
 240.1411&IRAM-30m&10.24&0.01&$-3.00\times 10^{-5}$&high&1.5976 (1.6473)&25.1 min (25.1 min)\\
 240.1525&IRAM-30m&10.24&0.01&$-1.3\times 10^{-4}$&high&2.5592 (2.2764)&25.1 min (25.1 min)\\
 240.2683&IRAM-30m&10.24&0.02&\nodata&\nodata&0.0450 (0.0418)&39.2 min (39.2 min)\\
 283.3487&IRAM-30m&8.68&0.33&$-0.0117$&high&\nodata (\nodata)&\nodata (\nodata)\\
 283.5868&IRAM-30m&8.67&0.32&$-2.01\times10^{-5}$&high&1.7843 (1.8441)&47.6 min (47.6 min)\\
 283.5932&IRAM-30m&8.67&0.32&$-3.00\times10^{-5}$&high&1.7176 (1.6969)& 47.6min (47.6min)\\
 283.7776&IRAM-30m&8.67&0.32&$-1.00\times10^{-5}$&high&1.0301 (1.3038)&47.7 min (47.7min)\\
 283.7854&IRAM-30m&8.67&0.33&$-4.01\times10^{-5}$&high&1.9453 (1.7935)&47.7 min (47.7min)\\
 283.9125&IRAM-30m&8.66&0.33&\nodata&\nodata&0.0245 (0.0233)&7.5 hr (7.5 hr)\\
 \enddata
 \tablecomments{$^a$ ATRAN (\url{https://atran.arc.nasa.gov/cgi-bin/atran/atran.cgi}) program is used for the calculation of the atmospheric transmission.\\ 
$^b$ For the diffuse cloud region, a frequency resolution of $1$ km/s and a signal to noise ratio$\geq$ 3 are used with IRAM-30m.\\
 $^c$ For the hot core/hot corino region,  a frequency resolution of 1 km/s and a signal to noise ratio$\geq$ 3  are used  with IRAM 30m.\\
 $^\star$Observed by \cite{font16}}
 \end{deluxetable*}

\begin{deluxetable*}{ccccccccc}
 \tablecaption{The telescopic parameters (IRAM and SOFIA) of some transitions of HCP toward hot core/corino  or diffuse cloud regions obtained with the RATRAN model. The footnote shows the adopted parameters for these calculations. \label{table:HCP_observation}}
 \tablewidth{0pt}
 \tabletypesize{\scriptsize}
 \tablehead{\colhead{Frequency}&\colhead{Telescope}&\colhead{ Beam size}&\colhead{Atmospheric}&\colhead{T$_{MB}$}&\colhead{Integration time$^c$}&\colhead{T$_{MB}$(K)}&\colhead{Integration time$^c$} \\
 \colhead{(GHz)}&\colhead{}&\colhead{($^{''}$)}&\colhead{transmission$^a$}&\colhead{(Diffuse)}&\colhead{(Diffuse)$^b$}&\colhead{Hot core(Hot corino)}&\colhead{Hot core(Hot corino)}}  
  \startdata
 79.9033 & IRAM-30m &30.79 &0.63&0.0005&high&0.0035 (0.0127)&high (6.1 hr)\\
 159.8025 & IRAM-30m &15.39 &0.55&-$3.01\times10^{-5}$&high&0.0098 (0.0359)&10.3 hr (55.6 min)\\
 279.6349 & IRAM-30m &8.80 &0.33&\nodata&\nodata&0.0179 (0.0658)&12.4 hr (46.5 min)\\
 \enddata
 \tablecomments{$^a$ ATRAN (\url{https://atran.arc.nasa.gov/cgi-bin/atran/atran.cgi}) program is used for the calculation of the atmospheric transmission.\\ 
$^c$ In this case, we use a frequency resolution of 1 km/s and a signal to noise ratio $\geq$ 3 with IRAM 30m.\\}
 \end{deluxetable*}
 
\begin{deluxetable*}{ccccccccc}
 \tablecaption{The telescopic parameters (IRAM and SOFIA) of some transitions of PH$_3$ toward hot core/corino  or diffuse cloud regions obtained with the RATRAN model. The footnote shows the adopted parameters for these calculations. \label{table:PH3_observation}}
 \tablewidth{0pt}
\tabletypesize{\scriptsize}
 \tablehead{\colhead{Frequency}&\colhead{Telescope}&\colhead{Beam size}&\colhead{Atmospheric}&\colhead{T$_{MB}$}&\colhead{Integration time}&\colhead{T$_{MB}$(K)}&\colhead{Integration time} \\
 \colhead{(GHz)}&\colhead{}&\colhead{($^{''}$)}&\colhead{transmission$^a$}&\colhead{(Diffuse)}&\colhead{(Diffuse)}&\colhead{Hot core/Hot core$^x$(Hot corino/Hot corino$^y$)}&\colhead{Hot core/Hot core$^x$ (Hot corino/Hot corino$^y$)}}
 \startdata
  266.9445&IRAM-30m&9.22&0.23&\nodata&\nodata&0.008/0.8634 (0.039/0.9395)&31.5 hr$^c$/32.0 min$^c$(127.8 min$^c$/ 32.0 min$^c$)\\
 \hline
 533.7946&SOFIA-GREAT&18.2&0.95&\nodata&\nodata&0.0032/0.3415 (0.0157/0.3702)&high$^l$/7.24 min$^m$(5.1 hr$^l$/6.16 min$^m$)\\
 533.8153&SOFIA-GREAT&18.2&0.95&\nodata&\nodata&0.003/0.2794 (0.0125/0.3029)&high$^l$/11.3 min$^m$(8.5 hr$^l$/9.6 min$^m$)\\
 \enddata
 \tablecomments{$^a$ ATRAN (\url{https://atran.arc.nasa.gov/cgi-bin/atran/atran.cgi}) program is used for the calculation of the atmospheric transmission.\\
 $^x$ second values are considering PH$_3$ abundance $n_m$ = $2.1 \times 10 ^{-10}$ for the case when initial P$^+$ abundance is $ 1.8 \times 10^{-9}$ mentioned in Table \ref{tab:abundances} not considering the destruction reactions.\\
$^y$ second values are considering PH$_3$ abundance $n_m$ = $2.3 \times 10 ^{-10}$ for the case when initial P$^+$ abundance is $ 5.6 \times 10^{-9}$ mentioned in Table \ref{tab:abundances} not considering the destruction reactions.\\
 $^c$ In this case, we use frequency resolution 1 km/s and a signal to noise ratio $\geq$ 3 with IRAM 30m.\\
 $^l$ In this case, we use a frequency resolution of 1 km/s with a signal to noise ratio $3$ with SOFIA. \\
 $^m$ In this case, we use a frequency resolution of 1 km/s with a signal to noise ratio $10$ with SOFIA.}
 \end{deluxetable*}

\clearpage
\restartappendixnumbering

\section{Radiative transfer model for the hot corino} \label{sec:append_c}

In Section \ref{sec:RATRAN}, we have used the spatial distribution of the physical input parameters for RATRAN, relevant for the hot core region. There, we have employed the same physical parameters to study the hot corino region's line profiles. To avoid any ambiguity for considering the same physical parameters for the low mass analog, we have used a spatial variation of the physical input parameters used in \cite{mott13} for low mass protostar IRAS4A. 

\begin{figure*}
\centering
\includegraphics[width=7cm, height=11cm]{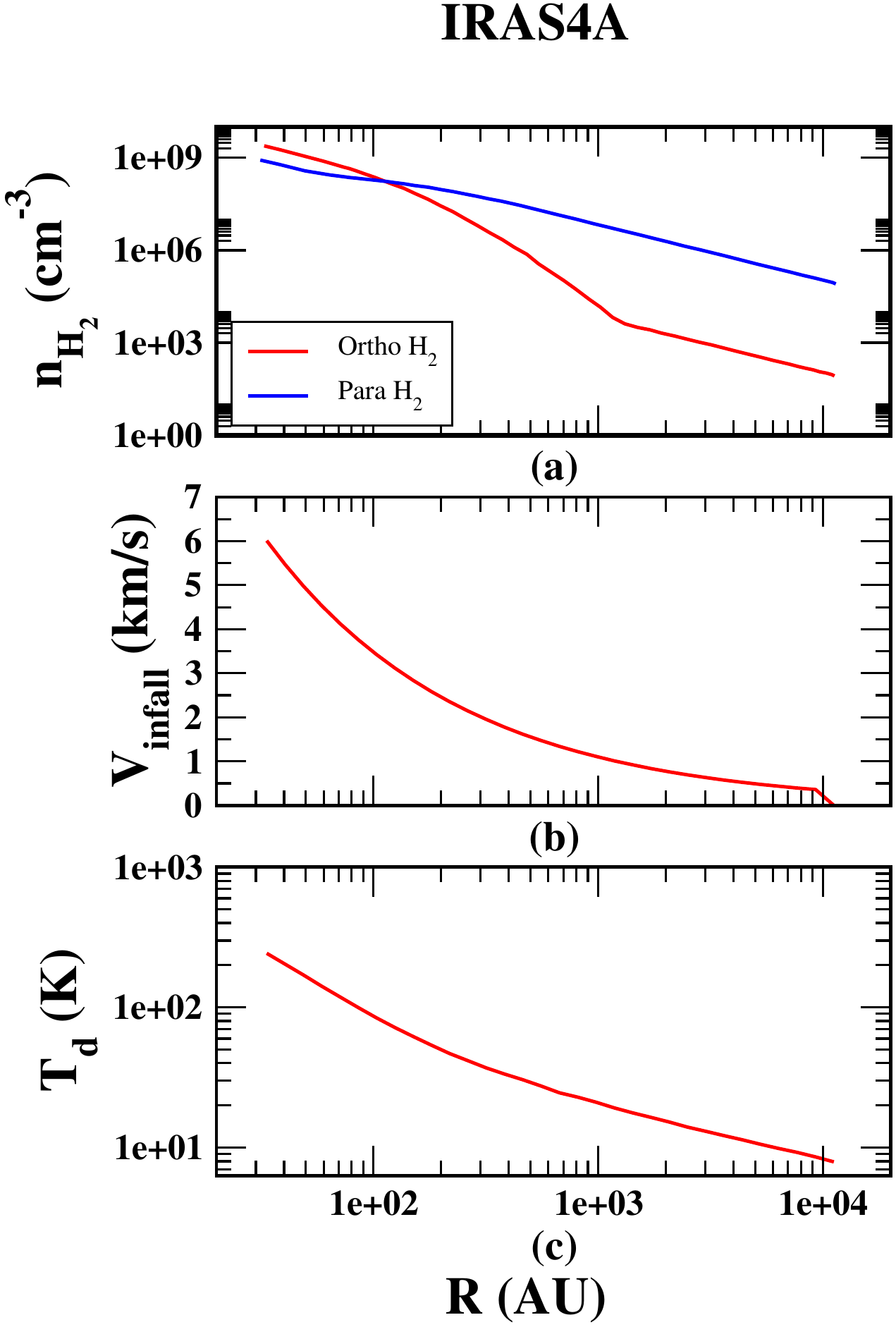}
\includegraphics[width=7cm, height=11cm]{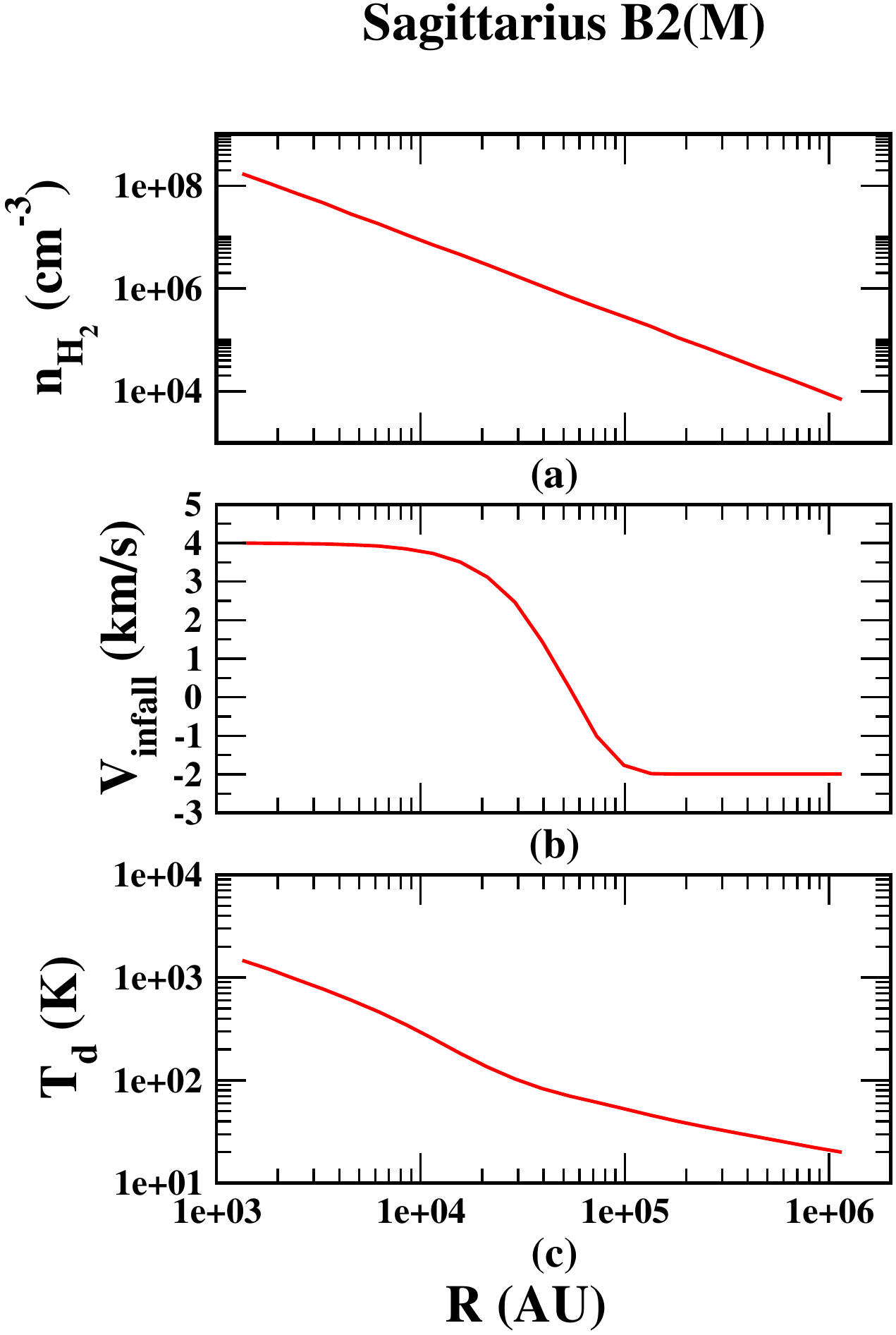}
\caption{Spatial distribution of physical parameters: (a) H$_2$ density, (b) In-fall velocity, and (c) Dust temperatures for hot corino (IRAS4A) and hot core (Sagittarius B2) are taken from \cite{mott13} and \cite{rolf10} respectively.
\label{fig:physical}}
\end{figure*}

\begin{figure*}
\centering
\includegraphics[width=8cm, height=10cm, angle=270]{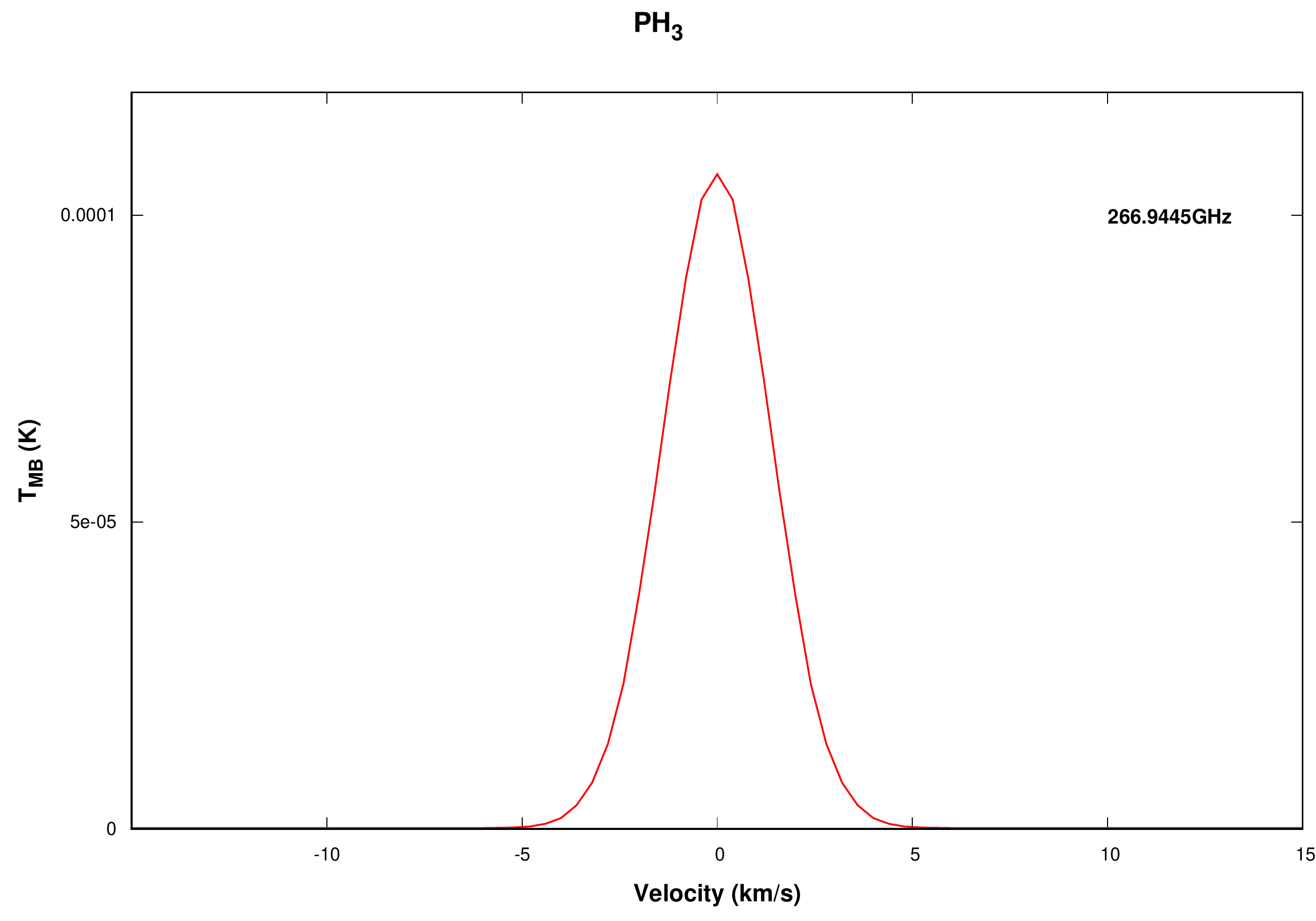}
\caption{The line profile of the transition of PH$_3$, which could be observed with the IRAM-30m toward IRAS4A is shown with the RATRAN model. Abundance of PH$_3$ is taken from the CMMC model of hot corino noted in Table \ref{tab:abundances}.
\label{fig:irasph3}}
\end{figure*}

In Figure \ref{fig:physical} the adopted spatial distribution of the physical input parameters are shown both for hot corino (IRAS4A, used here) and hot core (Sagittarius B2(M), used in section \ref{sec:RATRAN}). Along with these physical input parameters, we further have used the abundances obtained from our CMMC model (Table \ref{tab:abundances}) to obtain the main beam temperature of the P-bearing molecules (PN, PO, HCP, PH$_{3}$) in the hot corino region. 
Intensities of other transitions are convolved using the appropriate beam sizes for IRAM-30m telescope and SOFIA-GREAT instrument noted in Table \ref{tab:IRAS4Ap}. No inverse P-Cygni type or red-shifted absorption spectral feature (see Figure \ref{fig:irasph3}) is obtained for the $266.9445$ GHz transition of PH$_3$ molecule with this physical condition.

\begin{deluxetable*}{cccccccc}
 \tablecaption{The telescopic parameters (IRAM-30m and SOFIA) of some transitions of PN, PO, HCP, PH$_3$ toward hot corino IRAS4A. \label{tab:IRAS4Ap}}
 \tablewidth{0pt}
\tabletypesize{\scriptsize}
 \tablehead{\colhead{Molecule}&\colhead{Frequency}&\colhead{Telescope}&\colhead{Beam size}&\colhead{Atmospheric}&\colhead{T$_{MB}$}&\colhead{Integration time$^{b,c}$}\\
 \colhead{ }&\colhead{(GHz)}&\colhead{ }&\colhead{($^{''}$)}&\colhead{transmission$^a$}&\colhead{(K)}&\colhead{}}
 \startdata
 &$^{\star \star}$93.9798 & IRAM-30m & 26.18 &0.7 &$2.83\times10^{-4}$&high\\
 &$^{\star \star}$140.9677 & IRAM-30m & 17.45 &0.65& $1.75\times10^{-3}$&high\\
 &234.9357 & IRAM-30m & 10.47 &0.4& $9.96\times10^{-3}$&19.0 hr\\
 &$^{\star \star}$281.9142 & IRAM-30m & 8.73 &0.35&$1.44\times10^{-2}$&15.3 hr\\
 &563.6590 &SOFIA-GREAT& 26.18 &0.72 &$3.22\times10^{-3}$&high\\
 &610.5880&SOFIA-GREAT&18.2&0.94& $3.06\times10^{-3}$&high\\
 &891.9210&SOFIA-GREAT&18.2&0.99& $2.88\times10^{-3}$&high\\
 &938.7622&SOFIA-GREAT&18.2& 0.98&$2.10\times10^{-3}$&high\\
 PN&985.5877&SOFIA-GREAT&18.2&0.17&$2.03\times10^{-3}$&high\\
 &1032.3968&SOFIA-GREAT&18.2&0.7& $1.93\times10^{-3}$&high\\
 &1079.1886&SOFIA-GREAT&18.2&0.81& $1.88\times10^{-3}$&high\\
 &1219.4526&SOFIA-GREAT&18.2&0.71& $1.8\times10^{-3}$&high\\
 &1266.1675&SOFIA-GREAT&18.2&0.95& $1.63\times10^{-3}$&high\\
 &1312.8613&SOFIA-GREAT&18.2&0.88& $1.56\times10^{-3}$&high\\
 &1359.5331&SOFIA-GREAT&18.2&0.96& $1.55\times10^{-3}$&high\\
 &1452.8076&SOFIA-GREAT&14&0.96&$2.27\times10^{-3}$&high\\
 &1499.4088&SOFIA-GREAT&14&0.98&2.25$\times10^{-3}$&high\\
 &1824.8703&SOFIA-GREAT&14&0.93&$1.84\times10^{-3}$&high\\
 \hline
 &108.7072&IRAM-30m&22.63&0.65& $1.13\times10^{-5}$&high\\
 &108.9984&IRAM-30m&22.57&0.65& $5.1\times10^{-4}$&high\\
 &109.0454&IRAM-30m&22.56&0.65& $4.59\times10^{-4}$&high\\
 &109.2062&IRAM-30m&22.53& 0.64&$2.24\times10^{-4}$&high\\
 &109.2714&IRAM-30m&22.51&0.65& $4.89\times10^{-5}$&high\\
 &109.2812&IRAM-30m&22.51&0.64& $6.12\times10^{-4}$&high\\
 &152.3891&IRAM-30m&16.14&0.6& $5.01\times10^{-3}$&high\\
 &$^\star$152.6570&IRAM-30m&16.11&0.6 &$5.64\times10^{-4}$&high\\
 &$^\star$152.6803&IRAM-30m&16.11&0.6& $5.15\times10^{-4}$&high\\
 &$^\star$152.8555&IRAM-30m&16.09&0.61& $2.36\times10^{-4}$&high\\
 &$^\star$152.8881&IRAM-30m&16.09&0.61& $6.73\times10^{-4}$&high\\
 &152.9532&IRAM-30m&16.08&0.61& $2.59\times10^{-5}$&high\\
 PO&239.7043&IRAM-30m&10.26&0.01& $1.34\times10^{-2}$&10.6 hr\\
 &239.9490&IRAM-30m&10.25&0.02& $6.19\times10^{-4}$&high\\
 &239.9581&IRAM-30m&10.25&0.01& $6.02\times10^{-4}$&high\\
 &240.1411&IRAM-30m&10.24&0.01& $1.59\times10^{-4}$&high\\
 &240.1525&IRAM-30m&10.24&0.01 &$4.62\times10^{-4}$&high\\
 &240.2683&IRAM-30m&10.24&0.02&$7.17\times10^{-6}$&high\\
 &283.3487&IRAM-30m&8.68&0.03 &$1.51\times10^{-2}$&19.8 hr\\
 &283.5868&IRAM-30m&8.67&0.32&$7.97\times10^{-4}$&high\\
 &283.5932&IRAM-30m&8.67&0.32 &$7.85\times10^{-4}$&high\\
 &283.7776&IRAM-30m&8.67&0.32& $6.68\times10^{-4}$&high\\
 &283.7854&IRAM-30m&8.67&0.33& $9.04\times10^{-4}$&high\\
 &283.9125&IRAM-30m&8.66&0.33& $1.04\times10^{-5}$&high\\
 \hline
 &79.9033 & IRAM-30m &30.79 &0.63& $4.04\times10^{-6}$&high\\
 HCP&159.8025 & IRAM-30m &15.39&0.55& $1.68\times10^{-5}$&high\\
 &279.6349 & IRAM-30m &8.80&0.33& $7.51\times10^{-6}$&high\\
 \hline
  &266.9445&IRAM-30m&9.22&0.23& $1.06\times10^{-4}$&high \\
 PH$_3$ &533.7946&SOFIA-GREAT&18.2&0.95& $5.53\times10^{-6}$&high\\ 
 &533.8153&SOFIA-GREAT&18.2&0.95& $4.90\times10^{-6}$&high\\
 \enddata
 \tablecomments{$^a$ ATRAN (\url{https://atran.arc.nasa.gov/cgi-bin/atran/atran.cgi}) program is used for the calculation of the atmospheric transmission.\\ 
$^b$ We use a frequency resolution of $1$ km/s and a signal to noise ratio $\geq$ 3 with IRAM-30m.\\
$^c$ We use a frequency resolution of 1 km/s and a signal to noise ratio $3$ with SOFIA.\\
$^*$Observed by \cite{font16}.\\
$^{**}$ Observed by \cite{mini18}}
 \end{deluxetable*}

 \clearpage
\bibliography{P-AD}{}
\bibliographystyle{aasjournal}

\end{document}